\documentclass[reprint,prb,superscriptaddress,footinbib]{revtex4-1}
\usepackage{graphicx}
\usepackage{amsmath}
\usepackage{amssymb}
\usepackage{wasysym}
\usepackage{xcolor}
\usepackage[abs]{overpic}
\newcommand{\ket}[1]{\left|{#1}\right\rangle}
\newcommand{\bra}[1]{\left\langle{#1}\right|}

\newcommand{\be}{\begin{eqnarray*}}
\newcommand{\ee}{\end{eqnarray*}}
\newcommand{\h}{\hat{H}}
\newcommand{\w}{\omega}
\newcommand{\an}[1]{\left(a\right)^{#1}}
\newcommand{\ani}[1]{\left(a_i^{\phantom{\dagger}}\right)^{#1}}
\newcommand{\anj}[1]{\left(a_j^{\phantom{\dagger}}\right)^{#1}}
\newcommand{\ad}[1]{\left(a^\dagger\right)^{#1}}

\newcommand{\adi}[1]{\left(a_i^\dagger\right)^{#1}}
\newcommand{\adj}[1]{\left(a_j^\dagger\right)^{#1}}

\newcommand{\adon}[1]{\left(a_1^\dagger\right)^{#1}}
\newcommand{\adtw}[1]{\left(a_2^\dagger\right)^{#1}}

\newcommand{\zup}[1]{\left(\hat{Z}^\dagger\right)^{#1}}
\newcommand{\yup}[1]{\left(\hat{Y}^\dagger\right)^{#1}}
\newcommand{\pion}[1]{\left(\frac{\pi}{N}\right)^{#1}}
\newcommand{\pionn}{\left(\frac{\pi}{N}\right)}

\newcommand{\bk}{\mathbf{k}}

\newcommand{\e}{\epsilon}

\begin{document}
\title{Perturbative Approach to Flat Chern Bands in the Hofstadter Model}
\author{Fenner Harper}
\author{Steven H. Simon}
\affiliation{Rudolf Peierls Centre for Theoretical Physics, University of Oxford, Oxford, United Kingdom}
\author{Rahul Roy}
\affiliation{Department of Physics and Astronomy, University of California, Los Angeles, California USA}
\date{\today}
\begin{abstract}
We present a perturbative approach to the study of the Hofstadter model for when the amount of flux per plaquette is close to a rational fraction. Within this approximation, certain eigenstates of the system are shown to be multi-component wavefunctions that connect smoothly to the Landau levels of the continuum. The perturbative corrections to these are higher Landau level contributions that break rotational invariance and allow the perturbed states to adopt the symmetry of the lattice. In the presence of interactions, this approach allows for the calculation of generalised Haldane pseudopotentials, and in turn, the many-body properties of the system. The method is sufficiently general that it can apply to a wide variety of lattices, interactions, and magnetic field strengths.
\end{abstract}
\maketitle
\section{Introduction}
The fractional (FQHE) and integer quantum Hall effects (IQHE) are striking examples of topological phases, and continue to receive a great deal of interest many years on from their discovery.\cite{Klitzing:428079,Tsui:1982yy} As their defining feature, quantum Hall systems exhibit vanishing longitudinal conductance and a precisely quantised transverse conductance given by
\begin{equation}
\sigma_H=\nu \frac{e^2}{h},\label{hall}
\end{equation} 
where $\nu$ is an integer (IQHE) or fraction (FQHE). Whilst the IQHE may be understood intuitively at a single-particle level, the FQHE is inherently a many-body effect that arises as a result of strong interactions between degenerate electrons in a partially filled Landau level.\cite{prange1987quantum} The FQHE is particularly notable for its quasiparticle excitations, which may have fractional charge and obey fractional or non-Abelian statistics.\cite{Kitaev:2003ul,Nayak:2008dp}

The ordinary quantum Hall effects can be described by continuum theories: lattice effects may be neglected at low enough densities when the lattice spacing is much smaller than the characteristic (magnetic) length scale of the wavefunctions. For example, typical magnetic field values\cite{Tsui:1982yy} of 10--15~T correspond to a magnetic length on the order of $10^{-8}$~m, which is far greater than the unit cell size in GaAs of $5.65$~\AA.\cite{BLAKEMORE:1982vy} The question naturally arises whether quantum Hall physics persists in the presence of a strong perturbation from the lattice---in particular, when there is a large magnetic flux per lattice plaquette. There have been many numerical and analytical studies of lattice quantum Hall systems over the past decades which have answered this question positively, at least in certain cases.\cite{Hofstadter:1976wt,Sorensen:2005bt,Moller:2009ir,Kol:1993wv,Kohmoto:1989vd,Thouless:1982wi,Assaad:1995wu,Hormozi:2012tn,Palmer:2006km,Hafezi:2007gz,Kjall:2012db}

In addition to these, many other lattice models---such as the Haldane model \cite{Haldane:1988uf}---have been developed that can lead to a quantised Hall conductivity even in the absence of a net external field. There is little physical difference between these models and those with an external field (since magnetic flux per plaquette is only defined modulo $2\pi$), and indeed, in some cases, both types of system have been shown to be adiabatically connected.\cite{Scaffidi:2012dx,Wu:2012ky}

When an electronic band is completely filled, the system has zero longitudinal conductance, and is hence an insulator.  However, in cases such as the Haldane model or Landau levels, such filled band systems can have nonzero quantised Hall conductance, and are then sometimes known as Hall insulators, or more frequently Chern insulators. The bands carrying the Hall current are correspondingly known as Chern bands.  Just as with Landau levels, one can consider what happens when a Chern band is fractionally filled with interacting particles.  If such a system has a quantised Hall conductance, it is then known as a fractional Chern insulator (FCI) in analogy to the FQHE.\cite{Kapit:2010ky,Neupert:2011db,Sun:2011dk,Sheng:2011iv,Parameswaran:2013pca,Bergholtz:2013ey}.

Further interest in this field has been generated with the advent of new experimental techniques. Using optical lattices, it has recently become possible to simulate complex tight-binding Hamiltonians using cold atoms and artificial magnetic fields.\cite{Celi:2014dg,Aidelsburger:2011hl,Baur:2013hj,Jaksch:611734,Palmer:2006km,Wang:2014ek,Dalibard:2011gg,Cooper:2008hx,Jotzu:2014vd} Of particular interest to us are the realisations of the Hofstadter model,\cite{Hofstadter:1976wt,Harper:1955uu} which has been experimentally established using cold atoms\cite{Miyake:2013jw,Aidelsburger:2013ew} and also in the solid state using graphene superlattices.\cite{Ponomarenko:2014hl,Dean:2014bv,Hunt:2013ef}

The observation of FQHE like states in bands that look little like Landau levels is quite surprising. While most of the literature in the field has been devoted to numerical studies, a number of theoretical pictures have also been proposed to explain the physics of  these FCI states.\cite{Qi:2011jo,McGreevy:2012ek,Parameswaran:2012uk,Roy:2012vo,Parameswaran:2013pca,Bergholtz:2013ey} However, we are still far from having as complete an understanding of FCI states as we do for the usual FQHE. As a step in the direction of understanding FCI states better, in this paper, we study FCI states in the Hofstadter model, where we can make some analytical progress by building on the large literature of analysis of the single particle Hofstadter problem. 

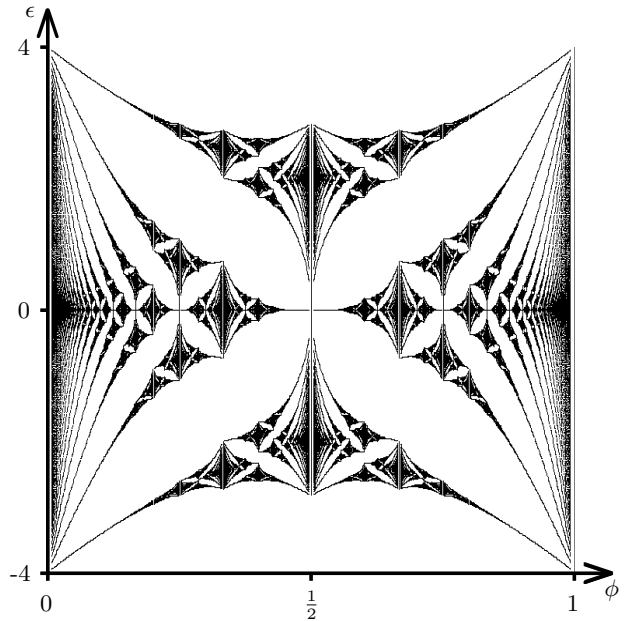
\begin{figure}[t]
\vspace{5mm}
\begin{overpic}[height=70mm,width=70mm, angle=0,clip,trim = 40mm 26mm 29mm 18mm]{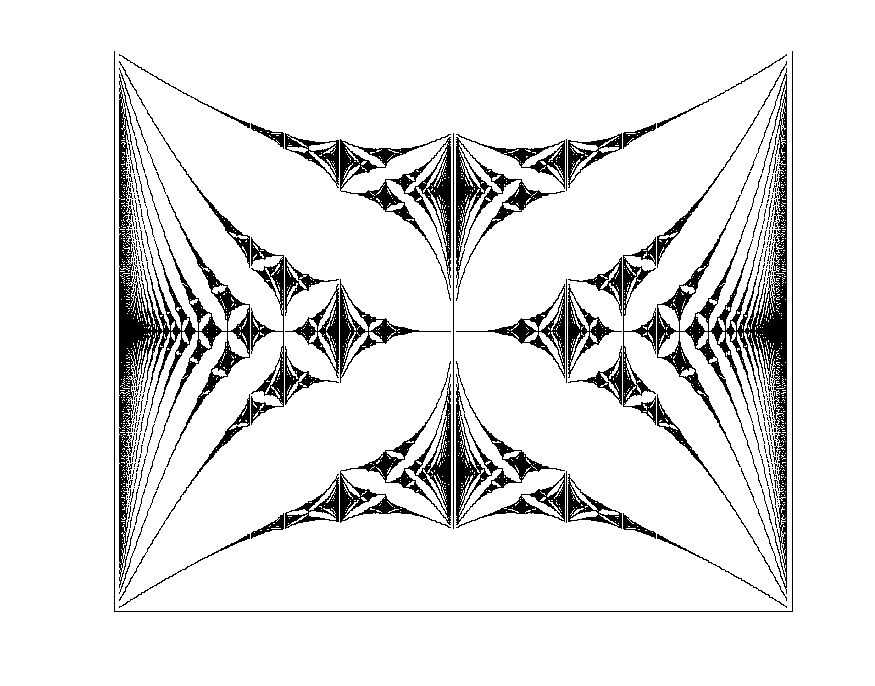}
\thicklines\put(0.1,-0.1){\color{black}\line(0,1){75}}
\thicklines\put(-0.1,-0.1){\color{black}\line(0,1){75}}
\thicklines\put(0,75){\color{black}\line(-1,-3){1.2}}
\thicklines\put(-0.1,75){\color{black}\line(-1,-3){1.2}}
\thicklines\put(0,75){\color{black}\line(1,-3){1.2}}
\thicklines\put(0.1,75){\color{black}\line(1,-3){1.2}}
\thicklines\put(-0.1,0.1){\color{black}\line(1,0){75}}
\thicklines\put(-0.1,-0.1){\color{black}\line(1,0){75}}
\thicklines\put(75,0.1){\color{black}\line(-3,-1){3.6}}
\thicklines\put(75,-0.1){\color{black}\line(-3,-1){3.6}}
\thicklines\put(75,0.1){\color{black}\line(-3,1){3.6}}
\thicklines\put(75,-0.1){\color{black}\line(-3,1){3.6}}
\thicklines\put(0,70.1){\color{black}\line(-1,0){1}}
\thicklines\put(0,69.9){\color{black}\line(-1,0){1}}
\thicklines\put(0,35.1){\color{black}\line(-1,0){1}}
\thicklines\put(0,34.9){\color{black}\line(-1,0){1}}
\thicklines\put(0,0.1){\color{black}\line(-1,0){1}}
\thicklines\put(0,-0.1){\color{black}\line(-1,0){1}}
\thicklines\put(-0.1,0){\color{black}\line(0,-1){1}}
\thicklines\put(0.1,0){\color{black}\line(0,-1){1}}
\thicklines\put(35.1,0){\color{black}\line(0,-1){1}}
\thicklines\put(34.9,0){\color{black}\line(0,-1){1}}
\thicklines\put(70.1,0){\color{black}\line(0,-1){1}}
\thicklines\put(69.9,0){\color{black}\line(0,-1){1}}
\put(-3,74){\color{black}$\epsilon$}
\put(74,-3){\color{black}$\phi$}
\put(-4,69){\color{black}4}
\put(-4,34){\color{black}0}
\put(-5,-1){\color{black}-4}
\put(-1,-5){\color{black}0}
\put(34,-5){\color{black}$\frac{1}{2}$}
\put(69,-5){\color{black}1}
\end{overpic}
\vspace{3mm}
\caption{Hofstadter's Butterfly shows the fractal spectrum of the Hofstadter model (with energy $\epsilon$) as a function of magnetic flux per plaquette, $\phi$. The entire pattern repeats for flux filling outside of the range 0 to 1. Landau level-like lines can be observed near to the points $(0,-4)$ and $(1/2,-2\sqrt{2})$, for example. }\label{butterfly}
\end{figure}

The Hofstadter model describes a charged, tight-binding particle hopping on a square lattice in the presence of an external magnetic field (see Section~\ref{model} for full details of the model and an account of its historical development). The tuneable parameters of the system are the hopping amplitude, $t$ (assumed, for most of this article, to be the same in all lattice directions), and the amount of magnetic flux per plaquette, $\phi$. We measure $\phi$ in units of the flux quantum $\phi_0=h/e$, which we set equal to one throughout. In the limit of weak flux per plaquette, the model reproduces the Landau levels (LL) of the continuum, which can be seen as lines emanating from the $\phi=0$ points of the energy spectrum in Figure~\ref{butterfly}. When $\phi$ is varied, the spectrum takes on the fractal structure known as Hofstadter's Butterfly.\cite{Hofstadter:1976wt}

In this article, we seek to analytically investigate the fractional Chern insulator (FCI) states that could be supported by the Hofstadter model in the presence of different many-body interactions. To find these, we first outline a systematic method to derive the single-particle solutions to the Hofstadter model for general flux $\phi$. The wavefunctions and energies we find then give us a description of the integer Chern insulator (ICI) states that would arise if we completely filled a whole number of Hofstadter Chern bands. To find the FCI states, we must partially fill a Hofstadter band and switch on interactions. The effects of the interaction can be fully characterised by the Haldane pseudopotentials (two-body interaction matrix elements) under the assumption of no band-mixing.\cite{Haldane:1983wb,Simon:2007bi}

The philosophy behind the approach will be to solve the system exactly near to a `simple' flux fraction $\phi=P/Q$ with $P$ and $Q$ small integers, and then to perturb about this point by adding or subtracting a small amount of flux $\delta\ll1/Q$ per plaquette. Near to $\phi=0$ this simply adds perturbative corrections to the continuum solution and moves us along the Landau level-like lines in the butterfly spectrum. However, motivated by similar Landau level-like lines emanating from other points in the butterfly (e.g. flux $\phi=1/2$, energy $\epsilon=-2\sqrt{2}\pm\delta$), we will show that continuum (LL) wavefunctions are a good description for Hofstadter wavefunctions \emph{in general}, provided an additional oscillating phase is included to account for the `large' part of the magnetic field. This insight was first exploited in the context of optical lattices in Ref.~\onlinecite{Palmer:2006km}, whose approach is equivalent to ours at zeroth order (i.e. with $\delta=0$). By perturbing in this way, we are able to exactly capture the wavefunction and energy corrections that are algebraically small in $\delta$. We will systematically ignore any exponentially small corrections ($\sim e^{-\alpha /\delta}$). In obtaining these wavefunctions, we correct a previous work in this area \cite{Hormozi:2012tn} and give a perturbative approach that is controlled and systematic for any given flux $\phi$.

After giving some historical and physical details of the Hofstadter model in Section~\ref{model}, we outline the perturbative procedure in detail for the case of small flux $\phi=1/N\ll1$ (with $N$ an integer) in Section~\ref{vanflux}. In Section~\ref{genbs} we generalise to $\phi=M/N\ll1$ (with $M$ and $N$ integers), and although there are some differences from the simpler case of $\phi=1/N$, we find that our main results remain the same for all values of $\phi\ll1$. We always obtain a structure very similar to that of Landau levels, but where rotational symmetry is broken by the lattice. 

Treating the cases of small $\phi=\delta$ as being a perturbation to $\phi=0$, we then consider the more complicated case where $\phi=P/Q\pm\delta$ in Section~\ref{genflux}. Much of the computational technique remains the same as in the above simpler case. We generally find that the system is described by a multi-species structure analogous to $Q$ copies of a Landau level, but again with broken rotational symmetry. We transform these states to the symmetric gauge in Section~\ref{symmgauge} and use these to calculate the explicit pseudopotentials for $\phi\approx1/2$ and $\phi\approx1/3$ in Section~\ref{pseudopotential_results}. 

We extend the method by considering different lattices and interactions in Section~\ref{extensions}, and discuss some applications of the results in Section~\ref{discussion}. Here we also consider how our approximate wavefunctions compare to the Wannier states of Qi et al.\cite{Qi:2011jo,Wu:2012uh} Finally, we give a summary and general conclusions in Section~\ref{conclusions}. We note that many of the calculations have been omitted from the main text to aid ease of reading. These can be found in the Appendices, in addition to tables of useful quantities such as perturbative energies and wavefunctions.

\section{The Hofstadter Model\label{model}}
\subsection{History\label{history}}
As a plentiful source of interesting physical phenomena, the Hofstadter model has received a great deal of attention since it was written down in the mid-20th Century. In this subsection, we aim to give a brief summary of those lines of investigation from the past decades which have directly influenced the present work. Where appropriate, we refer to sections later in the article where the ideas of the cited references are used or developed. This is not intended to be an exhaustive review, however, and will be confined to the scope of this article. 

Motivated by the Peierls substitution for charged particles in a magnetic field,\cite{Peierls:1933tv} Harper first considered the broadening effect of a uniform background field on a tight-binding conduction band of electrons.\cite{Harper:1955uu} The resulting discrete hopping equation has become known as the Harper Equation, which we introduce and explain in the next section (see Eq.~\eqref{harper}). This idea was developed further by Azbel\cite{ISI:A1964WM22700018} and Hofstadter\cite{Hofstadter:1976wt} who studied the detailed band structure of the model and noted the peculiar differences between the cases of rational and irrational flux per plaquette. These three authors are responsible for the model being variously called the Hofstadter model, Harper-Hofstadter model, or Azbel-Harper-Hofstadter model: in this document we will refer to the Hofstadter model, although we note the important contributions made in its development by the other authors.

Important studies of the Hofstadter model were carried out shortly after its discovery by Wannier and collaborators. In particular, these authors considered the model in the context of the magnetic translation group,\cite{Rauh:1974uo,Zak:1964un,Zak:1964vu} contrasted the cases of rational and irrational field values,\cite{Obermair:1976wp,Wannier:1978uf} calculated the density of states\cite{Wannier:1979vg} (to be compared with our Figure~\ref{ebandsthird}) and extended the model to the hexagonal lattice.\cite{Claro:1979tn} Aubry and Andr\'{e}\cite{Aubry:1980vm} studied the duality properties of the Harper equation and showed the existence of a localisation transition.

A key development in the understanding of the Hofstadter model then arose when Thouless, Kohmoto, Nightingale and den Nijs (known collectively as TKNN) calculated the Hall conductivity using the Kubo formula and found it to be related to an invariant integer.\cite{Thouless:1982wi} This result was subsequently derived through a thermodynamic approach by Streda,\cite{STREDA:1982vg,STREDA:1982uy} and the results extended to the hexagonal lattice by Macdonald.\cite{MacDonald:1984vt} The invariant integer was shown to be topological in origin and equal to the first Chern number of the Bloch band wavefunction\cite{Simon:1983wi,Avron:1983tj,Kohmoto:1985ts}---or the integral of the Berry curvature over the Brillouin zone (see Ref.~\onlinecite{Xiao:2010kw} for a review of the Berry phase applied to electronic bands). A decade later, it was shown\cite{Hatsugai:1993vn} that the invariant Chern number was also related to the winding number of the edge states\cite{Halperin:1982tb} in a finite system. 

In their seminal paper, TKNN also perturbatively derived a Diophantine equation, from which one can directly obtain the Chern number (see Ref.~\onlinecite{bernevig2013topological} for an expanded discussion of this derivation). The existence of such a Diophantine equation for \emph{any} system with magnetic translational symmetry was later proved by Dana et al,\cite{Dana:1985vg} although it should be noted that the equation only yields an unambiguous value for the Chern number in certain lattice systems.\cite{Avron:2014dx} We introduce the TKNN equation in Subsection~\ref{hofdetails} (Eq.~\eqref{tknn}) and apply it to a case of interest to us in Subsection~\ref{bandstrucMN}.

A semiclassical analysis of the Hofstadter model, at least for the small flux case, is an approach that has been in use for many decades, occurring even in Harper's original paper.\cite{Harper:1955uu} For a comprehensive account of the early literature, including a WKB analysis where the magnetic flux plays the role of the $\hbar$, we refer the reader to the review by Fischbeck.\cite{Fischbeck:1970vz} A semiclassical approach to the case of finite flux came later on when, in an important set of papers, Wilkinson\cite{Wilkinson:1984wa,Wilkinson:1984wd} (building on earlier work by Sokoloff\cite{Sokoloff:1981ug}) used the deviation from rationality as the small expansion parameter to derive new results about general bands in the Hofstadter model---we use some of these results in Appendix~\ref{wkbappendix}. Further work in this direction was carried out by Sokoloff,\cite{Sokoloff:1985um} Wang et al\cite{Wang:1987jw} and Watson,\cite{Watson:1991to} whose approach we follow closely in Appendix~\ref{wkbappendix}.

An operator algebraic study of the Harper equation was developed by Rammal and Bellissard\cite{Rammal:1990co} (see also Bellissard's chapter in Ref.~\onlinecite{Berg:2005ul}), which offers a complementary semiclassical approach to the WKB methods of Wilkinson et al. The perturbative expressions for Hofstadter energy bands (which we derive in Section~\ref{genflux}) are in agreement with results obtained using this algebraic approach. More recently, Gat and Avron\cite{Gat:2003hs} used a semiclassical approach to calculate the magnetisation and de Haas-van Alphen oscillations of Bloch electrons in the Hofstadter model.

In the mathematics literature, the Harper equation has been studied in the more general context of the \emph{almost Mathieu operator}, which extends the Hofstadter model to include anisotropic hopping. The major line of enquiry has been towards understanding the spectral properties of the system for different values of the coupling strength and magnetic flux---specifically, the comparison between rational and irrational field values. For an overview of research in this area, we refer the reader to an early review by Simon,\cite{Simon:1982ve} two more recent reviews by Last\cite{Last:1994up,Last:2005vl}, and a comprehensive study by Avila and Jitomirskaya.\cite{Avila:2009uz} Further investigations of the fractal and hierarchical fractal structure of Hofstadter's butterfly were also carried out in Refs.~\onlinecite{Stinchcombe:1987vo,Bell:1989vo,MacDonald:1983vn,Osadchy:2001jm}. In this article we are most interested in the case of rational flux, since any irrational number can be approximated to arbitrary accuracy by a nearby rational number. In this situation, Bloch solutions are shown to exist within a finite magnetic unit cell.

Most recently, as outlined in the introduction, studies have focussed on the relation between the Hofstadter model and other Chern insulators. There have also been significant advances in the development of experimental realisations of the model---in graphene multilayers and optical lattices, for example. A discussion of these areas, and accompanying references, may be found in the previous section.

\subsection{Details of the Model\label{hofdetails}}
The Hofstadter Hamiltonian \cite{Hofstadter:1976wt,Harper:1955uu} is
\begin{equation}\label{hofeq}
\h=-t\sum_{\langle mn\rangle}\left[c^\dagger_nc^{\phantom{\dagger}}_m\exp\left[2\pi i\int^n_m\mathbf{A}\cdot\mathrm{d}\mathbf{l}\right]+\mathrm{h.c.}\right],
\end{equation}
where $n$ and $m$ are site coordinates ($n=(n_x,n_y)$ etc.), $t$ is the hopping amplitude and $c^\dagger_n$ is an electron creation operator. The sum is over neighbouring sites of a square lattice with periodic boundary conditions and a (uniform) magnetic field described by the vector potential $\mathbf{A}$. If the Landau gauge is chosen ($\mathbf{A}=(0,Bx,0)$), the wavefunctions separate into pure Bloch waves in the $y$-direction so that
\be
\psi(x,y)&=&e^{ik_yy}\psi(x),
\ee
where $(x,y)=(n_x,n_y)$ and we have set the lattice spacing to one. The states in the $x$-direction are given by solutions to the discrete Harper equation,\cite{Harper:1955uu}
\begin{equation}\label{harper}
-\psi_{n-1}-\psi_{n+1}-2\cos\left(2\pi\phi n-k_y\right)\psi_{n}=\epsilon \psi_{n}.
\end{equation}
Here we have divided through by $t$ (so that $\epsilon\to\epsilon/t$) and $n$ is now the lattice index in the $x$-direction only.

Let us choose a rational $\phi=p/q$ with $p$ and $q$ coprime (we refer to the mathematical literature cited in Section~\ref{history} for a discussion of irrational flux). In this case, Eq.~\eqref{harper} is periodic under the substitution $n\to n+q$, and the magnetic unit cell is elongated in the $x$-direction to become $q\times1$ plaquettes in size. The wavefunctions are then Bloch solutions ($\psi_{\mathbf{k}}(\mathbf{r})=e^{i\mathbf{k}\cdot\mathbf{r}}u_{\mathbf{k}}(\mathbf{r})$), which form $q$ bands in $\mathbf{k}$-space. These are easily identified in the butterfly spectrum by considering a specific $\phi=p/q$. The $q$ different Hofstadter bands are gapped in all cases except for the two central bands when the denominator $q$ is even.\cite{Hofstadter:1976wt}

The Hall conductivity of a filled band is determined by its Chern number ($C$), an integer that plays the role of $\nu$ in Eq.~\eqref{hall}.\cite{Thouless:1982wi} This can be calculated directly from the wavefunction by integrating over the Brillouin zone,
\begin{equation}
C=\frac{1}{2\pi i}\int_{BZ}\mathrm{d}^2\mathbf{k}\,F,
\end{equation}
with the Berry curvature ($F$) \cite{Berry:1984ka,Xiao:2010kw} given by
\begin{equation}
F=\bigg\langle\frac{\partial u_{\bk}}{\partial{k_x}}\bigg|\frac{\partial u_{\bk}}{\partial{k_y}}\bigg\rangle-\bigg\langle\frac{\partial u_{\bk}}{\partial{k_y}}\bigg|\frac{\partial u_{\bk}}{\partial{k_x}}\bigg\rangle.\label{berrycurvatureequation}
\end{equation}
It can also be obtained simply from the TKNN Diophantine equation for the $r$th band,\cite{Thouless:1982wi}
\begin{equation}\label{tknn}
t_rp+s_rq=r
\end{equation}
which, under the constraint $|t_r|<q/2$, has a unique integer solution $(s_r, t_r)$. The Chern number of the $r$th band is then given by $C_r=t_{r}-t_{r-1}$. 

Similar Diophantine equations exist for any lattice model with magnetic translational symmetry,\cite{Dana:1985vg} including models that are defined on other lattices and for those which include further neighbour hopping terms (see, for example, Appendix~\ref{otherlattice}). However, only the Diophantine equation for the rectangular lattice (with nearest neighbour hoppings) determines the Chern number unambiguously.\cite{Avron:2014dx}
\section{Perturbative Approach for $\phi=1/N$\label{vanflux}}
\subsection{Band Flatness}
We will first solve Harper's Equation~(\ref{harper}) perturbatively for the case of small flux with $\phi=\delta\equiv1/N$ and $N$ large. As we will eventually consider the interacting system, there are three energy scales that we need to keep track of. These are the interaction strength itself ($V$), the single-particle bandwidth ($W$), and the band gap between the partially filled band of interest and the next empty band above this ($E_g$). For the lowest band to form a legitimate basis for the many-body states, we require the interaction strength to be much smaller than the bandgap. In addition, we require the interaction strength to be much larger than the bandwidth so that this is the dominant energy scale. Overall we need to satisfy the inequality
\begin{equation}
W\ll V\ll E_g,\label{energy_inequality}
\end{equation}
although we note that in practice Chern insulator states often persist at interaction strengths that would naively seem too large \cite{Sorensen:2005bt,Kourtis:2014hg,Sheng:2011iv}.

\begin{figure}[t]
\includegraphics[scale=0.35]{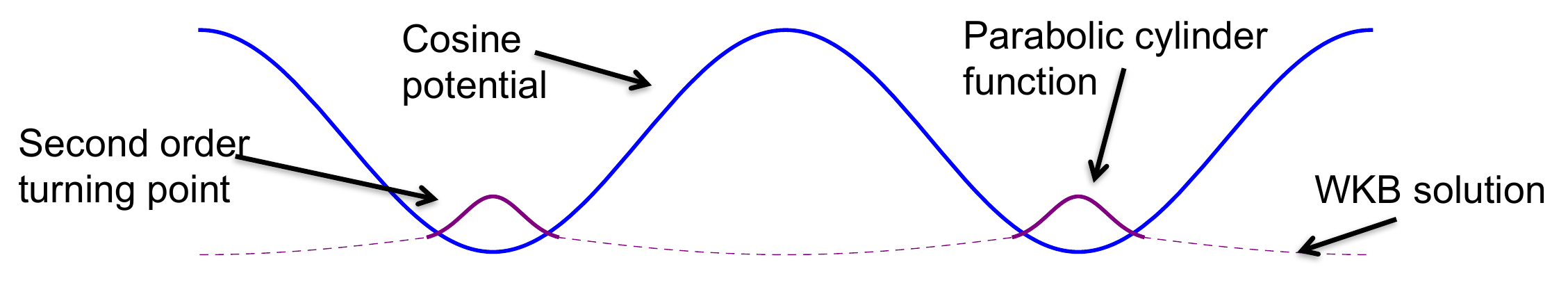}
\caption{Harper's equation may be thought of as describing a particle moving in a cosine potential. Semiclassically, we envisage a ground state wavefunction that consists of localised Gaussians in the troughs of the potential, coupled together by tunnelling. For the lowest energy states, the turning point is a quadratic turning point, since the first derivative of the potential is vanishingly small.}\label{wkbdiag}
\end{figure}

We can estimate the values of $N$ for which this inequality holds from Hofstadter's butterfly, where we require the Landau level line thickness (bandwidth) to be much smaller than the energy gap above it. Alternatively, we can find the condition analytically by calculating the semiclassical wavefunction. Assuming $N$ is large we may expand the discrete derivative in Eq.~\eqref{harper} to find
\be
-\frac{1}{2}\psi''(x)-\cos\left(\frac{2\pi x}{N}-k_y\right)\psi(x) = \frac{(2+\epsilon)}{2}\psi(x),
\ee
where we have substituted $\psi_n\to \psi(x)$ as if it were a continuum function. This equation now looks similar to the Schr\"{o}dinger equation for a particle moving in a cosine potential with period $N$. Intuitively, we would expect the ground state of this Hamiltonian to consist of localised Gaussian-like wavefunctions at each trough of the potential, coupled weakly together by tunnelling through the barrier, as shown in Figure \ref{wkbdiag}. We formalise this idea using the WKB approximation, where the WKB part of the solution describes tunnelling through the barrier, and the solution near each trough (second order turning point) is given by a parabolic cylinder function (see Appendix~\ref{wkbappendix}).

From this, we find that for large flux denominators $N$, the energy bands take the form $E(\mathbf{k})\sim E_0+Ae^{-\sigma N}\left(\cos(Nk_x)+\cos(Nk_y)\right)$ for some (approximately constant) $E_0, A, \sigma$ \footnote{Ref.~\onlinecite{Watson:1991to} shows this to be the case when there is a linear turning point in the WKB solution. We verify in Appendix~\ref{wkbappendix} that this also holds for the lowest states where there is a quadratic turning point.}. On the other hand, we find from perturbation theory that the bandgap is algebraic in $1/N$. For large enough $N$, the bandwidth is then exponentially small and negligible compared to the bandgap. For the lowest band we find $W/E_g<0.1$ for $N\geq3$, but this condition becomes more stringent if we consider higher bands. If we choose $V$ to lie between these energy scales then the inequality in (\ref{energy_inequality}) is satisfied.

It is interesting to note that the bands are also \emph{Berry flat} (i.e. they have an approximately uniform Berry curvature). A uniform Berry curvature leads to a  closure of the commutators of the projected density operator at lowest order and is expected to be an indicator of a better environment for fractional Chern insulators.\cite{Parameswaran:2012uk,Estienne:2012bma} We have found from numerics and the WKB approximation (Appendix~\ref{wkbappendix}) that the Berry curvature takes the same functional form as the energy bands, $F(\mathbf{k})\sim F_0+Be^{-\sigma N}\left(\cos(Nk_x)+\cos(Nk_y)\right)$ and so becomes exponentially flat in the limit of large $N$. The mean Berry curvature of the lowest band is $F_0=N/(2\pi)$ for $N\geq3$, which fixes $C=1$. In general, the $r$th lowest band also has Chern number $C_r=1$ for large enough $N>N_{c,r}$ but the critical values $N_{c,r}$ increase with $r$. This corresponds to our perturbative approximation (see Section~\ref{vanfluxpert}) breaking down for larger $r$. 

We will find that the discreteness of the lattice and the tunnelling between troughs each have only exponentially small effects on the energies and wavefunctions. In fact, the discreteness and tunnelling corrections are related through a Fourier transform. The Harper equation is self-dual under such a transformation,\cite{Aubry:1980vm} and the potential barrier and discrete derivative terms exchange places.

In our approximation we will ignore anything exponentially small in $N$ and assume that the bands are completely flat (and Berry flat). In the language of WKB, this amounts to ignoring tunnelling between neighbouring magnetic unit cells (where the sewing together of wavefunctions would introduce a $\mathbf{k}$-dependence) and treating the wavefunctions as continuous instead of discrete. We will find a localised, perturbative wavefunction in each unit cell and combine these into a Bloch solution by hand. We note that for an appropriately chosen finite size system, the dispersion and Berry curvature can be shown to be exactly flat.\cite{Scaffidi:2014tg}
\subsection{Perturbation Theory\label{vanfluxpert}}
To begin, we return to the Harper equation and represent $n$ as a continuum variable ($x$) by making the substitution $\psi_n\to \psi(x)$. We rewrite the discrete differences in terms of derivatives using translation operators,
\be
\begin{array}{ccccccc}
\hat{T}_+&=&e^{\partial_x},&&\hat{T}_-&=&e^{-\partial_x},
\end{array}
\ee
so that Harper's equation becomes
\be
-\left[\hat{T}_++\hat{T}_-+2\cos(2\pi\phi x-k_y)\right]\psi(x)=\epsilon \psi(x).
\ee

Up to this point, the transformed Harper equation is still exact. We now make an approximation by expanding the $\hat{T}$ operators and the cosine potential in powers of $1/N$. The left side of Harper's equation separates consistently into terms at successive orders in $1/N$, and to first order we find
\begin{equation}
\left[-\frac{1}{2}\partial_x^2+\frac{1}{2}\left(\frac{2\pi}{N}\right)^2x^2\right]\psi(x)=\frac{(4+\epsilon)}{2}\psi(x),\label{harper_first}
\end{equation}
where for consistency we note that $x\sim\sqrt{N}$ scales with the magnetic length. In the continuum, changing $k_y$ merely translates the wavefunction in the $x$-direction (as can be seen by considering the substitution $x'= x-N k_y/2\pi$) so we have set $k_y=0$ without loss of generality. 

The first order equation just describes a quantum harmonic oscillator, and so we can make the standard ladder operator substitutions,
\begin{eqnarray}
a&=&\sqrt{\frac{\w}{2}}\hat{x}+\frac{i}{\sqrt{2\w}}\hat{p},\nonumber\\
a^\dagger&=&\sqrt{\frac{\w}{2}}\hat{x}-\frac{i}{\sqrt{2\w}}\hat{p},\label{laddopp}
\end{eqnarray}
with
\be
\begin{array}{ccccccccccc}
\omega&=&\frac{2\pi}{N},&&\hat{p}&=&-i\partial_{x},&&\hat{x}&=&x.
\end{array}
\ee
The energy levels then form a harmonic ladder (with energies $\epsilon_l$), and the wavefunctions ($\ket{l}$) are just Gaussian functions multiplied by the appropriate Hermite polynomial. This is precisely what we would find if we solved the Schr\"{o}dinger equation for a free charged particle in the presence of a magnetic field, and so at first order the Hofstadter model recovers the Landau level states of the continuum. The differences in this case are that $k_y$ is now only defined modulo $2\pi$, the wavefunctions exist only at discrete lattice points, and in the $x$-direction the wavefunctions are Bloch periodic with period $N$. The magnetic length (which scales the width of the Gaussian) is given by $l_B=1/\sqrt{2\pi\delta}=\sqrt{N/(2\pi)}$, and so gets larger as the magnetic field gets weaker, but is much less than the period $N$ of the potential. In the limit of large $N$, the magnetic length is so large that the underlying lattice is not `seen' by the wavefunction, and the wavefunction is close to an ideal continuum state.

To go beyond the continuum states we must consider higher order terms in the Harper equation. In general, the $k$th order term on the left hand side of (\ref{harper_first}) will be of the form
\be
\hat{H}_k&=&-\frac{1}{(2k)!}\left[\partial_x^{2k}+\left(-1\right)^k\left(\frac{2\pi}{N}\right)^{2k}x^{2k}\right]\\
&=&-\frac{1}{(2k)!}\left(\frac{\pi}{N}\right)^{k}\left[\left(a-a^\dagger\right)^{2k}+\left(-1\right)^{k}\left(a+a^\dagger \right)^{2k}\right].
\ee
The Harper equation includes contributions at all orders, each of which will perturb the energy levels and wavefunctions at all orders of perturbation theory. Whilst the first few corrections can be calculated easily using elementary perturbation theory, the process is made more efficient by defining a (canonical) unitary operator that acts on an unperturbed state to give the corresponding perturbed state at a given order,
\begin{equation}
\ket{\tilde{l}}=U^\dagger\left(a^\dagger,a\right)\ket{l}\equiv e^{-f(a^\dagger,a)}\ket{l},
\end{equation}
where $f$ is anti-Hermitian in the bosonic ladder operators. By choosing an appropriate ansatz for $f$ and using the Baker-Campbell-Hausdorff formula, finding the corrections amounts to ensuring that the canonical transformation
\begin{equation}
\bra{m}e^{-f^\dagger(a^\dagger,a)}\h e^{-f(a^\dagger,a)}\ket{l}\equiv \epsilon_l\delta_{ml}
\end{equation}
is diagonal, a process which can be easily automated. In most cases we do not need to calculate to high order as $N$ is presumed large (and necessarily greater than one).

Since the perturbing Hamiltonians may be written entirely in terms of ladder operators, the wavefunction corrections will mix in higher and lower Landau levels, and will preserve the (assumed) degeneracy of the different $k_y$-states. Including contributions up to second order we find the energy levels are
\be
\epsilon_l&=&-4+\frac{4\pi}{N}\left(l+\frac{1}{2}\right)-\frac{1}{2}\pion{2}\left(2l^2+2l+1\right),
\ee
and the unitary operator, which generates the second order wavefunctions from the unperturbed states is
\be
U^\dagger&=&\exp\left[\left(\frac{1}{96}\pionn+\frac{1}{128}\pion{2}\right)\left(\ad{4}-\an{4}\right)\right.\\
&&\left.+\frac{1}{320}\pion{2}\left(\ad{5}a-a^\dagger\an{5}\right)\right].
\ee
We fix the wavefunction normalisation by integrating the wavefunction density as if it were a continuum function. Corrections to the normalisation due to the lattice would be exponentially small in $N$ (as seen from the Euler-Maclaurin formula), which we ignore.

Finally, if we write the $l$th perturbed $x$-direction wavefunctions as $\tilde{\psi}_l(x-k_yN/(2\pi))=\langle x,k_y|\tilde{l}\rangle$ then we can form Bloch solutions through
\begin{equation}
\Psi_{l,{\mathbf{k}}}(\mathbf{r})=\frac{1}{2\pi}\sum_{m}e^{ik_xmN}e^{ik_yy}\tilde{\psi}_l\left(x-\frac{k_yN}{2\pi}-mN\right).\label{blochstatesvanflux}
\end{equation}
We have normalised these wavefunctions for the infinite system so that
\be
\int\mathrm{d}x\,\mathrm{d}y\,\left[\Psi_{l',{\mathbf{k}'}}(\mathbf{r})\right]^*\Psi_{l,{\mathbf{k}}}(\mathbf{r})&=&\delta_{l'l}\delta(\mathbf{k}'-\mathbf{k}).
\ee
\subsection{Two-body Interactions}
Perturbation theory gives us the single-particle wavefunctions, but to find the many-body states we must turn on interactions. For now we consider only the simple two-body delta function interaction,
\be
 \hat{V}&=&\sum_{i<j}V(\mathbf{r}_i,\mathbf{r}_j)=\sum_{i<j}\delta(\mathbf{r}_i-\mathbf{r}_j),
 \ee
which acts between each pair of particles $(i,j)$ in turn and is relevant for bosonic systems.

For simplicity we will focus on the lowest Hofstadter band ($l=0$), which is analogous to the lowest Landau level. We project to this lowest single-particle band and form non-interacting many-body states from the perturbed wavefunctions found previously. Then, using these states as a basis, we calculate the overlap integrals between an initial two-body state, the interaction potential, and a final two-body state. Using the Bloch states defined in (\ref{blochstatesvanflux}), the two-body interaction matrix elements for the lowest band are
\begin{eqnarray}
V_{\mathbf{k}_1\mathbf{k}_2\mathbf{k}_3\mathbf{k}_4}&=&\bra{\Psi_{0,\mathbf{k}_1}(\mathbf{r}_1)\Psi_{0,\mathbf{k}_2}(\mathbf{r}_2)}\delta\left(\mathbf{r}_1-\mathbf{r}_2\right)\nonumber\\
&&\times\ket{\Psi_{0,\mathbf{k}_3}(\mathbf{r}_1)\Psi_{0,\mathbf{k}_4}(\mathbf{r}_2)}\label{vkkkk1N}\\
&\equiv&\frac{1}{(2\pi)^2}\delta\left({\Sigma k_x}\right)\delta\left({\Sigma k_y}\right) T_{k_{1y}k_{2y}k_{3y}k_{4y}},\nonumber
\end{eqnarray}
where we have defined the useful notation
\begin{equation}
\sum k_x=k_{1x}+k_{2x}-k_{3x}-k_{4x}\label{ksum}
\end{equation}
and similarly for $\sum k_y$. The matrix element $T_{k_{1y}k_{2y}k_{3y}k_{4y}}$ is given by
\begin{eqnarray}
T_{k_{1y}k_{2y}k_{3y}k_{4y}}&=&\int\mathrm{d}x\,\left[\tilde{\psi}_0\left(x-\kappa_{1}\right)\right]^*\left[\tilde{\psi}_0\left(x-\kappa_{2}\right)\right]^*\nonumber\\
&&\tilde{\psi}_0\left(x-\kappa_{3}\right)\tilde{\psi}_0\left(x-\kappa_{4}\right),\label{tkkkk1N}
\end{eqnarray}
where we have also defined $\kappa_i=k_{iy}N/(2\pi)$.

The $\delta$ functions in $V_{\mathbf{k}_1\mathbf{k}_2\mathbf{k}_3\mathbf{k}_4}$ ensure that crystal momentum is conserved, whilst the matrix element magnitudes are contained within the quantities $T_{k_{1y}k_{2y}k_{3y}k_{4y}}$. Explicitly, these are made up of Gaussian and polynomial factors that depend on the set of $\{k_{iy}\}$. 

From the two-body matrix elements we can in principle write down the matrix elements between states with any number of particles. If this larger matrix is then diagonalised, we can obtain the low-lying many-body wavefunctions and energies for a given system size.

Although it is possible to work in the Landau gauge, in order to interpret our results in terms of the usual continuum theories, we will switch to the symmetric gauge (and associated disk geometry). We will therefore calculate the equivalent symmetric gauge matrix elements, the Haldane pseudopotentials, which use eigenstates of relative angular momentum as their initial and final states. These are useful because the centre of mass angular momentum is conserved and the relative angular momentum is a good quantum number for a rotationally invariant interaction in the continuum. We define the two-particle state with relative angular momentum $L$ by
\begin{equation}
\ket{L;i,j}=C_L\left(z_i-z_j\right)^L,\label{2pbasis}
\end{equation}
with $C_L$ a normalisation constant and where we have used complex $z=x+iy$ to denote the two-dimensional particle position. Here, we have suppressed the Gaussian factors, and also the centre of mass degree of freedom [which would include factors of $\left(z_i+z_j\right)$], since this is conserved by the interaction. For a pairwise, rotationally invariant interaction in the continuum, Haldane pseudopotentials are then defined as
\be
V^{L}&=&\bra{L;i,j}V\left(z_i-z_j\right)\ket{L;i,j},
\ee
which allow us to write the many-body interaction as
\be
\hat{V}&=&\sum_{i<j}\sum_L V^{L}P_{ij}^L
\ee
with $P_{ij}^L$ a projector for particles $i$ and $j$ onto the state with relative angular momentum $L$.

In the Hofstadter model, the lattice breaks rotational symmetry and we no longer expect the angular momentum to be conserved: the pseudopotential coefficients should instead form a matrix in $L$ with elements $V^{LL'}$. We postpone a full discussion of the gauge transformation until Section~\ref{symmgauge} and for now merely state that the operator $U^\dagger$ defined earlier is gauge independent and has the same expression (in terms of inter-Landau level operators) in any gauge we choose. The pseudopotential matrix elements for the $\delta$-function interaction are then given by
\be
V^{LL'}&=&\bra{L}U_{i}U_{j}\delta\left(z_i-z_j\right)U^\dagger_{i}U^\dagger_{j}\ket{L'}.
\ee
Here we have dropped the labels $i$ and $j$ from the two-particle states $\ket{L}$ but added them to the $U^\dagger$ operators to show that each of the two particles is individually perturbed.

The pseudopotential matrix to second order is then (showing only rows and columns corresponding to $L,L'\in\{0,2,4,6,8\}$, and with $\delta=1/N$) 
\begin{widetext}
\be
V^{LL'}&=&\frac{1}{4\pi l_B^2}\left(\renewcommand\arraystretch{1.5}\begin{array}{ccccc}
1-\frac{7}{1536}\left(\pi\delta\right)^2 & 0 & \frac{\sqrt{6}}{96}\left(\pi\delta\right)+\frac{\sqrt{6}}{128}\left(\pi\delta\right)^2& 0 & \frac{\sqrt{70}}{3072}\left(\pi\delta\right)^2\\
0 & \frac{1}{256}\left(\pi\delta\right)^2 & 0 & 0 & 0\\
\frac{\sqrt{6}}{96}\left(\pi\delta\right)+\frac{\sqrt{6}}{128}\left(\pi\delta\right)^2 & 0 & \frac{1}{1536}\left(\pi\delta\right)^2 & 0 & 0\\
0 & 0 & 0 & 0 & 0 \\
\frac{\sqrt{70}}{3072}\left(\pi\delta\right)^2 & 0 & 0 & 0 & 0
\end{array}\right).
\ee
\end{widetext}
We note that the rotational symmetry-breaking (off-diagonal) terms arise at a lower order than the diagonal corrections and that all of the coefficients are small ($<0.03$). Angular momentum is only conserved modulo four, and so we expect any many-body spectrum to split into four sectors.

We emphasise that we are able to describe the interaction in terms of pseudopotentials only when the lattice can be treated as a small perturbation to the continuum. As $\delta$ is increased, the exponentially small terms that we previously neglected become important and the pseudopotential description would break down. We need to consider the full pseudopotential matrices because the two-body interaction now depends on $(z_i-z_j)$ rather than $|z_i-z_j|$. In general, for larger $\delta$, the interaction will depend also on $(z_i+z_j)$, which would make the pseudopotentials one step more complicated.
\section{General Band Structure\label{genbs}}
\subsection{Band Structure for $\phi=M/N$\label{bandstrucMN}}
In the previous section we limited the discussion to the case where the magnetic flux fraction per plaquette had the simple form $\phi=1/N$. To consider the full phase space of the Hofstadter model, we need to be able to treat a general $\phi=p/q$.

A natural extension to Section~\ref{vanflux} would include flux fractions $\phi=M/N$ with $N$ large (and $M$ and $N$ coprime). Provided $M\ll N$, the lowest energy levels should still lie within the narrow Landau level line of the Hofstadter spectrum that emanates from the point $(0,-4)$, and we might expect our perturbative approach to be applicable.

There is a significant difference for the case where $M\neq 1$, however: if we zoom in on the Landau level part of the spectrum we see that the Landau level `line' in fact splits into $M$ `mini-bands' which are exponentially close in energy ($\Delta\epsilon \sim e^{-\alpha N}$). We can explain the origin of these by returning to the Harper equation and looking at the cosine potential for $\phi=M/N$,
\be
V(n)&=&-2\cos\left(\frac{2\pi Mn}{N}-k_y\right).
\ee
Since $N$ is large, we can again expand perturbatively about the trough at $n=0$, and the wavefunctions and energy levels follow as in the previous discussion under the replacement $1/N\to M/N$. However, we can also make the replacement $n=n'+N/M$, which leaves the cosine potential unchanged. There is now a trough located at $n'=0$, and if we expand perturbatively about this, we obtain exactly the same energy levels and wavefunctions as before---except that the wavefunctions are now centred on $n=N/M$.

In total, we find $M$ wavefunctions centred at the positions $n=rN/M$, where $r=0,1,\ldots,M-1$. In our perturbative approximation, these have exactly the same energy and exactly the same perturbation series in terms of Landau levels. The true wavefunctions will have exponentially small corrections due to the discreteness of the lattice and tunnelling effects, which lead to the exponentially small separation in energy levels that we observe in the Hofstadter spectrum. In our approach we ignore this splitting, since it would easily be overcome by thermal excitations or the many-body interaction between particles. Instead, we group the set of $M$ mini-bands together and interpret them as collectively representing a single Landau level on the lattice. This is analogous to what we found in Section~\ref{vanflux} above, but it is now $\delta=M/N$ that parametrizes the deviation from the continuum.

Numerically we find that the total bandwidth of the set of $M$ mini-bands is exponentially flat and again given by $E(\bk)\sim E_0+Ae^{-\sigma N}\left(\cos(Nk_x)+\cos(Nk_y)\right)$, just as we found for the single band in Section~\ref{vanflux}. The \emph{total} Berry curvature of the $M$ mini-bands also takes this form, and the \emph{total} Chern number of the set of $M$ mini-bands is also equal to one, analogous to what we would expect for a single Landau level. 

This total Chern number is to be expected from the TKNN Diophantine Eq.~\eqref{tknn}, which, if we insert $p/q=M/N$, reads
\be
t_r M +s_rN = r.
\ee
Grouping the lowest $M$ mini-bands together by substituting $r=M$ yields the solution $(t_M=1, s_M=0)$. In general, grouping the lowest $lM$ mini-bands together (where $l$ is a positive integer) yields the solution $(t_{lM}=l,s_{lM}=0)$, so that each group of $M$ mini-bands has total Chern number of one. These solutions require $\left|t_{lM}\right|<N/2$ and so describe the lowest $\lfloor N/2\rfloor$ sets of $M$ mini-bands.

We also justify this value for the total Chern number from our perturbative wavefunctions in Appendix~\ref{perturbative_chern} (where we extend it to other field filling fractions). We expect the exact $\mathbf{k}$-dependent form of the bands and Berry curvature to follow from an extension to the semiclassical approach outlined in Appendix~\ref{wkbappendix}. 

In terms of our perturbation theory, the energies, the wavefunctions, and hence the pseudopotentials, depend on $\delta\equiv M/N$ as if it were a continuous parameter. The precise values of $M$ and $N$ only affect the periodicity properties of the wavefunctions, which we will find in Section~\ref{symmgauge} do not significantly change the many-body physics. Therefore, provided we neglect the tunnelling between troughs, we can treat $\delta$ generally and extrapolate even to irrational values.

\subsection{Band Structure for $\phi=P/Q+M/N$}
The band structure has further complications when we are not in the small flux regime. In general, we wish to consider a value of $\phi=p/q$ that is close (in magnitude) to a fraction with a small denominator. To this end, we write the simple fraction as $P/Q$ with $P$ and $Q$ coprime, and perturb around this point by adding or subtracting a small amount of flux $\delta$. For the most part we will choose $\delta=1/N$ for simplicity, but this can be extended to $\delta=M/N$ without loss of generality (see Appendix~\ref{fraccancel} for a discussion of general $P/Q+M/N$). To relate this general discussion to the small flux case considered previously, it is simplest to choose $P=0$, $Q=1$ in all that follows.
 
The first complication arises when we try to identify the size of the magnetic unit cell. In the previous section this was fixed at $N\times1$ plaquettes, but in general the cell length is given by the denominator of $\phi$, which may be affected by cancellation between $P$, $Q$ and $N$. With $Q$ prime, there are three possible cancelled forms for $\phi$ (see Appendix~\ref{fraccancel}),
\begin{equation}\label{cancellation}
\phi\equiv\frac{p}{q}=\frac{P}{Q}+\frac{1}{N}=\left\{
\renewcommand\arraystretch{1.5}
\begin{array}{cc}
\frac{PN+Q}{QN} & (a) \\
\frac{PN/Q + 1}{N} & (b) \\
\frac{(PN+Q)/Q^2}{N/Q} & (c)
\end{array}\right.
\end{equation}
which arise when
\be
\begin{array}{lcccll}
(a)&&N\mod{Q}&\neq&0\\
(b)&&N\mod{Q}&=&0&\\
&\mbox{and}&(PN+Q)\mod{Q^2}&\neq&0\\
(c)&&N\mod{Q}&=&0&\\
&\mbox{and}&(PN+Q)\mod{Q^2}&=&0.
\end{array}
\ee
Although our approach does not require $Q$ to be prime, this choice is varied enough to discuss the main complications in the band structure. The additional cancellation possibilities with $Q$ not prime and $M\neq1$ are discussed in Appendix~\ref{fraccancel}.

From Eq.~\eqref{cancellation}, we see that the unit cell in this case may be $QN$, $N$ or $N/Q$ sites in length. This integer also gives the number of sites within each cell and the number of bands we should observe in total. Although this number can vary, we will argue that the physics remains essentially the same in each of the different possible cases. To motivate this, we will first describe the band structure we find from exact diagonalisation, and then explain it according to our perturbation theory.

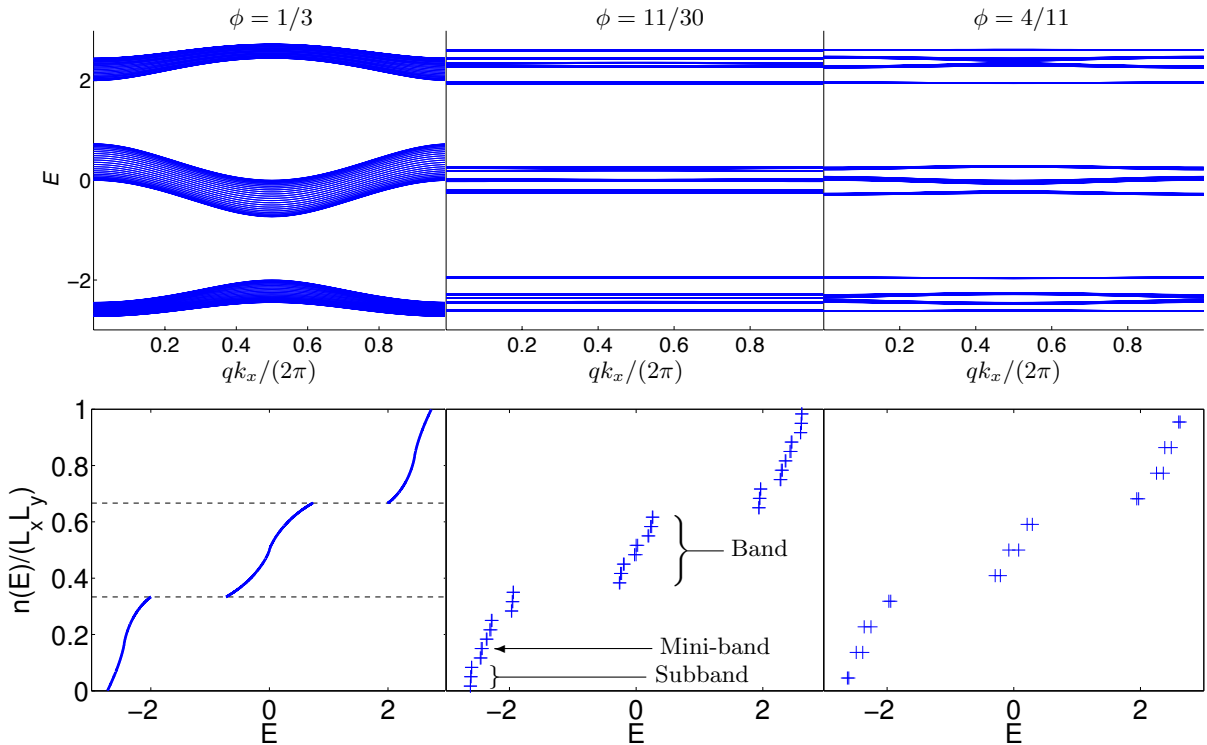
\begin{figure*}[t!]
\hspace{4mm}\begin{overpic}[scale=0.33,clip=true,trim=0 0 20 20]{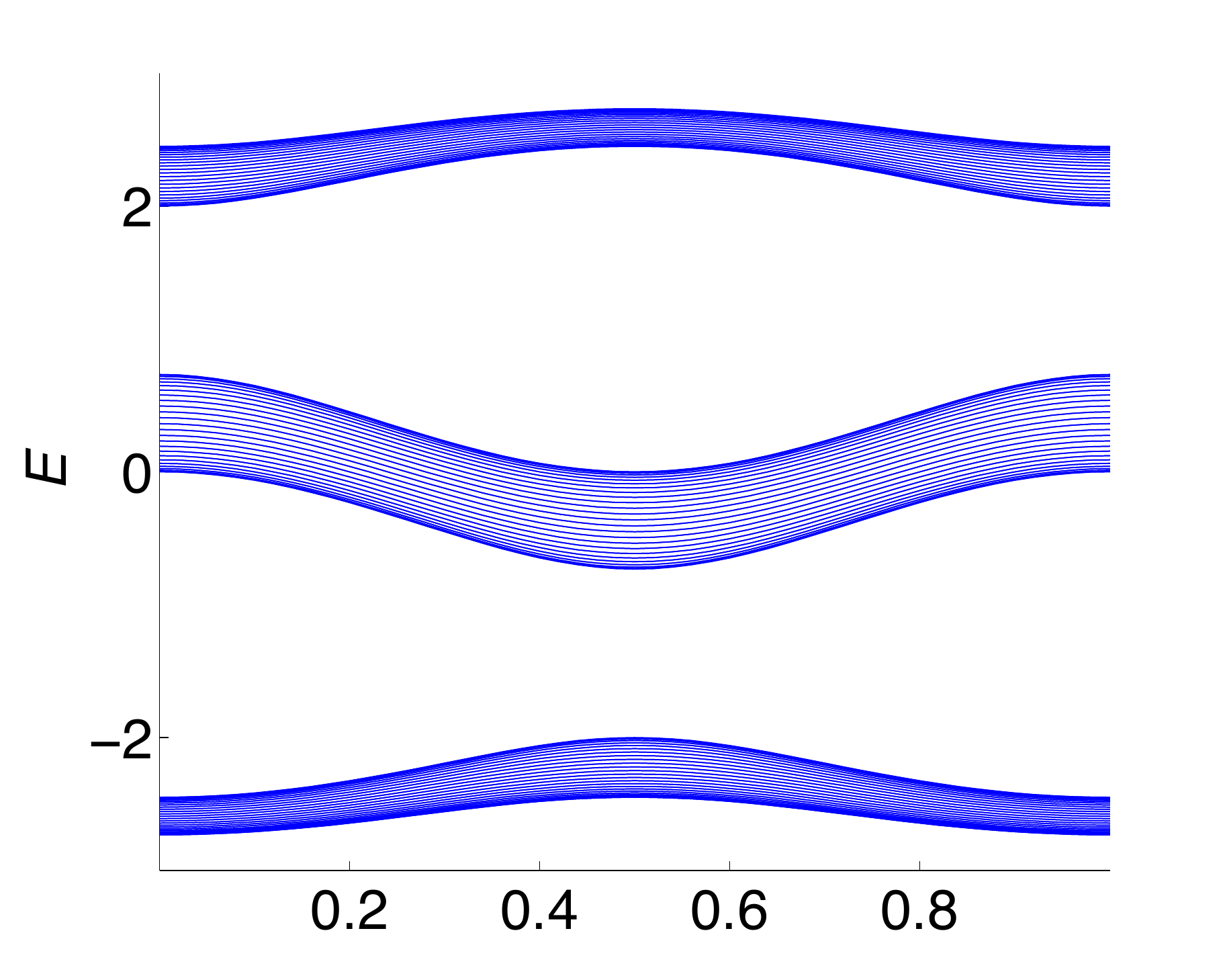}
\put(25,-1){$qk_x/(2\pi)$}
\put(26,46){$\phi=1/3$}
\end{overpic}
\!\!\!\!\!\!\!\!\!\!
\begin{overpic}[scale=0.33,clip=true,trim=72 0 20 20]{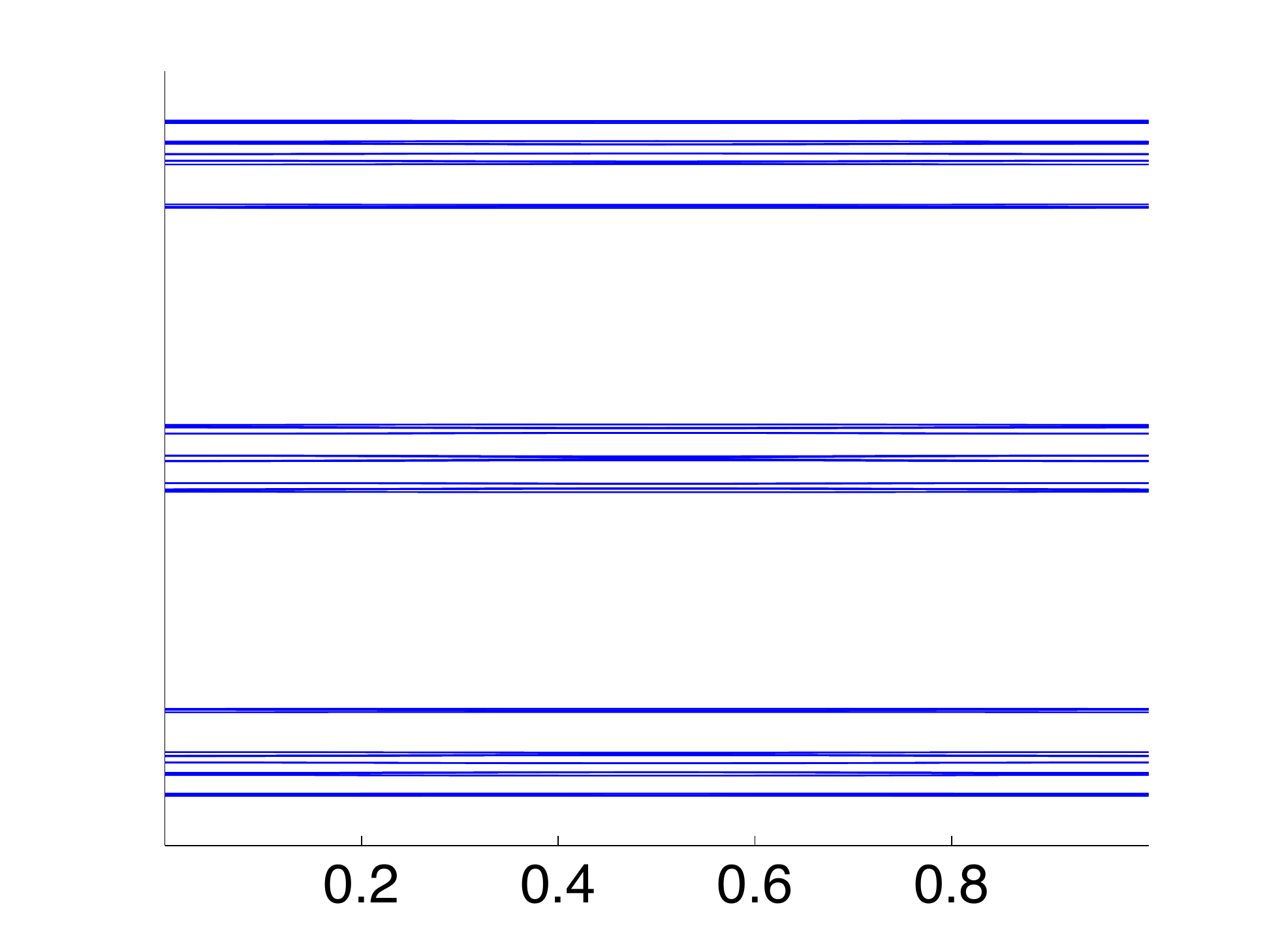}
\put(19,-1){$qk_x/(2\pi)$}
\put(20,46){$\phi=11/30$}
\end{overpic}
\!\!\!\!\!\!\!\!\!\!\!
\begin{overpic}[scale=0.33,clip=true,trim=72 0 20 20]{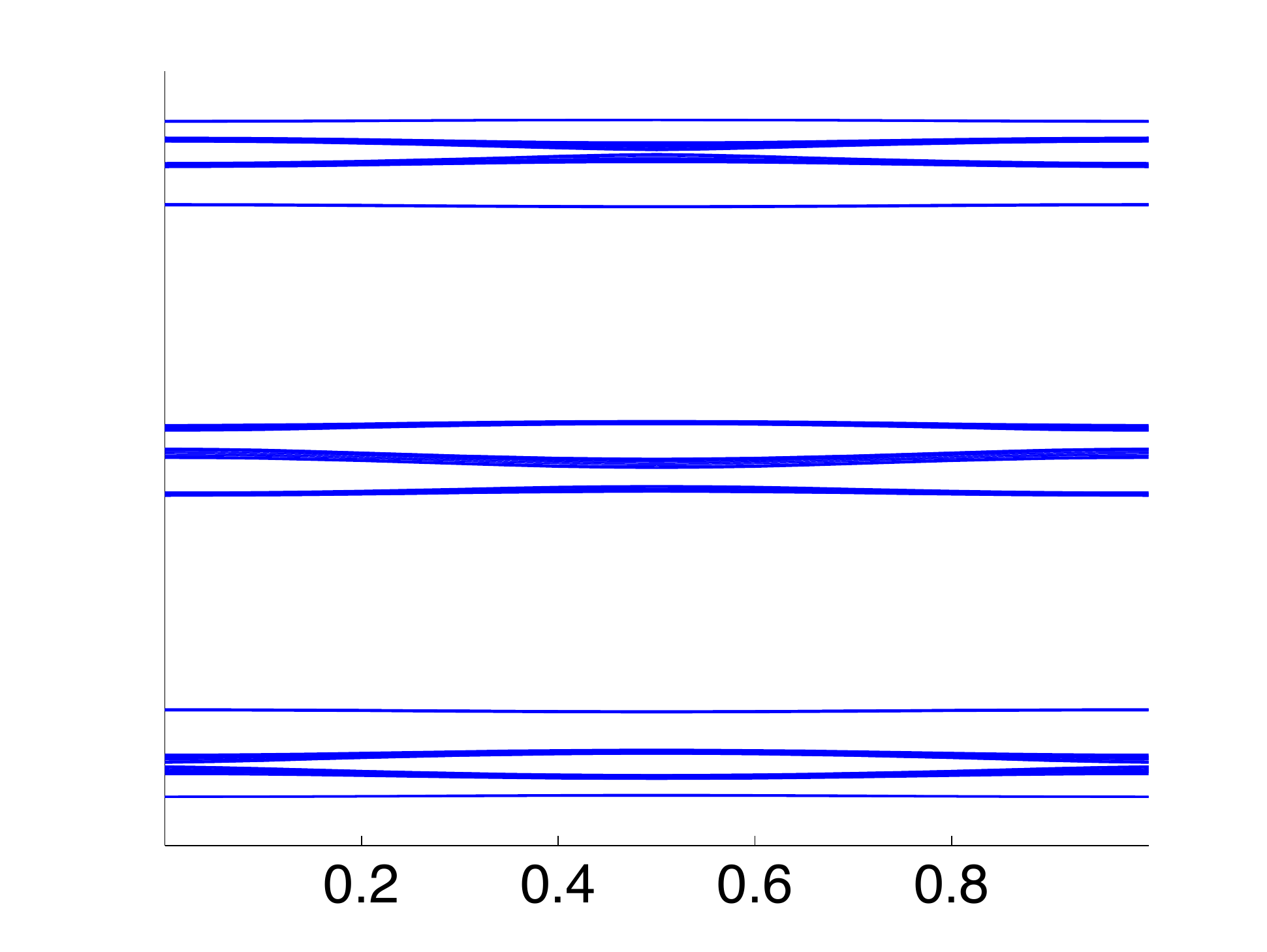}
\put(19,-1){$qk_x/(2\pi)$}
\put(20,46){$\phi=4/11$}
\end{overpic}

\vspace{4mm}

\begin{overpic}[scale=0.33,clip=true,trim=0 0 20 20]{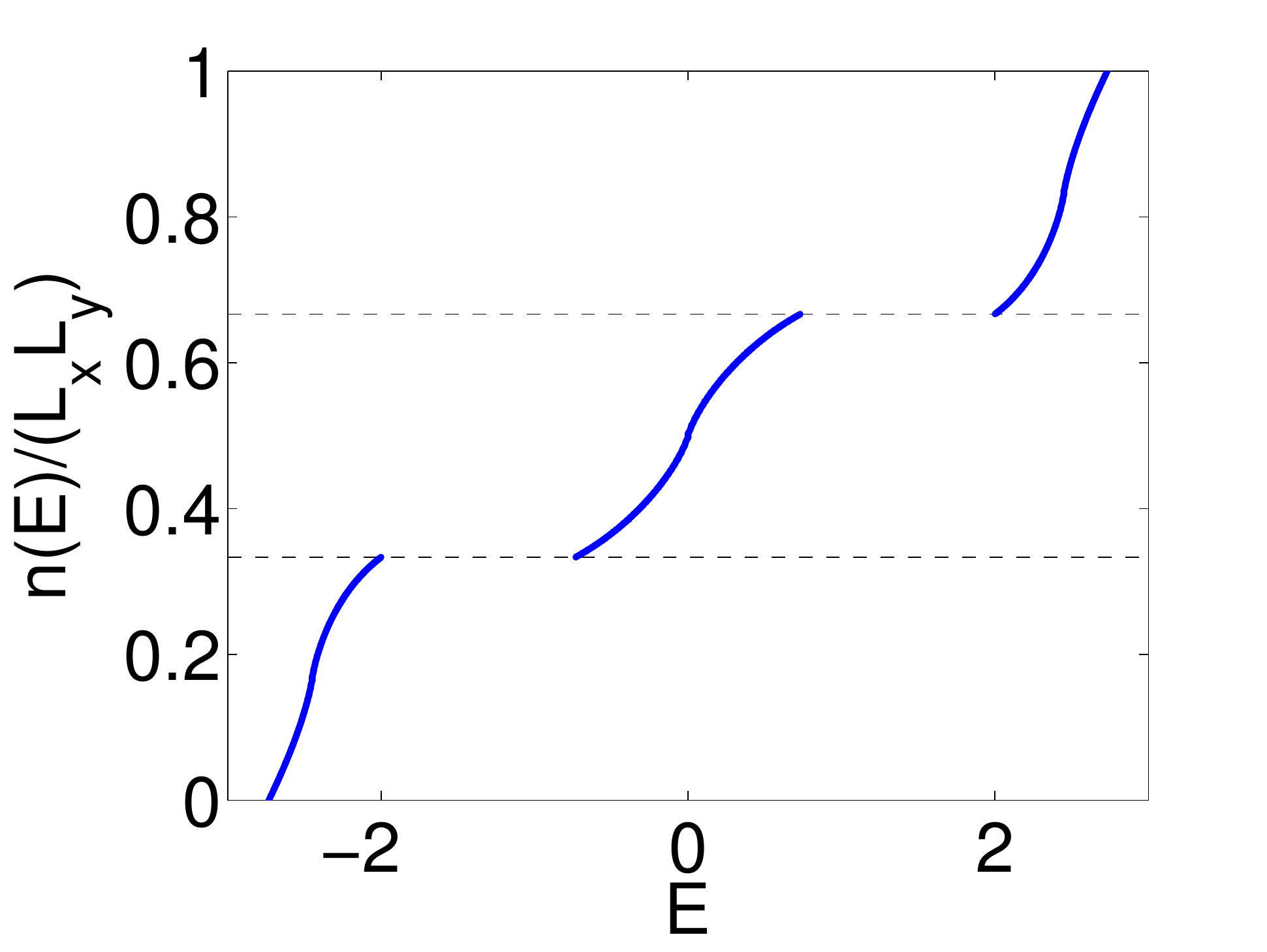}
\end{overpic}
\!\!\!\!\!\!\!\!\!\!
\begin{overpic}[scale=0.33,clip=true,trim=72 0 20 20]{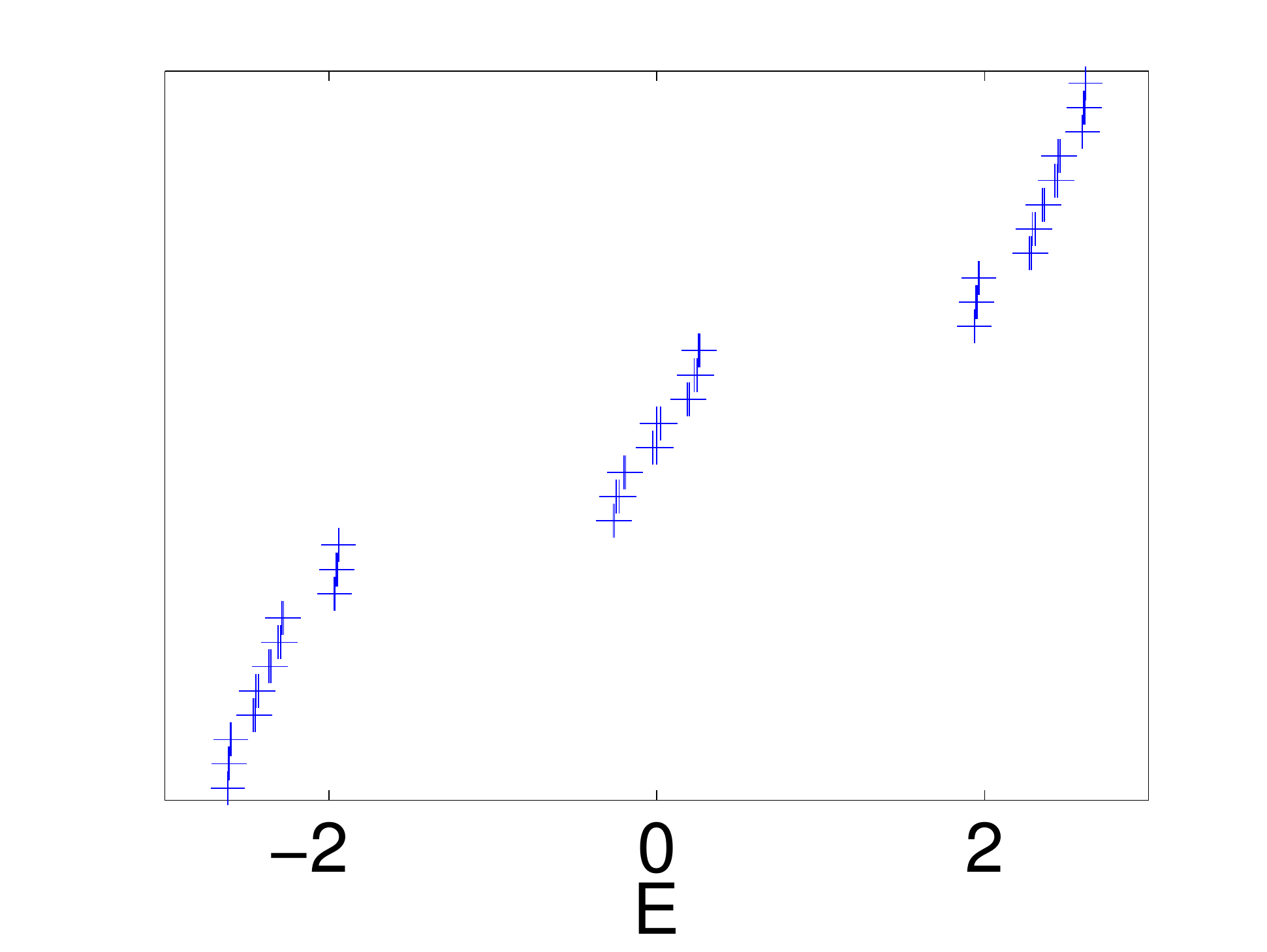}

\put(6.7,9.6){\color{black}\line(1,0){20}}
\put(5.7,8.8){\color{black}$\}$}
\put(27.5,13.4){\color{black}\vector(-1,0){21}}
\put(28.5,12.7){\color{black} Mini-band}
\put(27.9,9){\color{black} Subband}
\put(30,25.6){\color{black}$\Bigg\}$}
\put(32,26.393){\color{black}\line(1,0){5}}
\put(38,25.7){\color{black} Band}

\end{overpic}
\!\!\!\!\!\!\!\!\!\!\!
\begin{overpic}[scale=0.33,clip=true,trim=72 0 20 20]{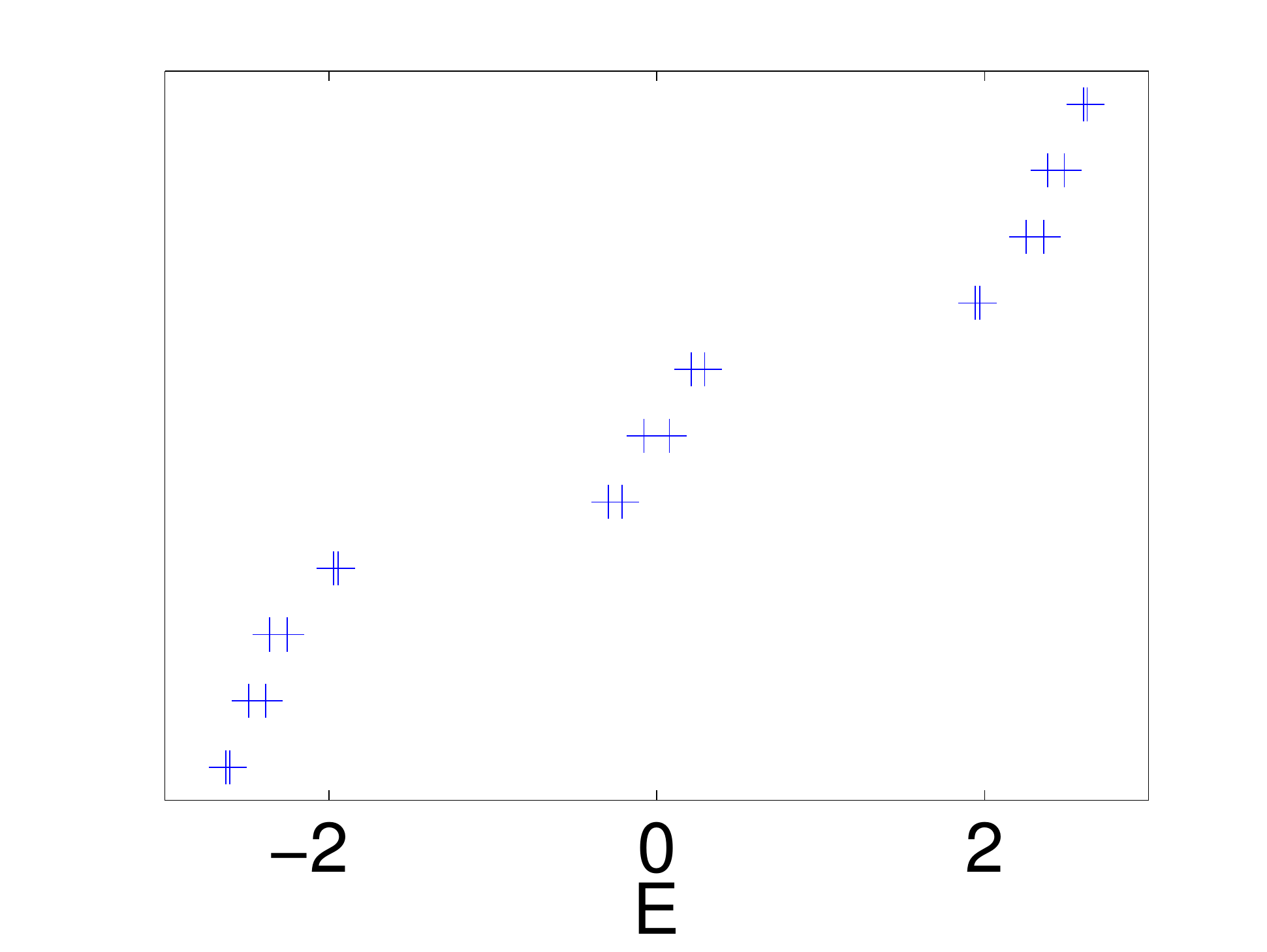}
\end{overpic}

\caption{Energy bands (top) and integrated density of states per unit area (bottom) for $\phi=1/3$, $\phi=1/3+1/30\equiv11/30$, and $\phi=1/3+1/33\equiv4/11$. \emph{Top}: The energy spectra are plotted against $k_x$ for fifty equally-spaced values of $k_y$ for each band. The spectra for the latter two field values show a clear energy gap above the lowest subband (which is just resolvable in each case). \emph{Bottom}: The density of states for the three complete bands is shown for $\phi=1/3$, whilst the individual mini-bands are plotted for $\phi\approx1/3$ with bandwidth given by the $x$-error bars. The band structure of the `pure' $\phi=1/3$ case carries over to the $\phi\approx1/3$ cases: the mini-bands arrange themselves into three bands and several subbands.}\label{ebandsthird}
\end{figure*}

In Figure~\ref{ebandsthird}, we plot the band structure and integrated density of states for $\phi=1/3$ and two nearby fractions $\phi=11/30$ (corresponding to $\delta=1/30$) and $\phi=4/11$ (corresponding to $\delta=1/33$). We notice that the three-band structure from the `pure' $\phi=1/3$ case is still apparent when $\phi\approx1/3$: even though there are now $q$ bands in total, these group themselves into three energy regions. In general, near to a flux fraction $P/Q$, the $q$ band solutions arrange themselves into $Q$ energy regions that correspond to the $Q$ original bands of the pure fraction. This reproduction becomes more accurate as $p/q$ gets closer to $P/Q$. Since the $q$ band solutions are generally close together in energy, we will refer to them as `mini-bands' as we did in the previous section. We refer to the $Q$ large groups of mini-bands as `bands' to emphasise that these are the energy structures that derive from the pure $P/Q$ case. We remark that the lowest band is formed of $p$ mini-bands if $p<q/2$ and is formed of $(q-p)$ mini-bands if $p>q/2$.

There is an additional, hidden structure to the energy levels for general $\phi\approx P/Q$: some of the mini-bands are separated by exponentially small energy gaps ($\Delta\epsilon\sim e^{-\alpha N}$) and others are separated by algebraically small energy gaps ($\Delta\epsilon\sim 1/N$). For $\phi=11/30$ shown in Figure~\ref{ebandsthird},  the three lowest mini-bands are exponentially close in energy but are separated algebraically from the fourth lowest mini-band. For the case with $\phi=4/11$ all of the gaps are algebraically small.

We group all of the mini-bands that are exponentially close in energy together in a \emph{subband}. We will find that these subbands are similar to the set of $M$ mini-bands we grouped together in the previous subsection and best represent the Landau levels of the continuum on the lattice. In this way, the subbands correspond to the rungs of the harmonic ladder discussed previously in Section~\ref{vanflux} and give rise to the linear Landau level-like lines that emanate from the various points of the Butterfly spectrum. As $N$ is increased, more and more subbands become apparent in the band structure.

The number of mini-bands in each subband depends on the cancellation of the fraction $p/q$ and is given in general by $n_s=qQM/N$. If $Q$ is prime and $M=1$, $n_s$ takes values $\{1,Q,Q^2\}$, and in general $n_s$ must lie between $1\leq n_s\leq Q^2$. For the two cases shown in Figure~\ref{ebandsthird}, we see that $n_s=3$ for $\phi=11/30$ and $n_s=1$ for $\phi=4/11$. Crucially, we find that the total Chern number of all of the mini-bands in a subband is equal to $Q$, so the subbands near a flux fraction $\phi=P/Q$ act like $Q$ copies of a Landau level.

\begin{figure}[h!]
\includegraphics[scale=0.7]{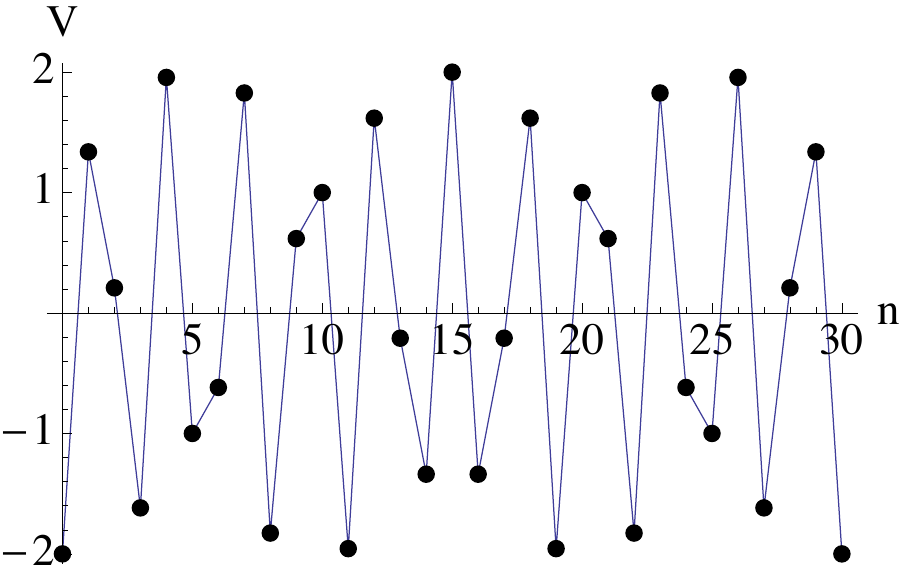}
\includegraphics[scale=0.7]{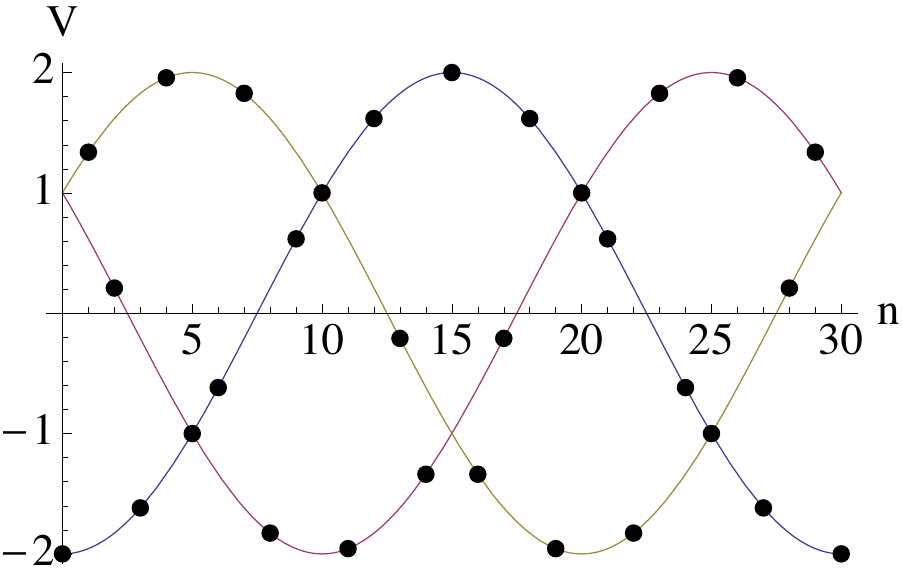}
\caption{\emph{Top}: Rapidly oscillating cosine potential for $\phi=1/3+1/30=11/30$. The eleven lattice sites in the troughs form the eleven mini-bands of the lowest band. \emph{Bottom}: Three effective potentials for $\phi=1/3+1/30=11/30$. The three troughs of the effective potentials give rise to the three mini-bands which form the lowest subband.}\label{potential_third}
\end{figure}

We will now explain these observations using our perturbative method, and in doing so justify why the subbands are the useful structures to consider in a real system. We recall that we can interpret Harper's equation as the Schr\"{o}dinger equation with a cosine potential,
\be
-\psi_{n-1}-\psi_{n+1}-2\cos\left(\frac{2\pi p n}{q}-k_y\right)\psi_{n}=\epsilon \psi_{n}.
\ee
This potential has $p$ periods fitted between $n=0$ and $n=q$, which are evaluated at lattice points (integer values of $n$). The lattice effectively `samples' the cosine potential, and if $p<q/2$ we can resolve all $p$ periods, whilst if $p>q/2$ we can only resolve $q-p$ effective periods. The lattice sites at the $p$ or ($q-p$) resolved troughs of the cosine potential form the $p$ or ($q-p$) mini-bands that comprise the lowest band (see Figure~\ref{potential_third}).

If we substitute $\phi=P/Q+M/N$, the cosine potential becomes
\begin{equation}
V(n)=-2\cos\left[\left(\frac{2\pi nM}{N}+\frac{2\pi Pn}{Q}\right)-k_y\right].\label{cosinepotential}
\end{equation}
Written like this, the potential looks like a single cosine curve with period $N/M$ but with an additional offset $2\pi Pn/Q$. This offset only takes $Q$ different values modulo $2\pi$ (depending on $(n\mod{Q})$), and so in total there will be $Q$ offset cosine potentials which we call \emph{effective potentials}, each lying on a different sublattice. These are entirely consistent with the single rapidly oscillating potential that comes from considering $\phi=p/q$, as shown in Figure \ref{potential_third}.

The subbands are formed from the states that lie in the troughs of the effective potentials. As the troughs are separated by $N/(MQ)$, but the length of the unit cell is given by $q$, the number of troughs per unit cell varies and is given by $n_s=qQM/N$.

When we carry out the perturbation theory, we will expand about the troughs of the effective potentials---these are equivalent to the troughs of the single cosine potential that we considered for the small flux case with $\phi=M/N$. As in this previous case, perturbation theory will yield a ladder of harmonic oscillator states with leading energy separation $O(1/N)$. We can also expand about troughs in the other effective potentials, which we will find lead to the same energy levels. The wavefunction perturbation series will in general be different, however. 

This is in contrast to the small flux case with $\phi=M/N$ where we found that expanding about the $M$ troughs led to the \emph{same} wavefunction perturbation series. Indeed, this is a general feature: replacing $\delta=1/N\to M/N$ leads to more troughs per unit cell corresponding to the same effective potential, and so this replacement does not introduce distinct perturbation series. The number of distinct perturbation series is always given by $Q$, the number of effective potentials or sublattices. This will be illustrated when we carry out the perturbation theory explicitly in Section \ref{genflux}.

Overall, we will obtain a ladder of states (mini-bands) for each effective potential trough in a unit cell. The total number of troughs depends on cancellation between $P,Q,M$ and $N$ and is denoted by $n_s$, but there will only ever be $Q$ distinct wavefunction perturbation series. In our approximation, the ladder of states from each trough will contain the same energy levels. We group together states from each trough which correspond to equal energies to form a subband.

An exact solution would show an exponentially small energy splitting between equivalent rungs due to the tunnelling effects that we neglect. Ignoring these is a good approximation in most cases where a finite temperature or the interaction strength would easily overcome the exponentially small splitting (and render it unresolvable).

Since we can only resolve subbands, these are the useful structures in the energy spectrum and should be considered as a whole. We find numerically that the total bandwidth and Berry curvature of the subband again decay exponentially, and that the total Chern number of the subband is given by $Q$. We justify the Chern number calculation in Appendix~\ref{perturbative_chern} and expect the band flatness can be verified analytically using a generalisation of the WKB approach discussed in Appendix~\ref{wkbappendix}.

Finally, we note that for a given total system size the number of $\mathbf{k}$-states in a subband is given by the same expression, no matter what size the unit cell is and no matter how many mini-bands the subband is comprised of. If the unit cell is extended so that there are more mini-bands, the Brillouin zone is correspondingly reduced in the $k_x$-direction (and vice versa). These effects cancel out so that the number of $\mathbf{k}$-states in each subband is always $QL_xL_yM/N$ where $L_x\times L_y$ is the area of the system. We should think of this as a mapping to $Q$ copies of a Landau level, which are then each individually perturbed by the lattice.
\section{Perturbative Approach for $\phi=P/Q\pm M/N$\label{genflux}}
\subsection{Preliminary Algebra and Definitions\label{genfluxprelim}}
We begin this section by carrying out some preliminary algebra on the Harper equation and then carry out the perturbation theory in subsection \ref{genfluxperturb}. We will find that the perturbed wavefunctions may still be expressed in terms of Landau levels, but the amplitudes and precise perturbation series now oscillate from site to site with period $Q$, as originally suggested in Ref.~\onlinecite{Palmer:2006km}.

We first recall that there are $Q$ possible offsets to the cosine potential $V(n)$ in Harper's equation \eqref{cosinepotential}, and so the equation itself can take $Q$ different forms. The form it takes depends on three variables: the lattice index modulo $Q$ (which affects the offset term $2\pi Pn/Q$ in the cosine potential); the value of $k_y$ (which translates the effective cosine potentials); and the trough that we are expanding about, which we will choose by shifting $n$ by a multiple of $N/(MQ)$.

To keep track of these contributions, we first note that for a given $0\leq k_y<2\pi$, the troughs will be located at
\begin{equation}
n=\frac{k_yN}{2\pi M}+\frac{sN}{QM},\label{n_offset}
\end{equation}
where $s$ is an integer that is constrained so that $n$ lies within the unit cell $0\leq n<q$. (Recall that the unit cell size is determined by $q$ which depends on the cancellation between $N$, $M$, $P$ and $Q$.) To expand about a given trough, we choose a $k_y$ and $s$ and let
\be
n=\frac{k_yN}{2\pi M}+\frac{sN}{QM}+n',
\ee
where $n'$ is small.

With this substitution we may write the cosine potential in Eq.~\eqref{cosinepotential}  as
\be
V(n)&=&-2\cos\left[\frac{2\pi}{Q}\left(Pn+s\right)+\frac{2\pi Mn'}{N}\right].
\ee
We have not substituted for the first $n$, since in this instance we are interested in its value modulo Q. Defining $\lambda=n\mod{Q}$, the cosine potential for the $\lambda$th sublattice (all sites with indices satisfying $n\mod{Q}=\lambda$) is then
\be
V(n)&=&-2\cos\left[\frac{2\pi}{Q}\left(P\lambda+s\right)+\frac{2\pi Mn'}{N}\right].
\ee
We see that the form the potential takes depends only on the integer
\be
j(\lambda,s)&=&(P\lambda+s) \mod{Q},
\ee
so that no matter what size the unit cell is, there are always $Q$ different forms for Harper's equation. We will find that these lead to the $Q$ distinct perturbation series described before. We might worry that the variable $n'$ is not necessarily an integer, but this is allowed within our approximation: the lattice discreteness means that there is a different wavefunction on each sublattice, but within each sublattice the true wavefunctions differ from their continuum forms by only exponentially small factors.

To proceed, we obtain a discrete equation that links together sites on a single sublattice only. We achieve this by writing out Harper equations for several neighbouring lattice sites, and algebraically eliminating amplitudes $\psi_{n'}$ on all but one of the sublattices. 

For example, let us consider a trough with $\phi=1/2+1/N$ and $s=0$. On the even sublattice ($\lambda=0$), the Harper equation for $n'$ is
\be
-\psi_{n'-1}^{1}-\psi_{n'+1}^{1}-2\cos\left(\frac{2\pi n'}{N}\right)\psi_{n'}^{0}=\epsilon \psi_{n'}^{0},
\ee
where the superscripts give the value of $j(\lambda,s)$ on each site. In this case, the value of $j(\lambda,s)=j(\lambda,0)$ just gives the sublattice index (which could be read off from the value of $n$), but we introduce the superscript here for comparison with later cases when $s\neq0$. Correspondingly, the Harper equation for $(n'+1)$ is
\be
-\psi_{n'}^{0}-\psi_{n'+2}^{0}+2\cos\left(\frac{2\pi (n'+1)}{N}\right)\psi_{n'+1}^{1}=\epsilon \psi_{n'+1}^{1},
\ee
where we note that the sign has changed in front of the cosine term because $(n'+1)$ lies on the other sublattice to $n'$. If we also write down the Harper equation for $(n'-1)$ then we can cancel out $\psi_{n'-1}^{1}$ and $\psi_{n'+1}^{1}$ to obtain 
\begin{eqnarray}
\frac{\psi^{0}_{n'-2}+\psi^{0}_{n'}}{\epsilon-2\cos\left(\frac{2\pi (n'-1)}{N}\right)}+\frac{\psi^{0}_{n'+2}+\psi^{0}_{n'}}{\epsilon-2\cos\left(\frac{2\pi (n'+1)}{N}\right)}\\
=\left[\epsilon+2\cos\left(\frac{2\pi n'}{N}\right)\right]\psi^{0}_{n'},\nonumber
\end{eqnarray}
which is written entirely in terms of the even sublattice. We can obtain a similar equation for the odd sublattice by setting $\lambda=1$ in the Harper equation for $n'$.

For general $P/Q$ we obtain a discrete difference equation for each of the $Q$ different sublattices. The equations near to other troughs and for different values of $k_y$ can be found by choosing an appropriate $s$ and $k_y$ and then changing the offset of $n'$ according to Eq.~\eqref{n_offset}. However, there will only ever be $Q$ distinct equations corresponding to the $Q$ possible values of $j$. 

If we had chosen $\phi=P/Q-1/N$ (with a negative sign) instead, then the troughs would move to the left as we increase $k_y$ rather than to the right. However, we can obtain the same discrete difference equations as before if we instead define
\be
n=-\left(\frac{k_yN}{2\pi M}+\frac{sN}{QM}+n'\right).
\ee
The definition of $n$ should therefore take the same $\pm$ sign as the magnetic field $P/Q\pm 1/N$.

Discrete difference equations near to some simple $\phi=P/Q$ are given in Appendix~\ref{example_harper}.
\subsection{Perturbation Theory\label{genfluxperturb}}
Once we have obtained a discrete difference equation for a single sublattice, we again make the continuum substitution $\psi^{j}_{n'}\to \psi^{j}(x')$, where $x'$ is a continuous variable corresponding to $n'$. As before, we rewrite the discrete differences using the operators
\be
\hat{T}_m=e^{m\partial_{x'}},
\ee
where $\hat{T}_m$ is a discrete translation of $m$ lattice sites in the $x$-direction. We expand the cosine terms (and generically also sine terms) in powers of $1/N$ and collect everything together order by order. This step is more complicated than for the case of small flux, and for $Q>2$ there will be terms present at half-integer powers of $1/N$.

At first order in $1/N$, we always recover a harmonic oscillator equation for $\psi^j(x')$ with characteristic frequency $\omega=2\pi M/N\equiv1/l_B^2$. This is what we found in Section~\ref{vanflux}, and we have checked that it is also the case for all values of $P/Q$ with $Q\leq5$---we expect it to hold in general. This permits the use of the QHO ladder operators defined in Eq.~\eqref{laddopp} and allows us to write the perturbed wavefunctions in terms of Landau levels. We again define unitary operators which generate the perturbation series from the unperturbed oscillator state,
\be
\big|\tilde{l};j\big\rangle&=&U^\dagger_{j}\big|l\big\rangle
\ee
with
\be
\tilde{\psi}^{j}_l\left(x'\right)&=&\big\langle x'\big|\tilde{l};j\big\rangle\equiv\tilde{\psi}^{j}_l\left(x-\frac{k_yN}{2\pi M}-\frac{sN}{QM}\right).
\ee
These quantities now have the additional label $j(\lambda,s)$ that tells us which discrete difference equation we have expanded. 

For a given $s$ and $k_y$, the complete $x$-wavefunction has amplitudes on every lattice site and we must write it as a sum over sublattice wavefunctions,
\begin{eqnarray}
\tilde{\psi}^{s}_{l,k_y}\left(x'\right)&=&\sum_{\lambda=0}^{Q-1} A_{j(\lambda,s)}\tilde{\psi}^{j(\lambda,s)}_l\left(x'\right)\delta_{x,\lambda}^{(Q)}.\label{fullxwf}
\end{eqnarray}
Here, $A_{j(\lambda,s)}$ is the coefficient for the wavefunction on the $\lambda$th sublattice, and the $\delta$ function $\delta_{x,\lambda}^{(Q)}$ is equal to one only when $x\mod{Q}=\lambda$. It therefore picks out the wavefunction only on the correct sublattice.

To find the coefficients $A_j$, we return to the original Harper equation (\ref{harper}) which connects amplitudes from different sublattices. By substituting the full wavefunction in from (\ref{fullxwf}), we obtain a set of simultaneous equations for the $A_j$ that can be solved using standard methods (up to normalisation). In fact, at lowest order these coefficients coincide with the amplitudes $\psi_n$ we would find for the $Q$ lattice sites in the simple $\phi=P/Q$ case.

Eq.~\eqref{fullxwf} gives the general $x$-wavefunction near a specified trough with momentum $k_y$. If we want to consider a different trough we simply change $s$ accordingly. This will cause the correspondence between the sublattices and the functions $\psi^{j}(x')$ to permute through a change in the function $j(\lambda,s)\equiv (P\lambda+s)\mod{Q}$, but the set of underlying functions will remain the same.

\begin{figure}[t]
\includegraphics[scale=0.8]{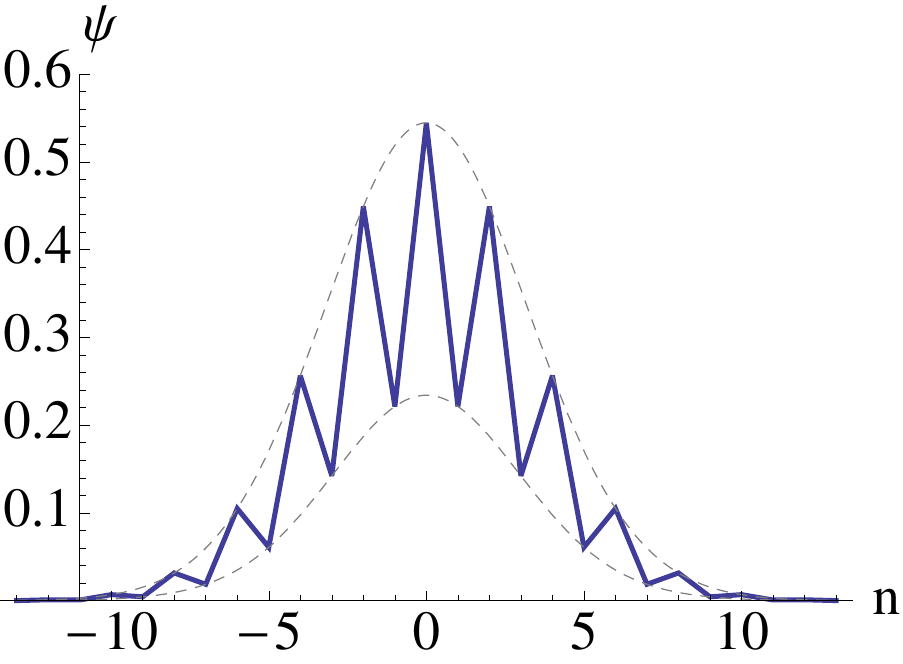}
\caption{First order perturbative wavefunction plotted near to the origin for $\phi=31/60$. The grey dashed lines indicate the continuum functions $A_j\psi^j(x)$.}\label{wf3160}
\end{figure}

In Figure~\ref{wf3160}, we plot the perturbative wavefunction $\tilde{\psi}^{0}_{0,0}(x)$ near to the origin for $\phi=1/2+1/60=31/60$ (i.e. $l=s=k_y=0$). The discrete wavefunction oscillates between the continuous functions $A_j\psi^j(x)$, which are also shown as a guide to the eye. We have normalised the wavefunction by integrating over each sublattice component as if it were a continuum function (see the similar more detailed discussion of integrals in Section~\ref{two-body} below). Exponentially small corrections to this normalisation can be found from the Euler-Maclaurin formula (and in general require multiplying by an appropriate elliptic theta function). We note that the effective magnetic length of the wavefunction for any $P/Q$ is always $l_B^2=N/(2\pi M)$, considering each sublattice separately.

Including the $y$-dependence, the perturbative wavefunctions are written
\begin{eqnarray}
\tilde{\psi}^{s}_{l,k_y}\left(x',y\right)&=&e^{ik_yy}\tilde{\psi}^{s}_{l,k_y}\left(x'\right),\label{wfxy}\\
&\equiv&e^{ik_yy}\sum_{\lambda=0}^{Q-1} A_{j(\lambda,s)}\tilde{\psi}^{j(\lambda,s)}_l\left(x'\right)\delta_{x,\lambda}^{(Q)}\nonumber
\end{eqnarray}
which we will find in Section~\ref{discussion} are related to the Wannier orbitals of Qi et al.\cite{Qi:2011jo} Finally, we form Bloch wavefunctions by introducing $k_x$,
\begin{eqnarray}
\Psi_{l,\mathbf{k}}^{s}(\mathbf{r})&\sim&\sum_{m}e^{ik_xmq}e^{ik_yy}\times\label{finalblochstates}\\
&&\tilde{\psi}_{l,k_y}^{s}\left(x-\frac{k_yN}{2\pi M}-\frac{sN}{QM}-mq\right).\nonumber
\end{eqnarray}
On the right hand side we have used the general unit cell length $q$ in the formation of the Bloch solution. Once the cell size is fixed, there are always $n_s$ possible choices of $s$ for each value of $k_y$.

We will obtain a final Bloch wavefunction of this form no matter what flux fraction $\phi=P/Q+M/N$ we choose (for suitably small $M/N\ll1$ and with $P$ and $Q$ co-prime). The only effect of different choices of $M$ and $N$ will be in the cancellation of the field fraction $\phi$ and, in turn, the unit cell size. We will always find that there is a $Q$-fold structure described by the labels $s$.

There are also perturbative solutions at very high energies, and near to the extrema of intermediate bands in the $\phi=P/Q$ band structure---we outline how these can be found in Appendix~\ref{otherkpoints}. We show how to calculate the Chern number of these perturbative wavefunctions in Appendix~\ref{perturbative_chern}, which we show is consistent with the TKNN equation.\cite{Thouless:1982wi}

Finally, we note that for large $Q$ it becomes increasingly difficult to rearrange Harper's equation into a discrete difference equation for a single sublattice. In these cases the perturbation theory can also be carried out by choosing an ansatz wavefunction for each sublattice that is a sum of Landau level wavefunctions,
\be
\tilde{\psi}^j_{0}(x')&=&\sum_{l,p}C^j_{l,p}\left(\frac{1}{N}\right)^p\psi_{l}(x').
\ee
Here the sum over $l$ is over all Landau levels and the sum over $p$ is up to as high an order in $1/N$ as required. The $\psi_l(x')$ on the right hand side is an unperturbed Landau level wavefunction and $C^j_{l,p}$ is the perturbation series coefficient for the $l$th Landau level at $p$th order in $1/N$. These ansatz wavefunctions may be substituted in Harper's equation directly, and the coefficients equated at each order in $1/N$. However, if this approach is used, in order to fix the coefficients consistently at order $(1/N)^p$ we must expand Harper's equation to order $(1/N)^{p+1}$---otherwise the solution is unconstrained. We believe that this leads to an error in the perturbative wavefunction given in Ref.~\onlinecite{Hormozi:2012tn}.
\subsection{Two-body Interactions\label{two-body}}
To conclude this section, we look at the two-body matrix elements for the $\delta$-function interaction using Landau gauge wavefunctions from the lowest Hofstadter band,
\be
V_{\mathbf{k}_1\mathbf{k}_2\mathbf{k}_3\mathbf{k}_4}^{s_1s_2s_3s_4}&=&\bra{\Psi_{0,\mathbf{k}_1}^{s_1}(\mathbf{r}_1)\Psi_{0,\mathbf{k}_2}^{s_2}(\mathbf{r}_2)}\delta\left(\mathbf{r}_1-\mathbf{r}_2\right)\\
&&\times\ket{\Psi_{0,\mathbf{k}_3}^{s_3}(\mathbf{r}_1)\Psi_{0,\mathbf{k}_4}^{s_4}(\mathbf{r}_2)}.\\
\ee
These are now labelled by $s$ indices in addition to $\mathbf{k}$ labels (for the previously considered case of $\phi=1/N$, the only allowed value of $s$ was zero).

Two $x$-wavefunctions, $\psi_{l,k_{1y}}^{s_1}(x')$ and $\psi_{l,k_{2y}}^{s_2}(x')$, will obviously have non-zero overlap if $s_2=s_1$ and $k_{2y}\approx k_{1y}$. However, they will also have non-zero overlap if $s_2=s_1+r$ and $k_2\approx k_1-2\pi r/Q$ (with integer $r$), as can be seen by noting that each wavefunction centre is located at $k_yN/(2\pi M)+sN/(QM)$. Provided that these overlaps are not prohibited by momentum constraints, we should expect them to feature in the calculation of $V_{\mathbf{k}_1\mathbf{k}_2\mathbf{k}_3\mathbf{k}_4}^{s_1s_2s_3s_4}$.

We let $k_{1y}=K+k_1'$ where $\left|k_1'\right|\ll \pi/Q$ and write the other wavevectors in terms of this,
\be
\renewcommand\arraystretch{1.5}
\begin{array}{ccccccc}
k_{2y}&=&K-\frac{2\pi r_2}{Q}+k_2';&&s_2&=&s_1+r_2\\
k_{3y}&=&K-\frac{2\pi r_3}{Q}+k_3';&&s_3&=&s_1+r_3\\
k_{4y}&=&K-\frac{2\pi r_4}{Q}+k_4';&&s_4&=&s_1+r_4,
\end{array}
\ee
where again the $r_i$ are integers and the $\left|k_i'\right|\ll \pi/Q$. In this way, $K$ is the large momentum that centres the wavefunctions, and $k_i'$ are small changes that translate the wavefunctions away from this centre. For non-zero overlap the $r_i$ are fixed once we have chosen the $s_i$, and so the only leftover freedom is with the primed momenta.

The overlap integrals will require us to multiply wavefunction components from the same sublattice together, carry out the integration, and then sum over the contributions from each sublattice. This sum over different sublattices means that the absolute values of $s$ will be unimportant: what matters is only the relative values of the $s_i$ compared to $s_1$. For example, we expect $V_{\mathbf{k}_1\mathbf{k}_2\mathbf{k}_3\mathbf{k}_4}^{0101}$ to be the same as $V_{\mathbf{k}_1\mathbf{k}_2\mathbf{k}_3\mathbf{k}_4}^{1212}$ (up to a shift of $K$).

After carrying out the integration we find
\be
V_{\mathbf{k}_1\mathbf{k}_2\mathbf{k}_3\mathbf{k}_4}^{s_1s_2s_3s_4}&=&\frac{1}{(2\pi)^2}\delta\left({\Sigma k_x}\right)\delta\left({\Sigma k_y}\right)T_{\kappa_1\kappa_2\kappa_3\kappa_4}^{s_1s_2s_3s_4},
\ee
where the summation notation is as defined in Eq.~\eqref{ksum} and where the matrix elements $T_{\kappa_1\kappa_2\kappa_3\kappa_4}^{s_1s_2s_3s_4}$ will be defined below. As usual, the integration over space enforces conservation of Bloch momentum, but we are now able to separate the large and small contributions to $k_y$. The $y$-momentum conserving delta function $\delta\left({\Sigma k_y}\right)$ may be written out as
\be
\delta\left(k_{1y}'+k_{2y}'-k_{3y}'-k_{4y}'-\frac{2\pi}{Q}(r_2-r_3-r_4)\right).
\ee
Since the primed momenta are much smaller than the $r$-dependent terms, and since the $r_i$ can only take integer values, the large and small contributions in the delta function must be set to zero independently. We can also replace the $r_i$ with the equivalent $s_i$, since these just differ by an offset that will cancel, and so
\be
\delta\left({\Sigma k_y}\right)=\delta\left({\Sigma k_y'}\right)\delta\left(\frac{2\pi}{Q}(s_1+s_2-s_3-s_4)\right).
\ee
Finally, we note that the argument of the final delta function need only be satisfied modulo $2\pi$, and so
\be
\delta\left({\Sigma k_y}\right)&=&\delta\left({\Sigma k_y}\right)\delta^{(Q)}_{\Sigma s}.
\ee
The second delta function is now written as a Kronecker delta, and by adding the superscript $(Q)$ we mean that $\Sigma s=0$ modulo $Q$. In contrast to the case of small flux, this allows for the possibility of Umkapp (and non $s$-conserving) processes. The $T$ functions are defined as for the close to vanishing flux case but with additional $s$ labels,
\be
T_{\kappa_1\kappa_2\kappa_3\kappa_4}^{s_1s_2s_3s_4}&=&\int\mathrm{d}x\,\left[\tilde{\psi}_0^{s_1}\left(x-\kappa_{1}\right)\right]^*\left[\tilde{\psi}_0^{s_2}\left(x-\kappa_{2}\right)\right]^*\\
&&\tilde{\psi}_0^{s_3}\left(x-\kappa_{3}\right)\tilde{\psi}_0^{s_4}\left(x-\kappa_{4}\right)
\ee
and where now $\kappa_i=k_{iy}'N/(2\pi)$. We can expand this further by writing the integral over $x$ in terms of a sum over sublattices and a sum over coordinates belonging to each sublattice ($x_\lambda$), so that
\be
\int\mathrm{d}x\to\sum_\lambda\sum_{x_\lambda}\approx\sum_\lambda\int\mathrm{d}x_\lambda.
\ee
Then,
\be
T_{\{\kappa_i\}}^{\{s_i\}}&=&\sum_{\lambda=0}^{Q-1} A_{j_1}^*A_{j_2}^*A_{j_3}A_{j_4}\int\mathrm{d}x_\lambda\,\left[\tilde{\psi}_0^{j_1}(x_\lambda-\kappa_1)\right]^*\\
&&\left[\tilde{\psi}_0^{j_2}(x_\lambda-\kappa_2)\right]^*\tilde{\psi}_0^{j_3}(x_\lambda-\kappa_3)\tilde{\psi}_0^{j_4}(x_\lambda-\kappa_4),
\ee
where we let $j_i\equiv j(\lambda,s_i)=\left(Ps_i+\lambda\right)\mod{Q}$. Since we have split the space up into $Q$ sublattices, integration over each $x_\lambda$ yields only $1/Q$ of the continuum value of the integral. As before, the explicit form of the $T$ is a product of Gaussian and polynomial factors in the $\kappa_i$.

In principle, these quantities allow us to calculate the low-lying many-body states. We would first project to the lowest single-particle subband and form non-interacting many-body states. Then, using these states as a basis, we could calculate the $V_{\mathbf{k}_1\mathbf{k}_2\mathbf{k}_3\mathbf{k}_4}^{s_1s_2s_3s_4}$ and diagonalise. However, for the reasons discussed previously we will switch to the symmetric gauge and calculate the Haldane pseudopotentials instead.

\section{Symmetric Gauge\label{symmgauge}}
\subsection{Wavefunctions}
To calculate the Haldane pseudopotentials, we must first find the perturbed wavefunctions in the symmetric gauge. For the small flux case we could just interpret the operator $U^\dagger$ in the symmetric gauge, but for general $P/Q$ it is not immediately clear how to carry out the transformation: there is a $Q$-periodic oscillation in the $x$-direction.

One approach would be to write down a Harper-like discrete difference equation for the symmetric gauge and find the perturbative wavefunctions directly from this by expanding order by order in $1/N$. We have carried out this approach elsewhere but find that we can equivalently (and more simply) obtain the the symmetric gauge wavefunctions from the Landau gauge wavefunctions by taking a suitable linear combination of states.

First we consider the relationship between the Landau and symmetric gauges in the continuum. The normalised wavefunctions in each gauge are given by
\begin{eqnarray}
&\psi_{n,k}^{\mathrm{LG}}(x,y)&=\sqrt[4]{\frac{2M}{N}}\frac{1}{\sqrt{2\pi}}\frac{1}{\sqrt{2^nn!}}e^{iky}\label{wflgsg}\\
&&\times H_n\left[\sqrt{\frac{2\pi M}{N}}\left(x-\frac{kN}{2\pi M}\right)\right]e^{-\frac{\pi M}{N}\left(x-\frac{kN}{2\pi M}\right)^2}\nonumber\\
&\psi^{\mathrm{SG}}_{n,m}(x,y)&=(-1)^n\sqrt{\frac{n!M}{m!N}}\left((x+iy)\sqrt{\frac{\pi M}{N}}\right)^{m-n}\nonumber\\
&&\times L_n^{m-n}\left[\frac{\pi M}{N}(x^2+y^2)\right]e^{-\frac{\pi M}{2N}\left(x^2+y^2\right)},\nonumber
\end{eqnarray}
where the $H_n$ are Hermite polynomials and the $L_{n}^\alpha$ are generalised Laguerre polynomials. We have explicitly written out the magnetic length $l_B^2=N/(2\pi M)$. The Landau level is given by the index $n$, whilst $k$ and $m$ give the linear and angular momentum degrees of freedom respectively in each gauge. The Landau gauge wavefunctions are similar to the perturbed Hofstadter lattice wavefunctions considered earlier.

Both sets of wavefunctions are complete, and so a wavefunction written in the symmetric gauge can always be expressed as a linear combination of Landau gauge wavefunctions, multiplied by a gauge transform factor $e^{i\chi(x,y)}$. In this way, we can write the transformations
\be
e^{\frac{i\pi M xy}{N}}\psi_{n,m}^{\mathrm{SG}}(x,y)&=&\int\mathrm{d}k\,B_m(k)\psi^{\mathrm{LG}}_{n,k}(x,y),\\
e^{-\frac{i\pi M xy}{N}}\psi_{n,k}^{\mathrm{LG}}(x,y)&=&\sum_m\,B^*_m(k)\psi^{\mathrm{SG}}_{n,m}(x,y).
\ee
It is simple to verify that the required $B_m(k)$ functions are given by
\be
B_m(k)&=&\sqrt[4]{\frac{N}{2\pi^2M}}\frac{1}{\sqrt{2^m m!}}H_m\left[\sqrt{\frac{N}{2\pi M}}k\right] e^{-\frac{k^2N}{4\pi M}},
\ee
and hold for all Landau levels (i.e. they are independent of $n$). We represent this gauge transformation in terms of operators as
\be
\ket{n,m}&=&\hat{B}\ket{n,k}.
\ee
Using this, we can relate quantities calculated in the Landau gauge to the same quantities calculated in the symmetric gauge.

On the lattice, we have shown that the Bloch wavefunctions are not true continuum functions but rather are periodic, oscillate in amplitude and have corrections from higher Landau levels. However, we can obtain semi-localised Wannier functions that are analogous to the Landau gauge states of the continuum by simply taking the repeating part of the Bloch wavefunction. These are related to the Wannier functions of Qi in Ref.~\onlinecite{Qi:2011jo}---see Section~\ref{discussion} for a full discussion of this.

We return to the perturbed Hofstadter wavefunctions from Eq.~\eqref{wfxy} and centre them at the origin, to find
\begin{equation}
\tilde{\psi}_{0,k-2\pi i sy/Q}^{s}(\mathbf{r})=e^{-2\pi isy/Q}\sum_{\lambda=0}^{Q-1} A_j \tilde{\psi}_{0,k}^{j,\mathrm{LG}}(x,y)\delta_{x,\lambda}^{(Q)}.\label{lghofwf}
\end{equation}
In this expression, $\tilde{\psi}_{0,k}^{j,\mathrm{LG}}(x,y)$ is a Landau gauge wavefunction that has been perturbed according to the perturbation series given previously. The initial phase factor of $e^{-2\pi isy/Q}$ is necessary for the centred wavefunctions to agree with the standard definitions of $\psi^{\mathrm{LG}}_{n,k}(x,y)$ given in Eq.~\eqref{wflgsg}---it arises when we translate a wavefunction with $s\neq0$ to the origin.

In terms of operators Eq.~\eqref{lghofwf} may be written
\be
\ket{\tilde{n},k,s}&=&e^{-2\pi isy/Q}U^\dagger_s\ket{n,k}.
\ee
We can transform this to the symmetric gauge by acting with the operator $\hat{B}$ (which commutes with $U^\dagger$) so that
\be
\ket{\tilde{n},m,s}&=&e^{-2\pi isy/Q}\hat{B}U_s^\dagger\ket{n,k}
\ee
This is equivalent to multiplying Eq.~\eqref{lghofwf} by the function $B_m(k)$ and integrating over $k$. We find
\begin{eqnarray}
\tilde{\psi}_{0,k-2\pi i sy/Q}^{s}(\mathbf{r})&\to&\tilde{\psi}_{0,m}^{s}(\mathbf{r})\label{symmwfpert}\\
&=&e^{-2\pi isy/Q}\sum_{\lambda=0}^{Q-1} A_j \tilde{\psi}_{0,m}^{j,\mathrm{SG}}(x,y)\delta_{x,\lambda}^{(Q)}\nonumber
\end{eqnarray}
In this way, we have transferred the sublattice structure described by $\lambda$ to the symmetric gauge.

By introducing this transformation, we are free to use whichever gauge we choose. The Landau gauge basis ($\ket{\tilde{n},k}$) is useful for calculating the Chern number (see Appendix~\ref{perturbative_chern}), for justifying the no-tunnelling approximation, and for many-body calculations on a torus. The symmetric gauge basis ($\ket{\tilde{n},m}$) is useful for comparing the many-body calculations in the disk geometry and for calculating pseudopotentials. For small $m$, the symmetric gauge states are fairly localised, and so these are the states that will be of interest for short-ranged interactions. 
\subsection{Haldane Pseudopotentials}
We can use the symmetric gauge wavefunctions to directly calculate the Haldane pseudopotentials. We first form two-particle states of relative angular momentum $L$,
\be
\tilde{\psi}_0^{L,s_1s_2}(\mathbf{r}_1\mathbf{r}_2)&=&\sum_{\{m_1,m_2\}}^{m_1+m_2=L}D^L_{m_1m_2}\tilde{\psi}_{0,m_1}^{s_1}(\mathbf{r}_1)\tilde{\psi}_{0,m_2}^{s_2}(\mathbf{r}_2),
\ee
where
\be
D^L_{m_1m_2}&=&\bigg[\bra{m_1}\otimes\bra{m_2}\bigg]\ket{L}
\ee
are the appropriate Clebsch-Gordan coefficients (given explicitly in Appendix~\ref{symmpseudo}). We have set the centre of mass angular momentum $M=0$ without loss of generality, as this remains conserved even in the presence of the lattice. Next, we calculate the overlap integrals with our chosen interaction. From the form of the wavefunctions in Eq.~\eqref{symmwfpert}, we see that the overlap integral will involve a sum of terms corresponding to each $\lambda$, and that the combination of initial phase factors will ensure that $\Sigma s$ is conserved modulo $Q$.

In practice we can ignore the implicit transformation that is occurring between Landau and symmetric gauges and instead just read off our perturbation series in the symmetric gauge (remembering to separately enforce $\Sigma s=0\mod{Q}$). The pseudopotentials again have matrix indices that indicate $L$ and $L'$, but now also possess a species degree of freedom given by $s$,
\be
V^{LL'}_{\{s_i\}}&=&\delta^{(Q)}_{\Sigma s}\bra{L}U_{is_1}U_{js_2}V\left(z_i-z_j\right)U^\dagger_{is_3}U^\dagger_{js_4}\ket{L'}.\label{pseudosym}
\ee
The $U_s^\dagger$ are the single particle unitary operators defined earlier which apply the perturbation due to the lattice and the labels $s_i$ take at least $Q$ different values depending on the size of the unit cell.

For the purposes of calculation we use centre of mass and relative coordinates rather than particle coordinates---full definitions of these are given in Appendix \ref{twoparticles}. Since the unitary operators $U^\dagger$ mix in components from higher Landau levels, the general two-body state that we must consider is $\ket{L,M;Y,Z}$, where $L$ and $M$ are the relative and centre of mass angular momenta and $Y$ and $Z$ are the relative and centre of mass Landau levels. Terms in a pseudopotential calculation then involve the general overlap integral
\begin{eqnarray}
&&\bra{L,M;Y,Z}\delta\left(z_i-z_j\right)\ket{L',M';Y',Z'}\nonumber\\
&&=\frac{\left(-1\right)^{L+L'}}{4\pi l_B^2}\delta_{YL}\delta_{Y'L'}\delta_{ZZ'}\delta_{MM'}\label{overlap_integral}
\end{eqnarray}
where a proof of the right hand side can be found in Appendix \ref{twoparticles}. An explicit pseudopotential calculation is given in Appendix \ref{symmpseudo}.

In the $\phi=M/N$ case the $M/N\to0$ limit connects precisely to the continuum, and so we should recover continuum pseudopotential coefficients to zeroth order (i.e. only $V^{00}$ should be non-zero for the delta function interaction). As we switch on the lattice (i.e. as we go to non-zero $M/N$), other elements in the pseudopotential matrix will become non-zero, with each element given by a perturbation series in powers of $M/N$. For general $P/Q$ there is no exact continuum limit, but we can use what we know about pseudopotentials to predict which many-body states will be energetically favoured. In the next section we give the pseudopotential matrices for $\phi=1/2\pm M/N$ and $\phi=1/3\pm M/N$.

\section{Pseudopotential Matrices\label{pseudopotential_results}}
The pseudopotential matrix for a delta function interaction with $\phi=1/N$ was given in Section~\ref{vanflux}, and extends to a general small $\phi=M/N$ under the substitution $\delta=1/N\to M/N$. Below we give pseudopotential matrices near to other simple flux fractions and interactions, and assume that $\delta=M/N\ll1$ throughout.
\subsection{$\phi=1/2\pm \delta$}
Near $\phi=1/2$, there are four types of two-particle states, described by $s_1,s_2\in\{00,01,10,11\}$. However, as mentioned previously, only interactions which conserve $s$ modulo two are non-zero. Below, we give the three distinct pseudopotential matrices---all others can be derived from these. Equivalent matrices are defined under the substitutions $s_i\to s_i+r$ for integer $r$ (where the transformed $s_i$ are defined modulo $n_s$). The matrix $V^{LL'}_{0110}$ differs from $V^{LL'}_{0101}$ only in the $V^{11}$ element, which picks up a minus sign due to the symmetry of the two-particle basis states [Eq.~\eqref{2pbasis}].

We give the even columns and rows corresponding to $L,L'\in\{0,2,4,6,8\}$, but also give the $L=1$ and $L'=1$ elements (bordered by lines) since the $(1,1)$ entry can now be non-zero. Although we cannot usually form bosonic continuum states with $L=1$, we now have two distinct particle species, and so a two-particle state formed from both species is not prohibited by particle statistics.
\begin{widetext}
\be
V_{0000}^{LL'}&=&\frac{V}{8\pi l_B^2}\left(\renewcommand\arraystretch{1.5}\begin{array}{c|c|cccc}
3-\frac{11}{32}(\pi \delta)^2 &0& -\frac{\sqrt{2}}{4}(\pi \delta)-\frac{7\sqrt{2}}{16}(\pi \delta)^2 & \frac{\sqrt{6}}{8}(\pi \delta)+\frac{5\sqrt{6}}{16}(\pi \delta)^2& -\frac{\sqrt{5}}{16}(\pi \delta)^2 & \frac{\sqrt{70}}{64}(\pi \delta)^2\\
\hline
0 & 0 & 0 & 0 & 0 & 0\\
\hline
-\frac{\sqrt{2}}{4}(\pi \delta)-\frac{7\sqrt{2}}{16}(\pi \delta)^2 &0& \frac{5}{16}(\pi \delta)^2 & -\frac{\sqrt{3}}{48}(\pi \delta)^2 & 0 & 0\\
\frac{\sqrt{6}}{8}(\pi \delta)+\frac{5\sqrt{6}}{16}(\pi \delta)^2 &0&  -\frac{\sqrt{3}}{48}(\pi \delta)^2 & \frac{1}{32}(\pi \delta)^2 & 0 & 0\\
-\frac{\sqrt{5}}{16}(\pi \delta)^2 &0& 0 & 0 & 0 & 0 \\
\frac{\sqrt{70}}{64}(\pi \delta)^2 &0& 0 & 0 & 0 & 0
\end{array}\right)
\ee
\be
V_{0101}^{LL'}&=&\frac{V}{8\pi l_B^2}\left(\renewcommand\arraystretch{1.5}\begin{array}{c|c|cccc}
1-\frac{67}{96}(\pi \delta)^2 &0& \frac{\sqrt{2}}{4}(\pi \delta)+\frac{7\sqrt{2}}{16}(\pi \delta)^2 & \frac{\sqrt{6}}{24}(\pi \delta)+\frac{\sqrt{6}}{16}(\pi \delta)^2& \frac{\sqrt{5}}{16}(\pi \delta)^2 & \frac{\sqrt{70}}{192}(\pi \delta)^2\\
\hline
0 & \frac{1}{2}(\pi \delta)^2 & 0 & 0 & 0 & 0\\
\hline
\frac{\sqrt{2}}{4}(\pi \delta)+\frac{7\sqrt{2}}{16}(\pi \delta)^2 &0& \frac{3}{16}(\pi \delta)^2 & \frac{\sqrt{3}}{48}(\pi \delta)^2 & 0 & 0\\
\frac{\sqrt{6}}{24}(\pi \delta)+\frac{\sqrt{6}}{16}(\pi \delta)^2 &0&  \frac{\sqrt{3}}{48}(\pi \delta)^2 & \frac{1}{96}(\pi \delta)^2 & 0 & 0\\
\frac{\sqrt{5}}{16}(\pi \delta)^2 &0& 0 & 0 & 0 & 0 \\
\frac{\sqrt{70}}{192}(\pi \delta)^2 &0& 0 & 0 & 0 & 0
\end{array}\right)
\ee
\be
V_{1100}^{LL'}&=&\frac{V}{8\pi l_B^2}\left(\renewcommand\arraystretch{1.5}\begin{array}{c|c|cccc}
1-\frac{91}{96}(\pi \delta)^2 &0& \frac{\sqrt{2}}{4}(\pi \delta)+\frac{7\sqrt{2}}{16}(\pi \delta)^2 & \frac{\sqrt{6}}{24}(\pi \delta)+\frac{3\sqrt{6}}{16}(\pi \delta)^2& \frac{\sqrt{5}}{16}(\pi \delta)^2 & \frac{\sqrt{70}}{192}(\pi \delta)^2\\
\hline
0 & 0 & 0 & 0 & 0 & 0\\
\hline
\frac{\sqrt{2}}{4}(\pi \delta)+\frac{7\sqrt{2}}{16}(\pi \delta)^2 &0& -\frac{1}{16}(\pi \delta)^2 & \frac{\sqrt{3}}{48}(\pi \delta)^2 & 0 & 0\\
\frac{\sqrt{6}}{24}(\pi \delta)+\frac{3\sqrt{6}}{16}(\pi \delta)^2 &0&  \frac{\sqrt{3}}{48}(\pi \delta)^2 & \frac{1}{96}(\pi \delta)^2 & 0 & 0\\
\frac{\sqrt{5}}{16}(\pi \delta)^2 &0& 0 & 0 & 0 & 0 \\
\frac{\sqrt{70}}{192}(\pi \delta)^2 &0& 0 & 0 & 0 & 0
\end{array}\right).
\ee
\end{widetext}

Motivated by Ref.~\onlinecite{Palmer:2006km}, we note that pseudopotential matrices are simplified if we define the new basis states
\be
\psi^\pm_{0,m}&=&\frac{1}{\sqrt{2}}\left(\psi^0_{0,m}\pm i\psi^1_{0,m}\right).
\ee
The pseudopotential matrices in this basis are given below using the same notation as before. Equivalent matrices may now be obtained under the exchange ${+\leftrightarrow-}$ and by noting that $V^{11}_{+-+-}=-V^{11}_{+--+}$ from the symmetry of the basis states:
\begin{widetext}
\be
V_{++++}^{LL'}&=&\frac{V}{4\pi l_B^2}\left(
\renewcommand\arraystretch{1.5}
\begin{array}{c|c|cccc}
 1-\frac{19}{96} (\pi \delta)^2 & 0 & 0 & \frac{\sqrt{6}}{24}(\pi \delta) +\frac{\sqrt{6}}{16} (\pi \delta)^2 & 0 & \frac{\sqrt{70}}{192} (\pi \delta)^2 \\
\hline
 0 & 0 & 0 & 0 & 0 & 0 \\
 \hline
 0 & 0 & \frac{3 }{16} (\pi \delta)^2& 0 & 0 & 0 \\
  \frac{\sqrt{6}}{24}(\pi \delta) +\frac{\sqrt{6}}{16} (\pi \delta)^2 & 0 & 0 & \frac{1}{96} (\pi \delta)^2& 0 & 0 \\
 0 & 0 & 0 & 0 & 0 & 0 \\
\frac{\sqrt{70}}{192} (\pi \delta)^2 & 0 & 0 & 0 & 0 & 0 \\
\end{array}
\right),
\ee
\be
V_{+-+-}^{LL'}&=&\frac{V}{4\pi l_B^2}\left(
\renewcommand\arraystretch{1.5}
\begin{array}{c|c|cccc}
 1-\frac{31 }{96}(\pi \delta)^2 & 0 & 0 & \frac{\sqrt{6}}{24}(\pi \delta) +\frac{\sqrt{6}}{16} (\pi \delta)^2 & 0 & \frac{\sqrt{70}}{192} (\pi \delta)^2 \\
 \hline
 0 & \frac{1}{4}(\pi \delta)^2 & 0 & 0 & 0 & 0 \\
 \hline
 0 & 0 & \frac{1}{16}(\pi \delta)^2 & 0 & 0 & 0 \\
 \frac{\sqrt{6}}{24}(\pi \delta) +\frac{\sqrt{6}}{16} (\pi \delta)^2 & 0 & 0 & \frac{1}{96}(\pi \delta)^2 & 0 & 0 \\
 0 & 0 & 0 & 0 & 0 & 0 \\
\frac{\sqrt{70}}{192} (\pi \delta)^2  & 0 & 0 & 0 & 0 & 0 \\
\end{array}
\right),
\ee
\be
V_{++--}^{LL'}&=&\frac{V}{4\pi l_B^2}\left(
\renewcommand\arraystretch{1.5}
\begin{array}{c|c|cccc}
 \frac{1}{2}(\pi \delta)^2 & 0 & -\frac{\sqrt{2}}{4}(\pi \delta)-\frac{7 \sqrt{2}}{16}(\pi \delta)^2 & 0 & -\frac{\sqrt{5}}{16}  (\pi \delta)^2 & 0 \\
\hline
 0 & 0 & 0 & 0 & 0 & 0 \\
 \hline
 -\frac{\sqrt{2}}{4}(\pi \delta)-\frac{7 \sqrt{2}}{16}(\pi \delta)^2 & 0 & 0 & -\frac{\sqrt{3}}{48} (\pi \delta)^2 & 0 & 0 \\
 0 & 0 & -\frac{\sqrt{3}}{48} (\pi \delta)^2& 0 & 0 & 0 \\
 -\frac{\sqrt{5}}{16}  (\pi \delta)^2  & 0 & 0 & 0 & 0 & 0 \\
 0 & 0 & 0 & 0 & 0 & 0 \\
\end{array}
\right).
\ee
\end{widetext}
At zeroth order we see that only the species-conserving terms $V_{++++}^{00}$, $V_{+-+-}^{00}$ and $V_{+--+}^{00}$ are non-zero, and are equal. In general, angular momentum is conserved modulo 4 for species-conserving interactions, but is only conserved modulo 2 for an umklapp interaction. We discuss possible ground states for these pseudopotentials in Section~\ref{discussion}.

\subsection{$\phi=1/3+1/N$}
Below, we give the three distinct pseudopotential matrices for $\phi=1/3+1/N$ to first order in $1/N$. Other pseudopotential matrices equivalent to the ones given below may be obtained under the substitutions $s_i\to s_i+r$, where $r$ is an integer, and by noting that $V^{LL'}_{1200}=\left(V^{LL'}_{0012}\right)^T$. Only even columns (corresponding to $L,L'\in\{0,2,4\}$) are shown:
\begin{widetext}
\be
V^{LL'}_{0000}&=&\left(\renewcommand\arraystretch{1.5}
\begin{array}{ccc}
\frac{6+\sqrt{3}}{4} & -\frac{3\sqrt{2}+\sqrt{6}}{16}\left(\pi\delta\right) & \frac{9(\sqrt{2}+2\sqrt{6})}{128}\left(\pi\delta\right)\\
-\frac{3\sqrt{2}+\sqrt{6}}{16}\left(\pi\delta\right) &0&0\\
 \frac{9(\sqrt{2}+2\sqrt{6})}{128}\left(\pi\delta\right) &0&0
\end{array}
\right),
\ee
\be
V^{LL'}_{0101}&=&\left(\renewcommand\arraystretch{1.5}
\begin{array}{ccc}
\frac{6-\sqrt{3}}{8} & \frac{3\sqrt{2}+\sqrt{6}}{32}\left(\pi\delta\right) & \frac{9(2\sqrt{6}-\sqrt{2})}{256}\left(\pi\delta\right)\\
\frac{3\sqrt{2}+\sqrt{6}}{32}\left(\pi\delta\right) &0&0\\
\frac{9(2\sqrt{6}-\sqrt{2})}{256}\left(\pi\delta\right) &0&0
\end{array}
\right),
\ee
\be
V^{LL'}_{0012}&=&\left(\renewcommand\arraystretch{1.5}
\begin{array}{ccc}
\frac{\sqrt{3}}{4} & \frac{\sqrt{6}}{8}\left(\pi\delta\right) & \frac{9\sqrt{2}}{128}\left(\pi\delta\right)\\
\frac{3\sqrt{2}-\sqrt{6}}{16}\left(\pi\delta\right) &0&0\\
\frac{9\sqrt{2}}{128}\left(\pi\delta\right) &0&0
\end{array}
\right).
\ee
\end{widetext}
We have not been able to find a single-particle rotation that simplifies the pseudopotential matrices near to $\phi=1/3$. Nevertheless, we make some suggestions for the supported many-body ground state in Section~\ref{discussion}.
\subsection{$\phi=1/3-1/N$}
We again give the three distinct pseudopotential matrices to first order in $1/N$, but corrections now appear at half order in $1/N$. Columns and rows corresponding to $L,L'\in\{0,1,2,4\}$ are shown, with the $L=1$ and $L'=1$ elements bordered by lines. Equivalent pseudopotential matrices may be obtained under the substitutions $s_i\to s_i+r$ where $r$ is an integer. We again note that $V^{LL'}_{1200}=\left(V^{LL'}_{0012}\right)^T$, but now due to the symmetry of the basis states, the $L=1$ row picks up a minus sign under the interchange of the first two $s$ indices, and the column $L'=1$ picks up a minus sign under the interchange of the last two $s$ indices:
\begin{widetext}
\be
V^{LL'}_{0000}&=&\left(\renewcommand\arraystretch{1.5}
\begin{array}{c|c|cc}
\frac{6+\sqrt{3}}{4} &0& -\frac{5(7\sqrt{2}-3\sqrt{6})}{16}\left(\pi\delta\right) & \frac{9(\sqrt{2}+2\sqrt{6})}{128}\left(\pi\delta\right)\\
\hline
0&0&0&0\\
\hline
-\frac{5(7\sqrt{2}-3\sqrt{6})}{16}\left(\pi\delta\right) &0&0&0\\
\frac{9(\sqrt{2}+2\sqrt{6})}{128}\left(\pi\delta\right) &0&0&0
\end{array}
\right),
\ee
\be
V^{LL'}_{0101}&=&\left(\renewcommand\arraystretch{1.5}
\begin{array}{c|c|cc}
\frac{6-\sqrt{3}}{8} -\frac{(9-5\sqrt{3})}{2}\left(\pi\delta\right)&-\frac{(3\sqrt{2}-\sqrt{6})}{8}\sqrt{\pi\delta} &\frac{5(7\sqrt{2}-3\sqrt{6})}{32}\left(\pi\delta\right) & \frac{9(2\sqrt{6}-\sqrt{2})}{256}\left(\pi\delta\right)\\
\hline
-\frac{(3\sqrt{2}-\sqrt{6})}{8}\sqrt{\pi\delta}& \frac{9-5\sqrt{3}}{2}\left(\pi\delta\right) &0&0\\
\hline
\frac{5(7\sqrt{2}-3\sqrt{6})}{32}\left(\pi\delta\right) &0&0&0\\
\frac{9(2\sqrt{6}-\sqrt{2})}{256}\left(\pi\delta\right) &0&0&0
\end{array}
\right),
\ee
\be
V^{LL'}_{0012}&=&\left(\renewcommand\arraystretch{1.5}
\begin{array}{c|c|cc}
\frac{\sqrt{3}}{4} +\frac{(21-13\sqrt{3})}{4}\left(\pi\delta\right)&\frac{(2\sqrt{6}-3\sqrt{2})}{4}\sqrt{\pi\delta} &\frac{(2\sqrt{2}+\sqrt{6})}{8}\left(\pi\delta\right) & \frac{9\sqrt{2}}{128}\left(\pi\delta\right)\\
\hline
0& 0 &0&0\\
\hline
\frac{(31\sqrt{2}-17\sqrt{6})}{16}\left(\pi\delta\right) &0&0&0\\
\frac{9\sqrt{2}}{128}\left(\pi\delta\right) &0&0&0
\end{array}
\right).
\ee
\end{widetext}
\section{Extensions\label{extensions}}
\subsection{General Lattices}
A Harper-like discrete Schr\"{o}dinger equation can be derived for many closely related models, and the perturbative approach outlined above follows accordingly. In Appendix \ref{otherlattice} we consider the anisotropic square lattice, the square lattice with next-nearest-neighbour hopping and the triangular lattice, giving the discrete Schr\"{o}dinger equation in each case, along with the energy bands and wavefunction corrections for vanishing flux. In each case we find that the wavefunctions include the higher Landau Level corrections that allow them to adopt the symmetry of the lattice ($\ket{n\pm2}$ for the anisotropic square lattice and $\ket{n\pm6}$ for the triangular lattice). For the triangular lattice the wavefunction corrections occur at second order, and so perturbatively this lattice is `closer' to the continuum.
\subsection{Higher Hofstadter Bands}
Pseudopotentials extend easily to higher Hofstadter bands through the inclusion of additional ladder operators. The overlap integrals we are interested in are then
\be
V^{LL'}_{P\mathrm{th\,LL}}&=&\bra{N_L,0;0,0}\frac{\anj{P}}{\sqrt{P!}}\frac{\ani{P}}{\sqrt{P!}}U_{i,j}\delta\left(z_1-z_2\right)\times\\
&&U^\dagger_{i,j}\frac{\adi{P}}{\sqrt{P!}}\frac{\adj{P}}{\sqrt{P!}}\ket{N_L',0;0,0}.
\ee
We may rewrite the ladder operators in our relative and centre of mass coordinate basis, but the procedure in general remains the same as for the lowest band. We find that the wavefunction corrections have larger coefficients for higher bands, but that there are more stringent conditions on $N$ for the flat band limit to be valid.
\subsection{General Interactions}
In order to consider fermions and more realistic systems, we need to go beyond the simple delta function interaction. Since we are on a lattice, it is natural to generalise the on-site interaction to a site-site interaction, $V(z_1,z_2)=U\delta\left(z_1-z_2-(u+iv)\right)$, with $u+iv$ the relative lattice vector of the interaction. On its own, this extends the approach to systems with, for example, near-neighbour exchange interactions. For long-range (e.g., Coulombic) interactions, matrix elements can be calculated by summing over weighted site-site interactions using
\be
V(z_1,z_2)&=&\sum_{u,v}U(u,v)\delta\left(z_1-z_2-(u+iv)\right),
\ee
where $U(u,v)$ is the amplitude of the interaction between sites separated by a displacement $u+iv$. Pseudopotential matrix elements for a general interaction can then be calculated through
\be
V_{s_1s_2s_3s_4}^{LL'}&=&\sum_{u,v}U(u,v)\times\\
&&\bra{L;s_1,s_2}\delta\left(z_1-z_2-(u+iv)\right)\ket{L';s_3,s_4},
\ee
where the expectation value under the summation is a pseudopotential for a site-site interaction, $V_{s_1s_2s_3s_4}^{LL',u+iv}$. We outline the procedure to calculate these quantities below and give a full derivation of the final expression in Appendix \ref{genint}.

A shift in the delta function has three effects on the interaction matrix elements. First, the continuum-like wavefunctions are shifted relative to one another. We take this into account by acting on the wavefunctions with a translation operator $\hat{T}_{u+iv}$ that shifts the relative coordinate $z_R\to z_R+u+iv$, and by then setting $z_R$ to zero. This operator involves the complex derivatives,
\be
\renewcommand\arraystretch{1.5}
\begin{array}{cccccc}
\vec{\partial}_{z_R}&=&\left(\hat{L}-\hat{Y}^\dagger\right)/4l_B,&\vec{\partial}_{\bar{z}_R}&=&\left(\hat{Y}-\hat{L}^\dagger\right)/4l_B,
\end{array}
\ee
which mix relative angular momentum states (see Appendix \ref{twoparticles} for operator definitions). As long as $|u+iv|\ll l_B\sim\sqrt{N}$ then this gives a valid perturbation series for the initial and final wavefunctions. This operator should act in addition to the perturbation from the lattice and is written explicitly as
\begin{equation}
\hat{T}_{u+iv}=e^{(u+iv)\left(L-Y^\dagger\right)/(4l_B)+(u-iv)\left(Y-L^\dagger\right)/(4l_B)}.\label{tuiv}
\end{equation}

The second effect of the delta function (for $Q\geq2$) is to mix the different sublattice components of the wavefunction: amplitudes on different lattice sites are derived from different perturbation series. This effect is not captured by the (continuum) Taylor series approximation using $\hat{T}$, so we must enforce $\delta_{x_i,x_j+u}$ when expanding the wavefunction components instead of the usual $\delta_{x_i,x_j}$.

Finally, there is a phase factor which comes from shifting $y$ in each wavefunction. This is given by $e^{\pi i v(s_3-s_4+s_2-s_1)}$ as shown in Appendix~\ref{genint}. Overall, the pseudopotential matrix for a site-site interaction is given by the perturbative expression
\be
V_{s_1s_2s_3s_4}^{LL',u+iv}&=&e^{\pi i v(s_3-s_4+s_2-s_1)}\left[\hat{T}_{u+iv}U^\dagger_{s_1s_2}\ket{L',0;0,0}\right]^\dagger\times\\
&&\delta_{x_i,x_j+u}\delta(z_R)\delta(\bar{z}_R)\left[\hat{T}_{u+iv}U^\dagger_{s_3s_4}\ket{L,0;0,0}\right].
\ee
We give the pseudopotential matrix elements for a nearest-neighbour interaction with $\phi=M/N$ in Appendix~\ref{genint}.
\section{Discussion\label{discussion}}
\subsection{Many-body Energy Spectra}
In this section we discuss some possible uses for our perturbative method---although we leave a full investigation of these applications to elsewhere.

One natural use is in the calculation of energy spectra and wavefunctions. The single-particle energy levels we calculated previously describe the Landau level-like features of the (non-interacting) Hofstadter Butterfly very well. To find the many-body energy levels, we must diagonalise a suitable many-body Hamiltonian. For the case of Hofstadter bosons in an optical lattice, the Hamiltonian we should diagonalise is (up to a chemical potential shift)
\begin{eqnarray}
\hat{H}&=&\sum_{i<j}\sum_{LL'}\sum_{\{s\}}V^{LL'}_{s_1s_2s_3s_4}\ket{L;i,j;s_1s_2}\bra{L';i,j;s_3s_4}\nonumber\\
&&+V_h\sum_\mathbf{r}|\mathbf{r}|^2\hat{n}_\mathbf{r},\label{mbham}
\end{eqnarray}
where $\ket{L;i,j;s_1s_2}$ is a two-particle state with relative angular momentum $L$ and species indices $s_1,s_2$. The term $\hat{V}_h=V_h\sum_\mathbf{r}|\mathbf{r}|^2\hat{n}_\mathbf{r}$ is a harmonic potential which confines the bosons to the centre of the trap.

\begin{figure}[t]

\includegraphics[clip=true,trim=4 0 50 0,scale=0.22]{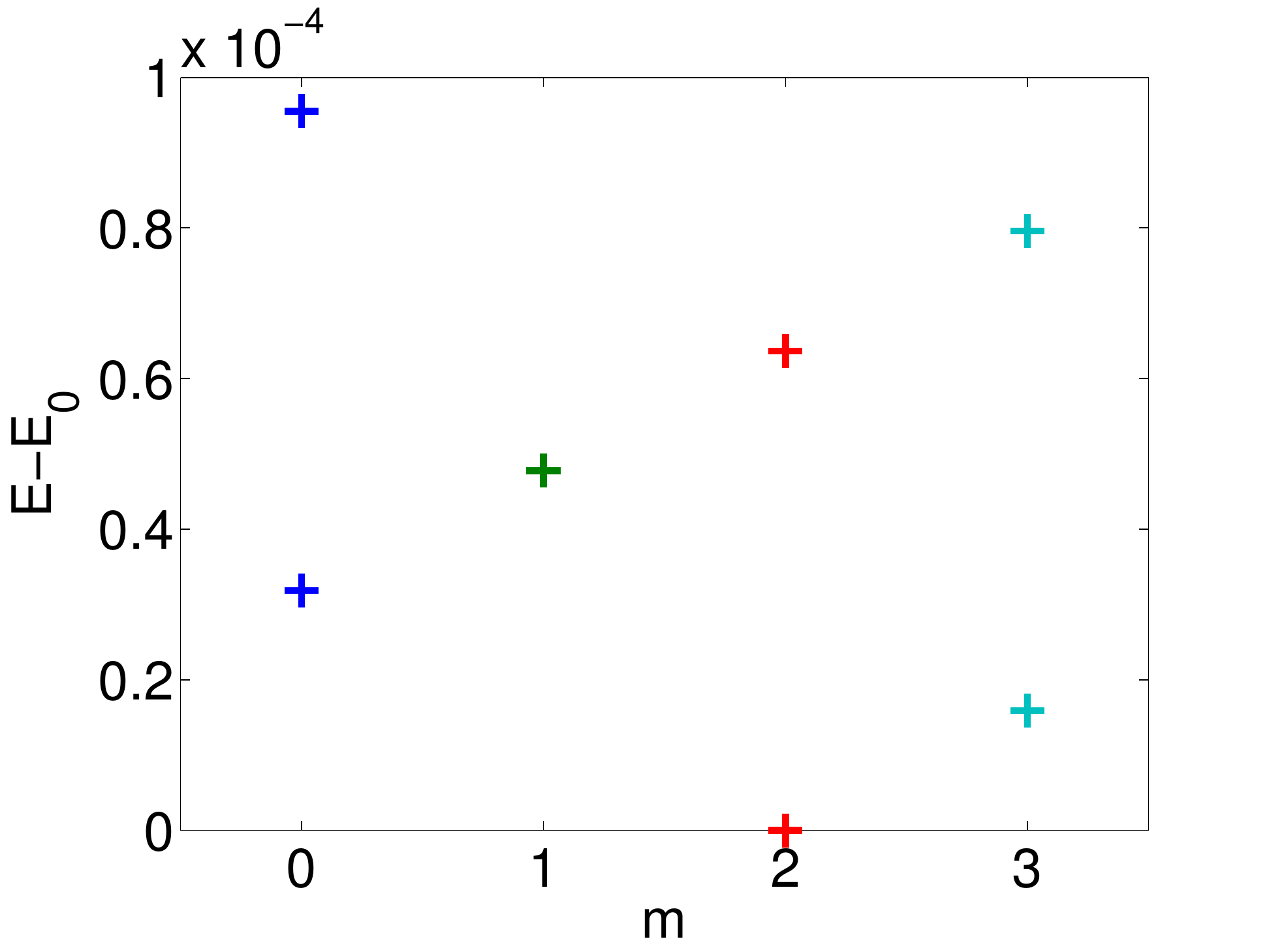}
\includegraphics[clip=true,trim=4 0 50 0,scale=0.22]{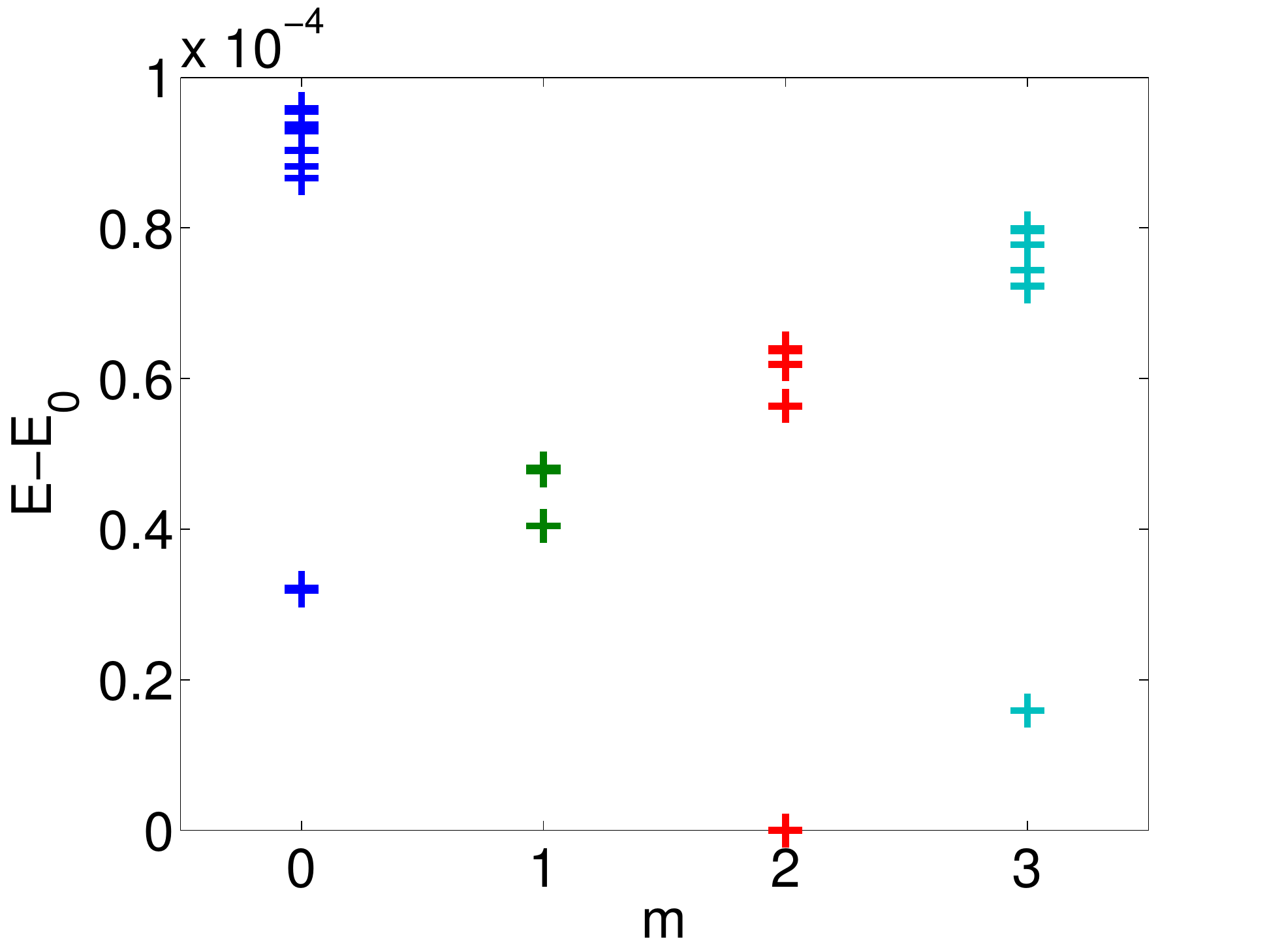}

\caption{Energy spectrum for three particles in the Hofstadter model with interaction strength ${V/(4\pi l_B^2)=10^{-3}}$, harmonic trap strength $V_h=10^{-5}$ and field strength given by $\delta=1/N=10^{-6}$ (left) and $\delta=1/N=1/5$ (right). The single particle basis states have a cutoff of $m_{\mathrm{max}}=10$, and only the lowest energy states are shown. The lowest energy linear part of spectrum starting at $m=2$ shows the Laughlin $\nu=1/2$ state and its edge excitations. The higher energy bulk excitations are not shown.}\label{threebspectrum}
\end{figure}

We numerically diagonalise this Hamiltonian for three particles in the small field  case and plot the result in Figure~\ref{threebspectrum}. Since angular momentum is no longer conserved, we must impose a cut-off on the maximum allowed single-particle momentum (otherwise the calculation would involve an infinite number of states): physically, the higher angular momentum states will be suppressed by the harmonic trap. Increasing this cutoff will allow more states to be observed in the spectrum, but this does not significantly affect the low-lying states. Angular momentum is conserved modulo four (as expected from the $C_4$ symmetry of the square lattice), and so there are only four possibilities on the $x$-axis in the figure.

The spectrum is shown for both $N=10^6$ (which is `close' to the continuum limit) and $N=5$---we recall that in our approximation we ignore the effect of the bandwidth, which is exponentially small in $N$. Both spectra have the bosonic Laughlin $\nu=1/2$ state as the ground state, which is located at momentum 6 ($=2$ modulo 4) for three particles. The other low-lying states are edge modes that have degeneracy $\{1,1,2,3,4,5,7\}$, as expected for three particles from the hydrodynamic Luttinger liquid edge theory.\cite{Wen:1995fg} The degeneracy is broken by the presence of the lattice, but it is likely that the states themselves are still well-described by the edge state picture due to the small coefficients in the pseudopotential matrix. The bulk states that lie above these edge states by the interaction energy scale are not shown here.  

This is the spectrum we would expect to be observed in experimental realisations of the Hofstadter model, where the system particles (cold atoms) are trapped by a weak harmonic potential. The spectra show good qualitative agreement with the exact diagonalisation studies of bosonic Laughlin edge states carried out in Ref.~\onlinecite{Kjall:2012db}, where the eigenstates are also shown to have a good overlap with the analytic edge states expected from the chiral Luttinger liquid theory. Similar angular momentum spectra have also been observed in numerical studies of other Chern insulator models. \cite{Luo:2013ul}

Since lattice calculations are computationally demanding for large numbers of particles, the pseudopotential approach may be useful for computing accurate numerical energy spectra with a much smaller set of (angular momentum) basis states. For $\phi\approx P/Q$, there will be $Q$ particle species to consider.

We are also able to apply our results to the torus geometry, as is more common in the context of Chern insulators. In this case it is useful to consider the linear momentum states $\ket{k_i}$ and interaction matrix elements $V_{\mathbf{k}_1\mathbf{k}_2\mathbf{k}_3\mathbf{k}_4}^{s_1s_2s_3s_4}$, as defined in the Landau gauge. 

A useful diagnostic tool in the study of Chern insulators is the two-particle energy spectrum, which exhibits bands that are analogous to the Haldane pseudopotentials.\cite{Lauchli:2013ck} This analogy can be seen directly through our pseudopotential matrices. For example, diagonalising the zeroth order pseudopotential matrix for $\phi\approx1/2$ yields three bosonic excited states with relative angular momentum $L=0$. These may be written in the rotated basis as 
\be
\{\ket{++}, \ket{--}, (\ket{+-}+\ket{-+})/\sqrt{2}\}
\ee
and have respective energies 
\be
V/(4\pi l_B^2)\times\{1,1,2\}.
\ee
In the two-particle spectrum on a torus, these excited states would correspond to bands in $\bk$-space with an energy ratio of $1:2$ and degeneracy ratio of $2:1$. As $N$ is made finite, these bands would gain structure from the lattice corrections, but may still be interpreted as arising from the $V^{00}$ pseudopotential element.

Similarly, for vanishing flux, the two-particle spectrum should show a single band, and for $\phi\approx1/3$ the spectrum should show two sets of bands with energy ratio $1:3$ and degeneracy ratio $3:3$. We leave a full study of the many-body Hofstadter spectrum on a torus to elsewhere.
\subsection{Many-body Wavefunctions\label{mbwf}}
The first order corrections to a known many-body wavefunction due to the effects of the lattice may be found simply by substituting the single particle perturbation series into the unperturbed wavefunction $\psi(\{z_i\})$ (provided any degeneracy is properly accounted for). For example, the $\nu=1/2$ state with $\phi=1/N$ becomes
\begin{equation}
\tilde{\psi}_{\nu=1/2}\left(\{z_i\}\right)=\prod_{1\leq j\leq D}U^\dagger_{j}\psi_{\nu=1/2}\left(\{z_i\}\right)\label{laughlinpert}
\end{equation}
where $\psi_{\nu=1/2}(\{z_i\})$ is the continuum $\nu=1/2$ state, $D$ is the number of particles, and $U^\dagger_{j}$ is the perturbation operator defined previously acting on particle $j$. This substitution works to first order because the only first-order elements in the pseudopotential matrix mix states with $L=0$ and $L=4$, and no pair of particles has relative angular momentum $L=0$ in the Laughlin ground state. In this way, the only corrections come from changes in the single-particle wavefunctions.

To find corrections at higher order, we must diagonalise the full many-body Hamiltonian (e.g. Eq.~\eqref{mbham}) and consider the resulting eigenstates. The general many-body wavefunctions therefore have corrections due to (i) perturbative corrections to the single-particle eigenstates and (ii) changes in the relative weights of the non-interacting states, which are characterised by the pseudopotential matrix. Our own preliminary numerical studies for small $\phi=1/N$ show that the wavefunctions produced in this way are very similar to the wavefunctions produced through exact diagonalisation of the Hofstadter lattice Hamiltonian.\footnote{We  considered three particles in the disk geometry with a parabolic confining potential and used either lattice sites or perturbed angular momentum eigenstates as basis states.} 

We note that the authors of Ref.~\onlinecite{Hafezi:2007gz} also considered bosonic Laughlin states on a lattice and compared the wavefunction overlap with the exact many-body state obtained from numerics. They observed substantial overlap (greater than $0.95$) up until $\phi\approx1/3$. We expect our perturbative wavefunctions to display similar overlap features with the exact lattice eigenstates, except that they should be algebraically more accurate (in $1/N$) in the large $N$ regime. According to the results given previously, the true many-body wavefunction should have both algebraically and exponentially small corrections, viz.
\be
\ket{\mathrm{GS}}&=&\mathcal{N}\left[\ket{\mathrm{GS}_{0}}+\frac{a}{N}\ket{\mathrm{E}_a}+ be^{-\tilde{\sigma} N}\ket{\mathrm{E}_e}\right].
\ee
In this expression, $\ket{\mathrm{GS}_0}$ is the unperturbed many-body (Laughlin) ground state, $\ket{\mathrm{E}_{a,e}}$ contains, respectively, all of the algebraically and exponentially small wavefunction corrections, $a$ and $b$ are some unspecified coefficients, and $\mathcal{N}$ is an overall normalisation constant. The overlap of the unperturbed Laughlin state with this true wavefunction will give
\be
\frac{\big\langle \mathrm{GS}_0 \big|{\mathrm{GS}}\big\rangle}{\sqrt{\big\langle \mathrm{GS}_0 \big|{\mathrm{GS}_0}\big\rangle\big\langle \mathrm{GS} \big|{\mathrm{GS}}\big\rangle}}&=&1-O\left(1/N^2\right)-O\left(e^{-\tilde{\sigma} N}\right),
\ee
where there is no correction at $O(1/N)$ because the first order perturbative correction is orthogonal to the ground state. 

Perturbation theory allows us to capture the algebraic corrections to the Laughlin wavefunction. At first order, by perturbing the component single-particle wavefunctions as described in Eq.~\eqref{laughlinpert}, the approximate many-body wavefunction exactly captures the first-order algebraic corrections,
\be
\ket{\widetilde{\mathrm{GS}}}&=&\ket{\mathrm{GS}_{0}}+\frac{a_1}{N}\ket{\mathrm{E}_{a_1}},
\ee
where $\ket{E_{a_1}}$ contains all of the first order corrections and $a_1$ is a coefficient. This state has overlap with the exact state of
\be
\frac{\big\langle \widetilde{\mathrm{GS}} \big|{\mathrm{GS}}\big\rangle}{\sqrt{\big\langle \widetilde{\mathrm{GS}} \big|\widetilde{\mathrm{GS}}\big\rangle\big\langle {\mathrm{GS}} \big|{\mathrm{GS}}\big\rangle}}&=&1-O\left(1/N^{3}\right)-O\left(e^{-\tilde{\sigma} N}\right),
\ee
and so is $O(1/N)$ more accurate than the unperturbed Laughlin state. In principle, higher order perturbative wavefunctions could be used to give even better overlap with the exact state, but this would require the diagonalisation of a many-body interaction matrix.

For small field values our perturbative wavefunctions should therefore be algebraically more accurate than the continuum Laughlin states (albeit with a small coefficient). For $\phi\approx1/3$ our flat-band approximation scheme would start to break down, the exponential corrections would become significant, and the overlap would decay as found in Ref.~\onlinecite{Hafezi:2007gz}. However, the topological nature of the ground state is expected to persist beyond the breakdown of the Laughlin description.\cite{Hafezi:2007gz}

For $\phi\approx P/Q$ the situation is more complicated. We noted that with $\phi\approx1/2$, a convenient basis change completely removes the umklapp terms to zeroth order. In this way, each species individually behaves like the $\nu=1/2$ Laughlin state, and the complete system behaves like the Halperin-221 state,\cite{Halperin:1983zz}
\be
\psi_{221}\left(\{z_i^\pm\}\right)&\sim&\prod_{i<j}\left(z_i^+-z_j^+\right)^2\left(z_i^--z_j^-\right)^2\prod_{i,j}\left(z_i^+-z_j^-\right)
\ee

where $z^\pm_i$ is the (complex) position of the $i$th particle with rotated species index $\pm$. In fact, to zeroth order, any unitary rotation of the species produces a Halperin-221 state with the same density and energy (zero), and so any of these states is an equally valid ground state. This includes a 221 state formed from the `unrotated' species, previously labelled by $0$ and $1$. 

Turning on the lattice breaks this degeneracy and also breaks rotational symmetry, meaning the 221 state is no longer an exact eigenstate. However, it is unlikely that the many-body state changes significantly for small $\delta$ due to the small coefficients in the pseudopotential matrix.

For $\phi\approx1/3$ there does not appear to be a single-particle species rotation that completely removes the zeroth-order Umklapp terms. Nonetheless, any Halperin-222111 state formed from three orthogonal species is an allowed zero-energy ground state at lowest order. We recall that in this case (and for larger values of the denominator $Q$) there will generally also be corrections to the wavefunction at half-integer powers of $1/N$.

The discussion above considers states in the disk geometry, but this may be adapted to the torus by instead considering perturbed linear momentum states $\ket{k_i}$ and interaction matrix elements $V_{\bk_1\bk_2\bk_3\bk_4}^{s_1s_2s_3s_3}$. In this case, there will be a subtle dependence on the size of the system, since species may map into one another across the boundary---the many body states will in general be analogous to the \emph{colour-entangled} Halperin states of Ref.~\onlinecite{Wu:2013ii}, with our species labels corresponding to colour. 

Reference~\onlinecite{Wu:2013wa} considers the specific colour-entangled states formed when layers of the Hofstadter model (each individually corresponding to small flux densities) are stacked together in different conformations. The resulting many-body states have a non-trivial ground state degeneracy that depends on the stacking arrangement and system size. These features also determine whether the states may be interpreted as a single-layer system or a multi-layer system. This is similar to our findings for flux values $\phi\approx P/Q$, where the non-interacting states act like $Q$ copies of a Landau level. The precise form of the fraction $\phi=p/q$ determines how the effective Landau levels are connected (in a similar way to the stacking of single Hofstadter layers in Ref.~\onlinecite{Wu:2013wa}). This feature, in conjunction with the length of the system modulo $Q$, will determine the degeneracy (and colour entanglement properties) of the many-body eigenstates on a torus.
\subsection{Relation to Wannier Orbitals and Lattice Dislocations}
Recent work by Qi and others\cite{Qi:2011jo,Wu:2012uh,Huang:2012ty} represents FCI states in terms of Wannier functions, which are wavefunctions that have been quasi-localised through a suitable unitary transformation acting on the Bloch states. In contrast to non-topological bands, where the wavefunction is single-valued in the Brillouin zone, Wannier states for bands with Chern number $C\neq0$ can only be localised in one direction. The perturbative wavefunctions we have described above are localised in the $x$-direction, and we might wonder how they relate to such Wannier functions, if at all.

Defining the Bloch state of a Chern band as $\psi_\mathbf{k}(\mathbf{r})=e^{i\mathbf{k}\cdot\mathbf{r}}u_\mathbf{k}(\mathbf{r})$ (with $u_\mathbf{k}(\mathbf{r})$ periodic in $\mathbf{r}$), the one-dimensional Wannier states are given by
\be
\ket{W(k_y,R)}&=&\int\frac{\mathrm{d}k_x}{\sqrt{2\pi}}e^{-ik_xR}e^{i\varphi(k_x,k_y)}\ket{\psi_\mathbf{k}},
\ee
where $R$ is the position of a unit cell in the $x$-direction (here assuming the unit cell spacing is one). Several choices of the phase $\varphi(k_x,k_y)$ are possible, but the choice that leads to maximum localisation in the $x$-direction is \cite{Barkeshli:2012vu}
\be
\varphi(k_x,k_y)=\frac{k_y}{2\pi}\int^{2\pi}_0a_y(0,p_y)\mathrm{d}p_y-\int^{k_y}_0a_y(0,p_y)\mathrm{d}p_y\\
+\frac{k_x}{2\pi}\int^{2\pi}_0a_x(p_x,k_y)\mathrm{d}p_x-\int^{k_x}_0a_x(p_x,k_y)\mathrm{d}p_x,
\ee
where $a_i(\mathbf{k})\equiv-i\bra{u_\mathbf{k}}\partial_{k_i}\ket{u_{\mathbf{k}}}$ is the (gauge-dependent) Berry connection. A notable feature of the Wannier states is that the centre of mass position of each (in terms of the unit cell spacing) is given by
\be
x_R(k_y)&=&\bra{W(k_y,R)}\hat{x}\ket{W(k_y,R)}\\
&=&R-\frac{1}{2\pi}\int^{2\pi}_0a_x(p_x,k_y)\mathrm{d}p_x.
\ee
Defining the single variable $K=k_y+2\pi R$, the centre of mass moves continuously to the right as a function of $K$ in a characteristic staircase manner (see, for example, Figure~1 of Ref.~\onlinecite{Qi:2011jo}).

In our approximation, we have ignored the exponentially small nonuniformity in the Berry curvature, which is what would lead to this staircase translation of the centre of mass. Instead, our centre of mass translates linearly with $K$ as in the pure Landau level case and the additional phase $\varphi(k_x,k_y)$ defined above is easily shown to vanish. If we were to form Wannier states from our Bloch wavefunctions given in Eq.~\eqref{finalblochstates}, the Wannier transformation would simply undo the Fourier transform we have put in by hand, and would return the local, perturbative wavefunctions that we derived in Section~\ref{genflux},
\be
\big\langle x,y\big|W(k_y,R)\big\rangle &=& e^{ik_yy}\tilde{\psi}^s_{l,k_y}\left(x-\frac{k_yN}{2\pi M}-\frac{sN}{QM}-R\right).
\ee
Here, $\tilde{\psi}^s_{l,k_y}$ is the $x$-dependent wavefunction from Eq.~\eqref{fullxwf} and $R\equiv mq$ is the location of the $m$th magnetic unit cell. In this way, the true Wannier states of the Hofstadter model tend towards our perturbative wavefunctions in the limit $N\to\infty$, assuming that some finite temperature or interaction energy overcomes the exponentially small splitting within the subband.

Since we have assumed flat Berry curvature, resulting in the trivial Wannier phase $\varphi(k_x,k_y)=0$, the perturbative Wannier states constructed above do not suffer from the finite-size orthogonalisation problems exhibited by the true Wannier states (see the discussion in Ref.~\onlinecite{Wu:2012uh}). Instead, perturbative Wannier functions corresponding to different $(k_y,R)$ are orthogonal up to exponentially small corrections (in $N$) due to the Gaussian tails permeating into neighbouring cells. The trade-off is that our perturbative states neglect to capture the interesting behaviour that accompanies Berry curvature fluctuations, and the exact form of the wavefunctions is only accurate in the large $N$ limit.

For $Q\geq2$ there will be several Wannier orbitals per unit cell, each labelled by a species index $s$ (this is in addition to the higher energy Wannier orbitals labelled by the Landau level index $l$). As noted previously, these states will take different smooth forms on each $x$-sublattice (defined by $\lambda$). 

\begin{figure}[t]

\includegraphics[scale=0.3]{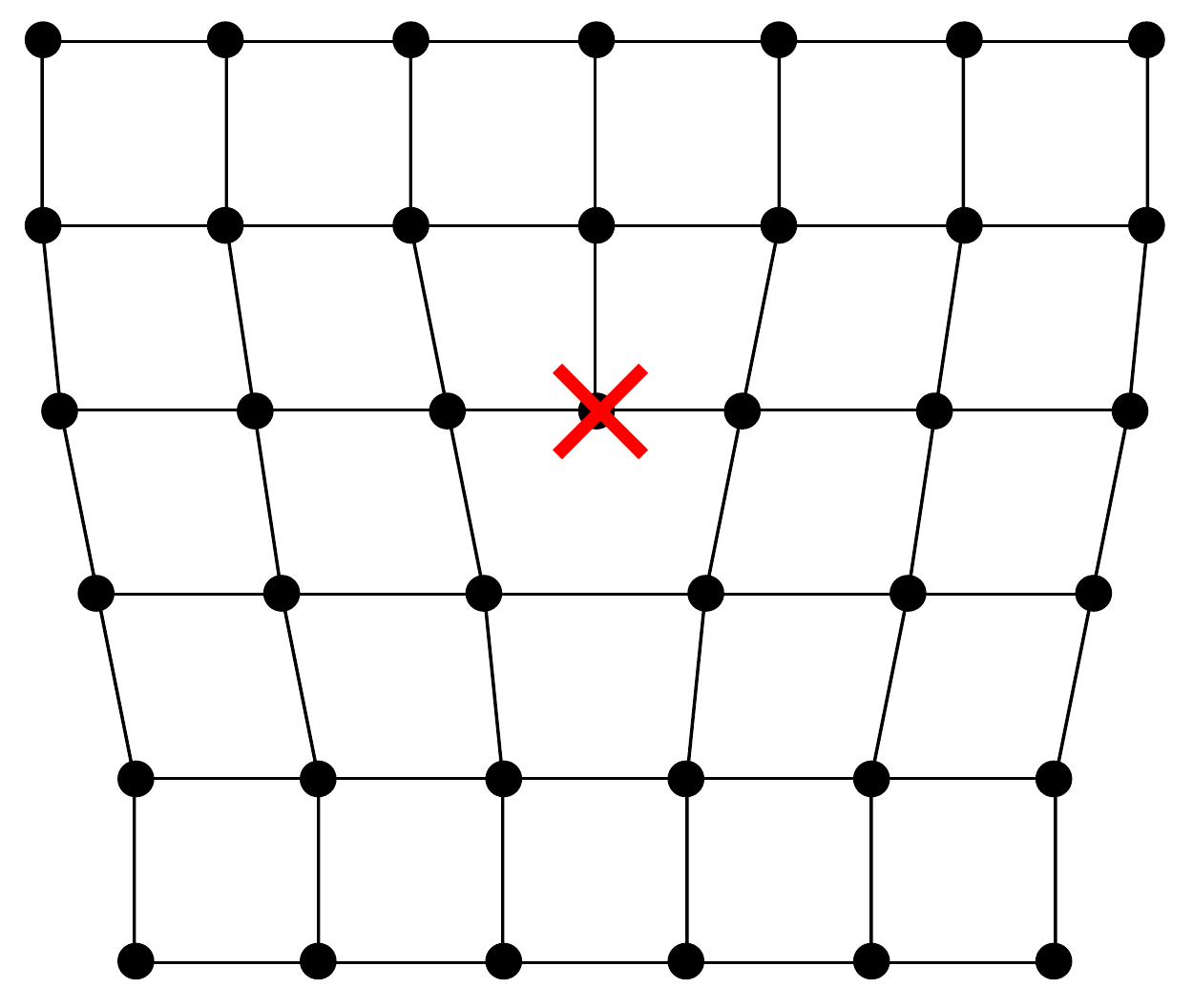}

\caption{Dislocation corresponding to a Burgers vector of $\mathbf{b}=-\hat{x}$.}\label{disloc}
\end{figure}

Barkeshli, Wen and Qi \cite{Barkeshli:2011ed,Barkeshli:2012vu} have considered multilayer systems analogous to these in the presence of lattice dislocations. Since the wavefunction amplitude depends sensitively on the site index in the $x$-direction, translating a Wannier state around a dislocation in the $x$-direction will permute the sublattices, and so permute one species into another (see Figure~\ref{disloc}). This property of the system changes its effective topology: if we have two such defects in a bilayer system, then the layers become connected through a wormhole and the system gains a topological degeneracy.\cite{Barkeshli:2011ed}

Multilayer systems formed from stacked Hofstadter layers were considered in this context in Ref.~\onlinecite{Wu:2013wa}, as discussed briefly at the end of Subsection~\ref{mbwf}. Such systems appear to form a natural setting in which to observe these topological defects and associated `topological nematic states':\cite{Barkeshli:2012vu} we expect these features to correspond directly to the multilayer states we observe near to $\phi\approx P/Q$.

In our case, we do not need to translate the wavefunction around an entire ($N\times1$) unit cell: we only need to translate it around the dislocation, which is on the order of the lattice spacing. Translation around a lattice dislocation with Burgers vector $\mathbf{b}=b_x\hat{x}+b_y\hat{y}$ will move the wavefunction weight formerly associated with sublattice $\lambda$ onto sublattice $\lambda'=\lambda+b_x$. If we consider the action of magnetic translation operators, we see that this is equivalent to changing $s\to s-Pb_x$ and $k_y\to k_y+2\pi b_x/Q$. Defining this translation around the dislocation by the operator $\hat{O}(\mathbf{b})$, we find
\be
\hat{O}(\mathbf{b})\big|\tilde{n},k_y,s\big\rangle&=&\big|\tilde{n},k_y+2\pi b_x/Q,s-Pb_x\big\rangle\\
\ee
where $\big|\tilde{n},k_y,s\big\rangle$ are the perturbative wavefunctions from Eq.~\eqref{lghofwf}.

If we wish to physically move the wavefunction around the defect, it makes more sense to consider the localised (for small enough $m$) symmetric gauge states discussed earlier in Section~\ref{symmgauge}, rather than the Wannier states which are only semi-localised. In the symmetric gauge, translation around a dislocation leads to
\be
\hat{O}(\mathbf{b})\big|\tilde{n},m,s\big\rangle&=&\big|\tilde{n},m,s-Pb_x\big\rangle,
\ee
and again permutes wavefunctions with different $s$-indices into one another. We leave a full discussion of dislocations in the Hofstadter model to elsewhere.

\section{Conclusions\label{conclusions}}
We have presented a straightforward and systematic approach to the study of generalised Hofstadter models near to simple, rational flux fractions. Using this method, the single-particle energy levels and (multi-component) wavefunctions at any field strength may be expressed as a perturbation series in the deviation from the nearby simple fraction. We find, using a combination of analytic and numerical methods, that the nonuniformity in the Berry curvature is exponentially small.  The main effect of the lattice in the context of fractional quantum Hall states is to break rotational invariance so that the wavefunctions take on the symmetry of the underlying lattice.

In deriving these perturbed wavefunctions we identify the subbands as the most useful structures of the Hofstadter model: with interactions turned on or at finite temperature, the constituent mini-bands will be blurred together and become unresolvable. Correspondingly, the total Chern number of the subband is the relevant topological quantity and in our approximation is shared equally between the constituent mini-bands.

From these single-particle states we can in turn calculate the Haldane pseudopotentials, which we find are connected smoothly to the pseudopotentials of the continuum. These interaction energies in principle give us all the information we need to find the supported many-body eigenstates. In particular, it is straightforward to generate numerical energy spectra for comparison with spectra that might be produced experimentally with cold atomic gases. More generally, knowledge of the pseudopotentials allows one to predict which of the known many-body FQHE states might be stabilised by the presence of a strong lattice.

We find near to $\phi=1/2$ that the system behaves like a quantum Hall bilayer to lowest order, and the system also seems to exhibit multilayer properties near to other fractions. In these situations, lattice dislocations may endow the states with novel topological properties, an area which requires further investigation.

As we have seen, by varying the flux in the Hofstadter model, one can tune the relative importance of the lattice in the fractional quantum Hall effect and make analytical progress in the study of fractional Chern insulators. We hope that our work will stimulate new investigations of the Hofstadter model, including further study of the multi-layer states and systems with more realistic interactions.

In this paper we have also investigated the Berry curvature of Hofstadter bands. The Berry curvature is the antisymmetric part of a tensor whose symmetric part, the quantum metric tensor, has also been suggested as playing a role in the physics of FCIs.\cite{Roy:2012vo,Dobardzic:2013we,Roy:u4ytErWz} This quantity is worthy of further investigation in the context of the Hofstadter model. Finally, the uniformity of the Berry curvature in the lowest Hofstadter band suggests that this would be an attractive venue for the experimental realisation of an FCI state.

\begin{acknowledgments}
We are grateful for many useful discussions with G.~M\"{o}ller, L.~Hormozi, N.~R.~Cooper, F.~J.~Burnell, and T.~Scaffidi, and thank the Aspen Center for Physics for its hospitality.  SHS and FH are supported by EPSRC grants EP/I032487/1 and EP/I031014/1. RR acknowledges support from the Sloan Foundation and UC startup funds.
\end{acknowledgments}

\appendix
\section{Flat Bands in the WKB Approximation\label{wkbappendix}}
In this appendix we use a semi-classical approach to explain the band-flattening exhibited by the Hofstadter model when $\phi=1/N$ and $N$ is large. The Harper equation has been studied many times before using the WKB approximation, as outlined in Section~\ref{history}; we are most interested in applying the approach of Ref.~\onlinecite{Watson:1991to} to the lowest band of the Hofstadter model for small flux. We extend Watson's original calculation to consider the second order turning points in the lowest band, and use the resulting wavefunction to calculate the Berry curvature. We note that the flattening we consider here is different from the \emph{exactly} flat bands and Berry curvature that arise in the Hofstadter model at certain special system sizes.\cite{Scaffidi:2014tg}

Beginning with the Harper equation (\ref{harper}), we follow Watson \cite{Watson:1991to} and define $x=m/N$ to rewrite the equation as
\be
\psi(x+1/N)+\psi(x-1/N)=2\cos\left[p(x)\right]\psi(x)
\ee
with
\be
p(x)&=&\cos^{-1}\left[-\epsilon/2-\cos(2\pi x)\right].
\ee
We have set $k=0$ here as in the continuum approximation it merely applies a translation in the $x$-direction: this will be reinstated later.

Assuming a semiclassical solution \footnote{We might worry that a discrete WKB approach is needed, but in his original paper \cite{Harper:1955uu}, Harper found this not to be the case.}
\be
\psi(x)&=&e^{NS_0(x)+S_1(x)},
\ee
we solve for $S_0$ and $S_1$ using Harper's equation to find the (oscillating and exponential) WKB solutions
\be
\psi^\pm_\mathrm{osc}(x)&=&\frac{1}{\sqrt{\sin p(x)}}\exp\left[\pm iN\int^xp(t)\,\mathrm{d}t\right]\\
\psi^\pm_\mathrm{exp}(x)&=&\frac{1}{\sqrt{\sinh\tilde{p}(x)}}\exp\left[\pm N\int^x\tilde{p}(t)\,\mathrm{d}t\right],
\ee
with
\be
\tilde{p}(x)&=&\cosh^{-1}\left[-\epsilon/2-\cos(2\pi x)\right].
\ee
Turning points occur whenever $p(x)=0$, or when $x=\pm\beta,1\pm\beta,2\pm\beta\ldots$ with
\be
\beta&=&\frac{1}{2\pi}\cos^{-1}\left(-\frac{\epsilon}{2}-1\right).
\ee

For the lowest band, the potential is so flat that two of these turning points coalesce to form a single, second order turning point. The exact solution in this region is a parabolic cylinder function ($U$ and $V$ in the notation of Ref.~\onlinecite{abramowitz1965handbook}), which sews together parts of the wavefunction from either side of the crossing. We expand this parabolic cylinder function to large values of its argument to arrive at the formulae that connect exponential WKB solutions from the left- and right-hand side of the trough,
\be
\left(\begin{array}{c}
U(a,-\xi)\\
V(a,-\xi)
\end{array}\right)
=
\left(\begin{array}{cc}
-\sin(\pi a) & \frac{\pi}{\Gamma\left(\frac{1}{2}+a\right)} \\
\frac{\cos(\pi a)}{\Gamma\left(\frac{1}{2}-a\right)} & \sin(\pi a)
\end{array}\right)
\left(\begin{array}{c}
U(a,\xi)\\
V(a,\xi)
\end{array}\right),
\ee
with definitions
\be
\epsilon&=&-4+\tilde{\epsilon}\\
b&=&4\pi^2\\
\xi&\equiv&\xi(x)=\left(4\phi^{-2}b\right)^{1/4}x\\
a&=&-(4b)^{-1/2}\phi^{-1}\tilde{\epsilon}\approx-\frac{1}{2}
\ee
From this, we propagate the WKB solution from the right-hand side of the trough across the classically forbidden region of the unit cell until we reach the next trough in the cosine potential. We can then impose Bloch periodicity on the entire wavefunction to obtain the band energy,
\begin{equation}\label{wkbeband}
\epsilon=-4+\frac{2\pi}{N}-\frac{4}{N}\sqrt{\frac{\pi}{e}}e^{-\sigma}\cos(k_xN),
\end{equation}
and periodic piecewise wavefunction
\begin{widetext}
\begin{eqnarray}
u(x,k_x,k_y)&=&\left\{\renewcommand\arraystretch{1.5}\begin{array}{rl}
(i)&e^{-ik_xn}A\left(\frac{N}{\pi}\right)^{\frac{1}{4}}\left[(2e)^{\frac{1}{4}}\left(1-\frac{\ln 2}{2}\left(\frac{1}{2}+a\right)\right)U\left[\xi(x-k_y/(2\pi)),a\right]\right.\\
&\left.+e^{-ik_xN}e^{-\sigma}(2e)^{-\frac{1}{4}}\left(1+\frac{\ln 2}{2}\left(\frac{1}{2}+a\right)\right)\sqrt{\frac{\pi}{2}}V\left[\xi(x-k_y/(2\pi)),a\right]\right]\\
(ii)&e^{-ik_xn}A\left[\psi_\mathrm{exp}^-(x-k_y/(2\pi))+e^{-ik_xN}e^{-\sigma}\psi_\mathrm{exp}^+(x-k_y/(2\pi))\right]\\
(iii)&e^{ik_x(N-n)}A\left(\frac{N}{\pi}\right)^{\frac{1}{4}}\left[(2e)^{\frac{1}{4}}\left(1-\frac{\ln 2}{2}\left(\frac{1}{2}+a\right)\right)U\left[\xi(x-k_y/(2\pi))-\xi_1,a\right]\right.\label{wkbwf}\\
&\left.+e^{-ik_xN}e^{-\sigma}(2e)^{-\frac{1}{4}}\left(1+\frac{\ln 2}{2}\left(\frac{1}{2}+a\right)\right)\sqrt{\frac{\pi}{2}}V\left[\xi(x-k_y/(2\pi))-\xi_1,a\right]\right],
\end{array}\right.
\end{eqnarray}
\end{widetext}
where the three pieces correspond to the regions
\be
(i)&:&0\leq x\leq\alpha\beta\\
(ii)&:&\alpha\beta\leq x\leq(1-\alpha\beta)\\
(iii)&:&(1-\alpha\beta)\leq x\leq1.
\ee

\begin{figure}[t]
\includegraphics[scale=0.7]{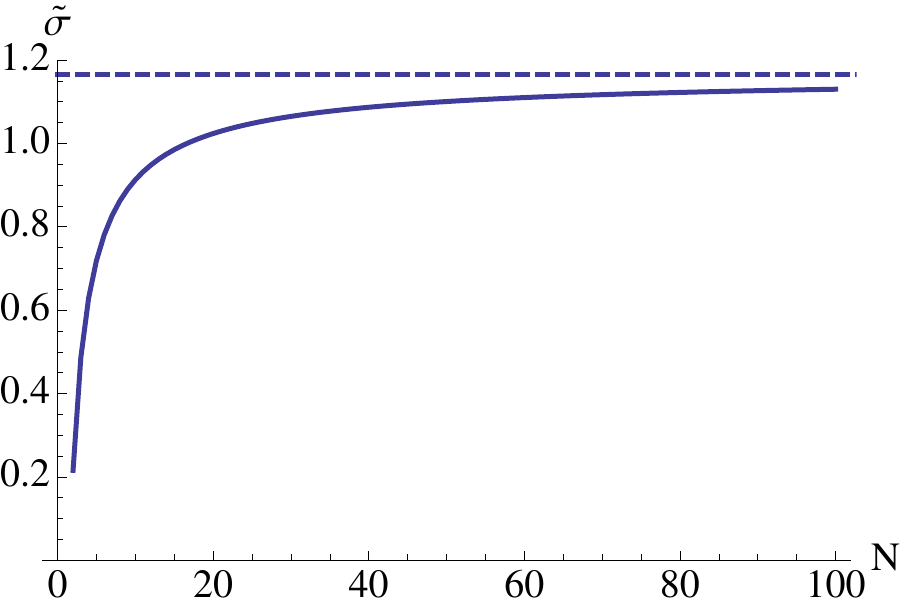}
\caption{The function $\tilde{\sigma}(N)$ (continuous line) and its limiting value (dashed line).}\label{sigplot}
\end{figure}

According to the definitions above, $\xi\equiv\xi(x)$ is an appropriately scaled version of $x$, $\xi_1\equiv\xi(1)$ and $a$ is very close to $-1/2$. In addition, $A$ is an overall normalisation constant and the quantity $\sigma$ describes the exponential decay of the bandwidth and Berry curvature through the implicit definition
\be
\sigma&=&N\int^{1-\beta}_{\beta}\cosh^{-1}\left(-\frac{\epsilon}{2}-\cos(2\pi t)\right)\mathrm{d}t\\
&\equiv&\tilde{\sigma}N\\
&\approx&1.166N.
\ee
The last line gives the value of $\sigma$ in the large $N$ limit, and we show $\tilde{\sigma}(N)$ in general in Figure~\ref{sigplot}.

In the expressions above, we have taken only the leading order algebraic and leading order exponential corrections [at $O(1/N)$ and $O(e^{-\tilde{\sigma}N})$ respectively]. The wavefunction is split into three parts: a parabolic cylinder solution near the troughs on the left and on the right of the magnetic unit cell, and a WKB solution in the classically forbidden region in the middle. 

There is some choice over where we choose to switch from the parabolic cylinder solution to the WKB solution, which we parameterise with $\alpha$. As the WKB solution is inaccurate close to the turning point, we allow the parabolic cylinder function to permeate a distance proportional to $\beta N$ beyond the crossing, until $n=\alpha\beta N$. This remains an unconstrained parameter in our wavefunction, but we shall see that it has a very limited effect on the results. Normally we choose $\alpha=2.3$.

\begin{figure}[t]
\includegraphics[clip=true,trim=21 0 0 0,scale=0.47]{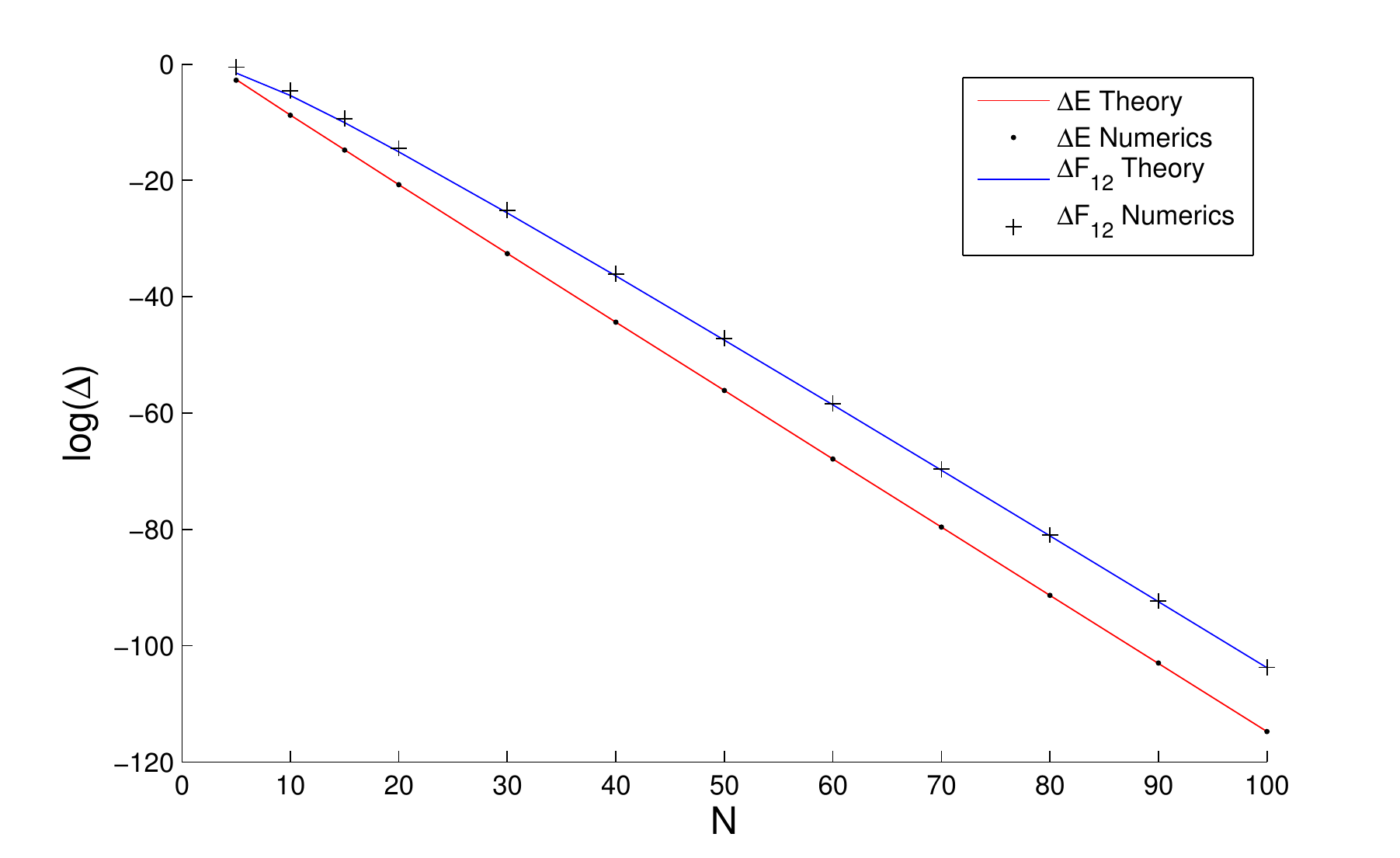}
\caption{Comparison of analytic bandwidth and Berry curvature with numerics ($\alpha=2.3$ here).}\label{devcomp}
\end{figure}

If we try to read off the bandwidth from Equation \eqref{wkbeband}, we notice that only the $k_x$-dependence has been captured: by assuming a continuous translational dependence on $k_y$ we have lost information. However, Harper's equation is self-dual under a Fourier transform,\cite{Aubry:1980vm} and it is possible to show from this that the bandwidth must be a function of $\cos(k_xN)+\cos(k_yN)$ (see, for example, Refs.~\onlinecite{Wilkinson:1984wd,Han:1994to}). The complete dispersion relation must therefore be
\begin{equation}\label{eband}
\epsilon=-4+\frac{2\pi}{N}-\frac{4}{N}\sqrt{\frac{\pi}{e}}e^{-\sigma}\left[\cos(k_xN)+\cos(k_yN)\right],
\end{equation}
giving a bandwidth ($\Delta E=E_\mathrm{max}-E_\mathrm{min}$) of
\begin{equation}
\Delta E=\frac{16}{N}\sqrt{\frac{\pi}{e}}e^{-\tilde{\sigma}N},
\end{equation}
where we make the $N$ dependence in the exponent explicit. This shows excellent agreement with the bandwidth obtained from exact diagonalisation, as demonstrated in Figure~\ref{devcomp}.

Similarly, we can calculate the Berry curvature from the semiclassical wavefunction. This calculation is lengthy and we do not give full details here, but one proceeds by inserting the wavefunction [Eq.~\eqref{wkbwf}] into the definition of the Berry curvature from Eq.~\eqref{berrycurvatureequation}. The leading exponentially small $\bk$-dependence stems from the classically forbidden region, and would not be captured by our perturbative calculation.

The WKB Berry curvature is found to be
\be
-iF(k_x)&=&\frac{N}{2\pi}+\left(\frac{\tilde{\sigma}P^2}{\sqrt{\pi e}\mathrm{Erf}(\alpha)}\right)N^2e^{-\tilde{\sigma}N}\cos(k_xN),
\ee
where the factor in brackets is of order unity and varies only very weakly with $N$. In this expression, $P$ is another calculable quantity that is close to unity, and is an average of the hyperbolic sine term in the WKB wavefunction,
\be
P&=&\Big\langle\frac{1}{\sqrt{\sinh\tilde{p}}}\Big\rangle_{\mathrm{WKB}}\\
&=&\Bigg\langle\frac{1}{\sqrt{\left(\frac{\epsilon}{2}-\cos(2\pi x)\right)^2-1}}\Bigg\rangle_{\mathrm{WKB}}.
\ee 
The average is taken over the range $\alpha\beta\leq x\leq 1-\alpha\beta$. Even though the Berry curvature depends on the free parameter $\alpha$, for any sensible value ($2\lesssim\alpha$), $\mathrm{Erf}(\alpha)$ and $P$ are very close to unity and vary by less than one percent. 

We see that only the $k_x$-dependence of the Berry curvature has been captured because we have again assumed a continuous translational dependence on $k_y$. However, as we show in Appendix~\ref{berrysymmetry}, the Berry curvature must be unchanged under the exchanges $k_x\to -k_y,k_y\to k_x$, and so can depend only on the expressions $\cos(k_xN)+\cos(k_yN)$ and $\cos(k_xN)\cos(k_yN)$. We argue that it must only be a function of the former as follows. 

First, if there were any dependence on the product $\cos(k_xN)\cos(k_yN)$, we would expect the coefficient of $\cos(k_xN)$ to depend on which value of $k_y$ we choose when we carry out the WKB analysis.  We would also expect this coefficient to be symmetric in the hopping amplitudes in the $x$- and $y$-directions (see Appendix~\ref{berrysymmetry}), but we find this not to be the case.

Secondly, we are able to calculate the Berry curvature of the lowest band analytically for certain simple values of $N$, which is achieved most easily using the method given in Ref.~\onlinecite{Barnett:2012id}. In the cited paper, the Hofstadter Berry curvature for $\phi=1/3$ is shown to be
\be
F_{N=3}(\bk)&=&\frac{3-2\cos\left(4\theta_\bk+\frac{2\pi}{3}\right)}{\sqrt{3}\left[1+2\cos\left(2\theta_\bk-\frac{2\pi}{3}\right)\right]^3}.
\ee
with
\be
\theta_\bk&=&\frac{1}{3}\arccos\left[-\frac{1}{2\sqrt{2}}\left(\cos(3k_x)+\cos(3k_y)\right)\right],
\ee
which is a function of $\cos(Nk_x)+\cos(Nk_y)$ only. From our own calculations we find the Hofstadter Berry curvature for $\phi=1/4$ to be
\be
F_{N=4}(\bk)&=&\frac{\sqrt{2}\left(6-\cos(4k_x)-\cos(4k_y)\right)}{\left(6+\cos(4k_x)+\cos(4k_y)\right)^{3/2}}
\ee
which is also only a function of $\cos(Nk_x)+\cos(Nk_y)$. In principle this could be carried out for higher values of $N$. In order for our WKB Berry curvature to connect to these analytic low-$N$ expressions, we would expect it also to be a function of $\cos(Nk_x)+\cos(Nk_y)$ only.

Finally, extensive numerical studies have verified that the leading Berry curvature deviation is exponentially small in $N$ and proportional to $\cos(Nk_x)+\cos(Nk_y)$. Subleading curvature deviations are suppressed by a further exponentially small factor and are proportional to $(\cos(Nk_x)+\cos(Nk_y))^2$. 

In this way, we replace $\cos(Nk_x)\to\cos(Nk_x)+\cos(Nk_y)$ in our WKB expression for the Berry curvature and obtain the leading Berry curvature deviation
\be
\Delta F&=&\frac{4\tilde{\sigma}N^2P^2e^{-\tilde{\sigma}N}}{\sqrt{\pi e}\mathrm{Erf}(\alpha)}.
\ee
This is plotted alongside the Berry curvature deviation obtained from exact diagonalisation in Figure \ref{devcomp}, and shows very good agreement. We find numerically that the corrections to both the bandwidth and Berry curvature expressions given above are suppressed by an additional exponential factor.

We expect that this WKB approach can be generalised to the other fractions discussed in the main text.\footnote{Ref.~\onlinecite{Watson:1991to} considers $\phi=M/N$ but assumes that the turning points are first order. The author finds that the bandwidth in this case still varies with the denominator $N$ according to $e^{-\tilde{\sigma}N}$.}

\section{Symmetries of the Berry Curvature}\label{berrysymmetry}
In this appendix we derive some of the symmetries of the Hofstadter Berry curvature by considering the system in different (Landau) gauges. To keep the conclusions as general as possible, we return to the Hofstadter Hamiltonian of Eq.~\eqref{hofeq}, but now allow the hopping amplitude to differ along each axis by letting $t\to(t_x,t_y)$.

 With the original gauge choice $\mathbf{A}^Y=(0,Bx,0)$, the discrete Harper equation is
 \begin{equation}\label{appendix_harperx}
- t_{x}\psi_{n-1}- t_{x}\psi_{n+1}-2t_{y}\cos\left(2\pi\phi n - k_y\right)\psi_{n}=\epsilon \psi_{n}.
\end{equation}
and the eigenstates take the form
\begin{eqnarray*}
\psi(x,y)&=&e^{ik_yy}\psi(x),
\end{eqnarray*}
as before. We have relabelled the lattice sites $(n_x,n_y)\to(x,y)$ in the eigenstates and set the lattice spacing to one.

If we instead choose the gauge field $\mathbf{A}^X=(-By,0,0)$, then the eigenvalue equation is
\begin{equation}\label{appendix_harpery}
- t_{y}\psi'_{n-1}- t_{y}\psi'_{n+1}-2t_{x}\cos\left(2\pi\phi n + k_x\right)\psi'_{n}=\epsilon \psi'_{n}.
\end{equation}
The Bloch wave is now in the $x$-direction and we may write the new eigenstates as 
\begin{eqnarray*}
\psi'(x,y)&=&e^{ik_x x}\psi'(y).
\end{eqnarray*}
We see that the substitutions 
\be
k_{x} &\rightarrow& k_{y}\\
k_{y} &\rightarrow& -k_{x}\\
t_{x} &\rightarrow& t_{y}\\
t_{y} &\rightarrow& t_{x}
\ee
transform Eq.~\eqref{appendix_harperx} to Eq.~\eqref{appendix_harpery}, and so it follows that the Berry curvature of the first system is related to that of the second through
\begin{eqnarray}
F^{Y}(k_{x},k_{y},t_{x},t_{y}) = F^{X}(-k_{y},k_{x},t_{y},t_{x}).  \label{berry-symm1}
\end{eqnarray}
The superscript on the Berry curvature $(X,Y)$ refers to the direction of the chosen Landau gauge vector potential $\mathbf{A}^{(X,Y)}$.

Next, we note that if $\psi_{n} $ is a solution to Eq.~\eqref{appendix_harperx} with $k_{y} = k$, then so is 
\be
\psi'_{n} = \psi_{n-1}
\ee
 with $k_{y} = k + 2\pi \phi$. We choose the magnetic flux per plaquette to be a general rational fraction $\phi=p/q$. From this, it follows that the Berry curvature is periodic as
\be
F^Y(k_{x} , k_{y})=F^Y(k_{x}, k_{y}+ 2\pi \phi).
\ee
However, we also know from Bloch periodicity that
\be
F^Y(k_{x}, k_{y})&=& F^Y(k_{x}, k_{y}+ 2\pi) \\
&=& F^Y(k_{x} + 2\pi/q,k_{y})
\ee
and so for a system with gauge choice $\mathbf{A}^Y$, the full periodicity of the Berry curvature  is 
\begin{eqnarray}
 F^Y(k_{x},k_{y})&=& F^Y(k_{x}, k_{y} + 2\pi/q) \nonumber \\
  &=&F^Y(k_{x} + 2\pi/q,k_{y}). ~\label{berry-relation1}
\end{eqnarray}

Next we consider a $q\times q$ unit cell and define the operators $\hat{T}_{q\hat{x}}$ and $\hat{T}_{q\hat{y}}$ which translate by $q$ plaquettes in the $x$- and $y$-directions respectively. Let $V^{\alpha}_{Y}(\mathbf{k})$ be the vector space generated by any set of simultaneous eigenstates of $ \hat{T}_{q\hat{x}}$ and $\hat{T}_{q\hat{y}}$, with respective eigenvalues $e^{iqk_{x}}$ and $e^{iqk_{y}}$, which belong to a given band $\alpha$ of the Hamiltonian with the gauge choice $\mathbf{A}^Y$. 

Let $\psi^{i}_{Y}(\bk,x,y)=\big\langle x,y\big|\psi^{i}_{Y}(\bk)\big\rangle$ be a set of orthonormal eigenstates which forms a basis for $V^\alpha_{Y}(\mathbf{k})$, and let
\be
u^{i}_{Y}(\bk,x,y) =\psi^{i}_{Y}(\bk,x,y)   e^{-i k_xx-ik_yy}
\ee
be the periodic parts of these Bloch states that satisfy
\be
u^{i}_{Y}(\bk,x,y)&=&u^{i}_{Y}(\bk,x+ q,y)\\
&=& u^{i}_{Y}(\bk,x,y+ q).
\ee
The Berry curvature corresponding to this set of basis vectors is then 
\be
F(V_Y^\alpha(\bk))&=&-i  \sum_{i} \sum_{x=0  }^{q-1}  \sum_{y=0  }^{q-1} \left (\frac{\partial u^{i*}_Y}{\partial k_{x}}\frac{\partial u^{i}_Y}{\partial k_{y}} - \mathrm{H.c.} \right).
\ee
This sum is independent of the particular choice of basis vectors $\psi^{i}_{Y}(\bk,x,y)$ and hence, we may regard this as the Berry curvature associated with the vector space $V^{\alpha}_{Y}(\bk)$.

The wavefunctions in band $\alpha$ at wavevectors $\bk, \bk+ 2\pi\hat{k}_x/q,\bk+ 4\pi\hat{k}_x/q, \ldots$ form a basis for $V^{\alpha}_{Y}(\bk)$. Using this basis to calculate $F(V^{\alpha}_{Y}(\bk))$, we find using Eq.~\eqref{berry-relation1} that
\be
F(V^{\alpha}_{Y}(\bk)) = q F^{Y}(k_{x},k_{y},t_{x},t_{y}).
\ee
Similarly, for the gauge choice, $\mathbf{A}^X$ we have 
\begin{eqnarray}
F(V^{\alpha}_{X}(\bk)) = q F^{X}(k_{x},k_{y},t_{x},t_{y}). \label{berry-y}
\end{eqnarray}

Let $U_{YX}$ be the unitary operator that transforms the Hamiltonian associated with gauge choice $\mathbf{A}^X$ to the one with gauge choice $\mathbf{A}^Y$. Then, the wavefunctions $U_{YX}\ket{\psi^{i}_{X}(\bk)} $ also form a basis for $V^{\alpha}_{Y}(\bk)$. Using this basis to calculate the Berry curvature, we find 
\begin{eqnarray}
F(V^{\alpha}_{Y}(\bk)) = q F^{X}(k_{x},k_{y},t_{x},t_{y}) \label{berry-x}
\end{eqnarray}
From Eqs.~\eqref{berry-x},\eqref{berry-y} and \eqref{berry-symm1}, we finally deduce that 
\begin{eqnarray}
F^{Y}(k_{x},k_{y},t_{x},t_{y}) = F^{Y}(-k_{y},k_{x},t_{y},t_{x}). \label{berry-relation2}
\end{eqnarray}

In Appendix~\ref{wkbappendix}, we found that the leading $k_x$ dependence of the Berry curvature was proportional to $\cos(k_xN)$ for small magnetic flux $\phi=1/N$. Equation \eqref{berry-relation2} then requires that the full $\bk$-dependent curvature involve only the expressions $\cos(k_xN)\cos(k_yN)$ and $\cos(k_xN)+\cos(k_yN)$.

We note that if the leading correction to the Berry curvature is of the form $ A(t_{x},t_{y}) \cos(k_{x}N)\cos(k_{y}N)$, then  Eq.~\eqref{berry-relation2} would imply that $A(t_{x},t_{y}) = A(t_{y},t_{x})$.  The WKB correction does not have this symmetry, which suggests that this cannot be the leading term.
\section{Further Band Structure Considerations for $\phi=P/Q\pm\delta$\label{fraccancel}}
First we prove that there are only three possible cancelled forms for $\phi=P/Q+\delta$ when $\delta=1/N$ and $Q$ is prime. We stated previously that the greatest possible cancellation occurs when $N\mod{Q}=0$ and $(PN+Q)\mod{Q^2}=0$, leading to
\begin{equation}
\phi=\frac{P}{Q}+\frac{1}{N}=\frac{(PN+Q)/Q^2}{N/Q}.
\end{equation}
We might think that we could cancel another factor of $Q$ to arrive at
\be
\phi=\frac{(PN+Q)/Q^3}{N/Q^2}.
\ee
For this to be the case we would require
\begin{eqnarray}
PN+Q&=&kQ^3,\\
N&=&lQ^2
\end{eqnarray}
with $k$ and $l$ integers. Combining these equations gives the requirement
\be
Pl+\frac{1}{Q}&=&kQ.
\ee
The above equation clearly has no solution with only integer variables, and hence it is only possible to cancel at most two factors of $Q$.

We also show that it is not possible to cancel factors other than $Q$ when $Q$ is prime. We write $P$ and $N$ in terms of prime factors as 
\be
P&=&p_1p_2p_3\ldots,\\
N&=&n_1n_2n_3\ldots,
\ee
where one or more of the $n_i$ and $p_i$ may be the same but none of the $p_i$ may equal $Q$. The magnetic field fraction is then
\be
\phi=\frac{PN+Q}{QN}&=&\frac{(p_1p_2p_3\ldots )(n_1n_2n_3\ldots )+ Q}{Q(n_1n_2n_3\ldots)}.
\ee
From the denominator, if we do not cancel a factor of $Q$, we might try to cancel one of the factors of $n_i$. This would lead to
\be
\phi&=&\frac{(p_1p_2p_3\ldots )(n_2n_3\ldots) +\frac{Q}{n_1}}{Q(n_2n_3\ldots)},
\ee
which for $Q$ prime does not have a valid numerator unless $n_1=Q$.

Next we consider the general case with $\phi=P/Q+M/N$ and $Q$ not necessarily prime. This can lead to different factors cancelling in the magnetic field fraction and, correspondingly, different numbers of troughs and mini-bands in each magnetic unit cell. Despite this, the perturbation theory and pseudopotential calculations follow as in the main text.

In terms of prime factors we may now write
\be
\phi&=&\frac{\left(p_1p_2\ldots\right)\left(n_1n_2\ldots\right)+\left(q_1q_2\ldots\right)\left(m_1m_2\ldots\right)}{\left(q_1q_2\ldots\right)\left(n_1n_2\ldots\right)},
\ee
where no $p_i$ may be the same as any $q_i$ and no $m_i$ may be the same as any $n_i$. If one of the $q_i$ is equal to an $n_i$ then we can cancel this factor directly (setting $n_1=q_1=\nu$)
\be
\phi&=&\frac{\left(p_1p_2\ldots\right)\left(n_2n_3\ldots\right)+\left(q_2q_3\ldots\right)\left(m_1m_2\ldots\right)}{\nu\left(q_2q_3\ldots\right)\left(n_2n_3\ldots\right)}.
\ee
Other $q_i$ factors can be cancelled similarly if they are equal to an $n_i$. Each factor can also be cancelled for a second time if, in addition to the above,
\be
&&\left(p_1p_2p_3\ldots\right)\left(n_2n_3\ldots\right)+\left(q_2q_3\ldots\right)\left(m_1m_2m_3\ldots\right)\\
&\equiv&\frac{PN+QM}{\nu}=l\nu
\ee
for some integer $l$. Using a similar argument to above, we cannot cancel a third factor unless $q_i$ is a repeated factor of $Q$. There are no other cancellation possibilities.

Overall, we can cancel each prime factor $q_i$ of $Q$ up to two times, depending on the value of $(PN+QM)\mod{q_i^2}$. The possible denominators of this cancelled fraction (which also give the magnetic unit cell size) are then $QN$ divided by any product of different $q_i$ and/or $q_i^2$. The largest possible denominator is $QN$ and the smallest $N/Q$ as in the prime $Q$ case. The number of spatial centres within a magnetic unit cell is given by the cell size divided by $N/(QM)$, which has the possible values $MQ^2$ divided by any product of different $q_i$ and/or $q_i^2$.

In the main text we show that, for the $Q$ prime case, the three different cancellation possibilities lead to the same subband structure. This remains true for the case of general $Q$ and $M$: there are more cancellation possibilities but the total Chern number, total Berry curvature and total bandwidth of the complete subband behave in the same way.

\section{Expanding Harper's Equation Near Other $\mathbf{k}$-points\label{otherkpoints}}
In the main text we found the lowest energy states by expanding near the Brillouin zone points $(k_x,k_y)=(0,2r\pi/Q)$. These are the correct points to expand about because they correspond to troughs in the lowest energy band of the pure $\phi=P/Q$ Hofstadter model from which the expansion is derived. From the symmetry of the energy bands, other troughs in the lowest energy band are located at $(k_x,k_y)=(2m\pi/Q,2r\pi/Q)$, while the peaks are located at $(k_x,k_y)=((2m+1)\pi/Q,(2r+1)\pi/Q)$---although in the Landau gauge only the $Q$ repeating units in the $y$-direction are distinct.

The $Q$ bands alternate the orientation of their curvature as they increase in energy, so the $\mathbf{k}$-points of the peaks of the highest band are located at even multiples of $\pi/Q$ if $Q$ is even and at odd multiples of $\pi/Q$ if $Q$ is odd. Peak and trough locations in intermediate bands can be worked out accordingly.

We find numerically that there are harmonic oscillator-like wavefunctions and energy levels near the peaks and troughs of \emph{all} of the $Q$ original bands. We considered only the lowest bands in the main text, but our approach should be applicable near to all $2Q$ band extremities. We simply need to set $(k_x,k_y)$ to appropriate values and choose the zeroth order solution for $\epsilon$ that corresponds to the desired original band energy.

\begin{figure}[t]

\includegraphics[scale=0.4]{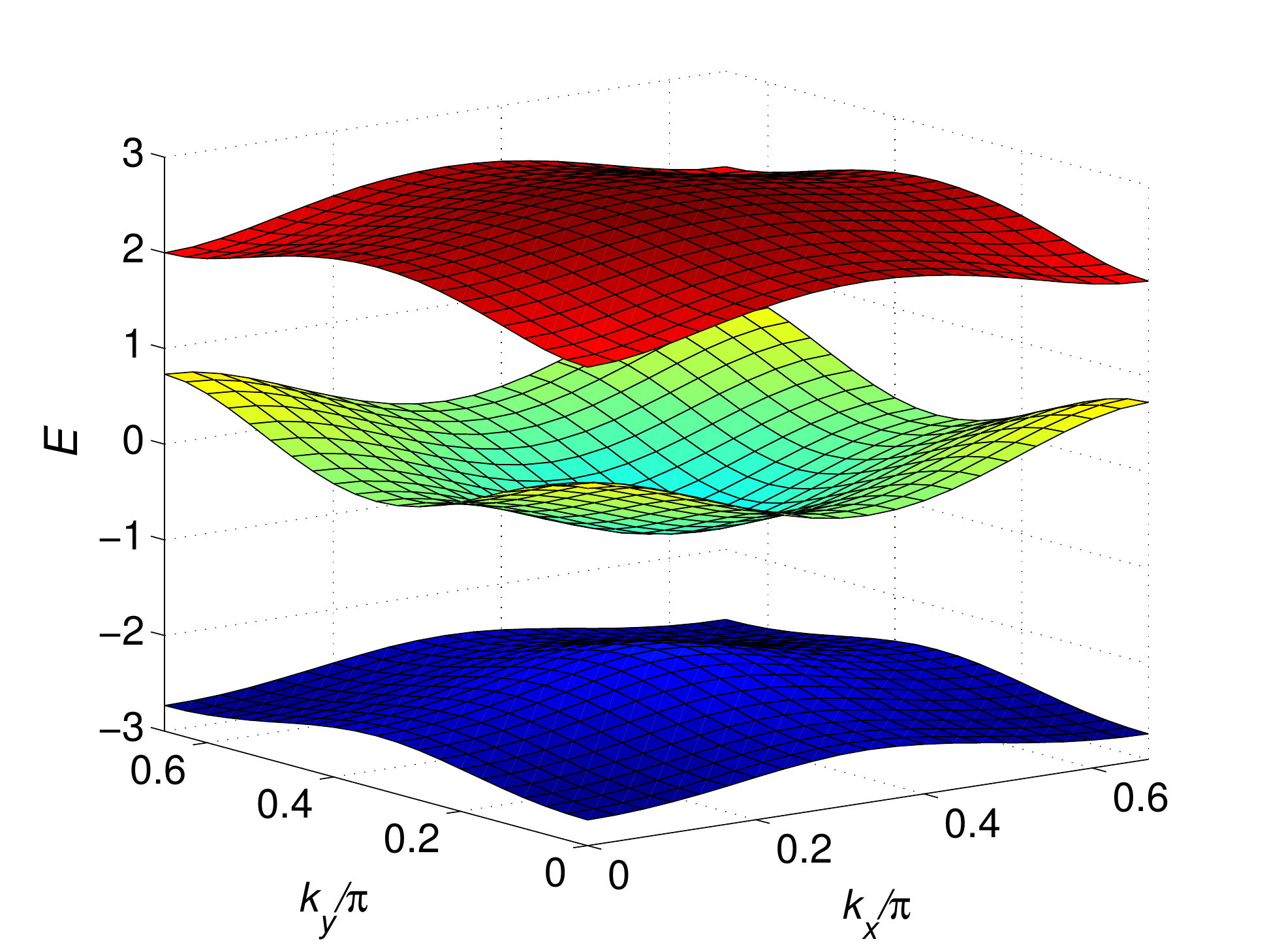}

\caption{Band structure for a repeating unit of the Brillouin zone for $\phi=1/3$ (the full Brillouin zone has two additional copies of this section in the $k_y$ direction). The bands alternate in orientation from top to bottom and the troughs and peaks are located at $(k_x,k_y)=(0,0)$ and $(k_x,k_y)=(\pi/3,\pi/3)$. We may expand about any of these extreme points.}\label{thirdbandsfull}
\end{figure}

For even multiples of $\pi/Q$ we can set $k_x=0$ and expand about the $Q$ different values $k_y=2r\pi/Q$ as described in the main text. For odd multiples of $\pi/Q$ we should be at $k_x=\pi/Q$ in the original system and cycle through values $k_y=(2r+1)\pi/Q$. To set $k_x=\pi/Q$ we change $\psi_n\to e^{i\pi n/Q}\psi_n$ in Harper's equation as if the $\psi_n$ were Bloch solutions. The true Bloch wavenumber of the perturbed system is fixed by the magnetic unit cell size to lie between $0\leq k_x\lesssim 2\pi/N$: the larger contribution of $\pi/Q$ ensures we are perturbing about the correct point in the original $P/Q$ system. Figure \ref{thirdbandsfull} shows the band structure for a repeating unit of the Brillouin zone for $\phi=1/3$, which provides six sets of extremities about which we may expand.

\section{Discrete Difference Equations for some $P/Q$ \label{example_harper}}
Below we give the discrete difference equations near to some simple flux filling fractions. As always, we will assume that $M/N$ is small.
\subsection{$\phi=M/N$}
For close to vanishing flux, there is only one sublattice, and we can use perturbation theory on Harper's equation directly,
\be
-\psi_{n-1}-\psi_{n+1}-2\cos\left(\frac{2\pi Mn}{N}-k_y\right)\psi_n=\epsilon \psi_n.
\ee

\subsection{$\phi=1/2\pm M/N$}
Close to $\phi=1/2$, the wavefunction depends on whether the site index is even or odd. Writing
\be
n=\pm\left(\frac{k_y N}{2\pi M}+\frac{sN}{QM}+n'\right),
\ee
where the sign is the same as in the definition of $\phi$, Harper's equation simplifies to
\be
-\psi^{j(\lambda-1)}_{n'-1}-\psi^{j(\lambda+1)}_{n'+1}&&\\
-2(-1)^{j(\lambda)}\cos\left(\frac{2\pi Mn'}{N}\right)\psi^{j(\lambda)}_{n'}&=&\epsilon \psi^{j(\lambda)}_{n'}.
\ee
(where $j(\lambda)\equiv j(\lambda,s)=(P\lambda+s)\mod{Q}$). We can eliminate sites of each parity in turn to find
\begin{eqnarray*}
\frac{\psi^{0}_{n'-2}+\psi^{0}_{n'}}{\epsilon-2\cos\left(\frac{2\pi M(n'-1)}{N}\right)}+\frac{\psi^{0}_{n'+2}+\psi^{0}_{n'}}{\epsilon-2\cos\left(\frac{2\pi M(n'+1)}{N}\right)}\nonumber\\
=\left[\epsilon+2\cos\left(\frac{2\pi Mn'}{N}\right)\right]\psi^{0}_{n'}
\end{eqnarray*}
for $j$ even and
\begin{eqnarray*}
\frac{\psi^{1}_{n'-2}+\psi^{1}_{n'}}{\epsilon+2\cos\left(\frac{2\pi M(n'-1)}{N}\right)}+\frac{\psi^{1}_{n'+2}+\psi^{1}_{n'}}{\epsilon+2\cos\left(\frac{2\pi M(n'+1)}{N}\right)}\nonumber\\
=\left[\epsilon-2\cos\left(\frac{2\pi Mn'}{N}\right)\right]\psi^{1}_{n'}
\end{eqnarray*}
for $j$ odd. We can use these equations to find both the lowest and highest energy states by choosing (respectively) $\epsilon=-2\sqrt{2}$ or $\epsilon=+2\sqrt{2}$ to zeroth order.
\subsection{$\phi=1/3\pm M/N$}
Close to $\phi=1/3$, there are three types of site index and Harper's equation for the lowest energy states takes the form
\be
-\psi^{j(\lambda-1)}_{n'-1} - \psi^{j(\lambda+1)}_{n'+1} &&\\
- 2\cos\left(\frac{2\pi j(\lambda)}{3}+\frac{2\pi Mn'}{N}\right)\psi^{j(\lambda)}_{n'}&=&\epsilon \psi^{j(\lambda)}_n.
\ee
Defining the functions
\be
A(n')&=&2\cos\left(\frac{2\pi Mn'}{N}\right)\\
B(n')&=& 2\cos\left(\frac{2\pi}{3}+\frac{2\pi Mn'}{N}\right)\\
&=&-\cos\left(\frac{2\pi Mn'}{N}\right)-\sqrt{3}\sin\left(\frac{2\pi Mn'}{N}\right)\\
C(n')&=& 2\cos\left(\frac{4\pi}{3}+\frac{2\pi Mn'}{N}\right)\\
&=&-\cos\left(\frac{2\pi Mn'}{N}\right)+\sqrt{3}\sin\left(\frac{2\pi Mn'}{N}\right),
\ee
the three effective Harper equations for $\phi=1/3+M/N$ may be written as
\begin{widetext}
\begin{eqnarray*}
\left[\epsilon+A(n)\right]\psi_{n}^{0}&=&\frac{\psi_{n-3}^{0}-\left[\epsilon+B(n-2)\right]\psi_{n}^{0}}{1-\left[\epsilon+B(n-2)\right]\left[\epsilon+C(n-1)\right]}+\frac{\psi_{n+3}^{0}-\left[\epsilon+C(n+2)\right]\psi_{n}^{0}}{1-\left[\epsilon+C(n+2)\right]\left[\epsilon+B(n+1)\right]}\\
\left[\epsilon+B(n)\right]\psi_{n}^{1}&=&\frac{\psi_{n-3}^{1}-\left[\epsilon+C(n-2)\right]\psi_{n}^{1}}{1-\left[\epsilon+C(n-2)\right]\left[\epsilon+A(n-1)\right]}+\frac{\psi_{n+3}^{1}-\left[\epsilon+A(n+2)\right]\psi_{n}^{1}}{1-\left[\epsilon+A(n+2)\right]\left[\epsilon+C(n+1)\right]}\\
\left[\epsilon+C(n)\right]\psi_{n}^{2}&=&\frac{\psi_{n-3}^{2}-\left[\epsilon+A(n-2)\right]\psi_{n}^{2}}{1-\left[\epsilon+A(n-2)\right]\left[\epsilon+B(n-1)\right]}+\frac{\psi_{n+3}^{2}-\left[\epsilon+B(n+2)\right]\psi_{n}^{2}}{1-\left[\epsilon+B(n+2)\right]\left[\epsilon+A(n+1)\right]}.\\
\end{eqnarray*}
\end{widetext}

For $\phi=1/3-M/N$ we should make the usual substitution for $n$,
\be
n=-\left(\frac{k_yN}{2\pi M}+\frac{sN}{QM}+n'\right).
\ee
However, since increasing $n'$ by one \emph{decreases} $n$ by one, we cycle through the sublattices in the opposite sense. This requires us to interchange $B\leftrightarrow C$ in the equations above, which we find leads to different corrections to the energies and wavefunctions. The corresponding equations for $\phi=2/3\pm M/N$ can be easily derived by setting $P=2$ in $j=(P\lambda+s)\mod{Q}$.

The highest energy states are located at odd multiples of $\pi/3$ in the original $\phi=1/3$ Brillouin zone. To find these states we substitute $\psi_n\to e^{i\pi n/3}\psi_n$ and
\be
n=\pm\left(\frac{k_yN}{2\pi M}+\frac{(s+1/2)N}{QM}+n'\right)
\ee
to arrive at the discrete difference equation
\be
-e^{-i\pi/3}\psi^{j(\lambda-1)}_{n-1} - e^{i\pi/3}\psi^{j(\lambda+1)}_{n+1} &&\\
- 2\cos\left(\frac{2\pi }{3}\left(j(\lambda)+\frac{1}{2}\right)+\frac{2\pi Mn'}{N}\right)\psi^{j(\lambda)}_{n}&=&\epsilon \psi^{j(\lambda)}_n.
\ee
This time we define the functions
\be
A(n')&=&2\cos\left(\frac{\pi}{3}+\frac{2\pi Mn'}{N}\right)\\
&=&\cos\left(\frac{2\pi Mn'}{N}\right)-\sqrt{3}\sin\left(\frac{2\pi Mn'}{N}\right)\\
B(n')&=& 2\cos\left(\pi+\frac{2\pi Mn'}{N}\right)\\
&=&-2\cos\left(\frac{2\pi Mn'}{N}\right)\\
C(n')&=& 2\cos\left(\frac{5\pi}{3}+\frac{2\pi Mn'}{N}\right)\\
&=&\cos\left(\frac{2\pi Mn'}{N}\right)+\sqrt{3}\sin\left(\frac{2\pi Mn'}{N}\right),
\ee
which differ from before. The discrete difference equations for the three sublattices are written in terms of these new functions $A, B, C$ and have some additional sign changes due to the $\pi/3$ phases,
\begin{widetext}
\begin{eqnarray*}
\left[\epsilon+A(n)\right]\psi_{n}^{0}&=&\frac{-\psi_{n-3}^{0}-\left[\epsilon+B(n-2)\right]\psi_{n}^{0}}{1-\left[\epsilon+B(n-2)\right]\left[\epsilon+C(n-1)\right]}+\frac{-\psi_{n+3}^{0}-\left[\epsilon+C(n+2)\right]\psi_{n}^{0}}{1-\left[\epsilon+C(n+2)\right]\left[\epsilon+B(n+1)\right]}\\
\left[\epsilon+B(n)\right]\psi_{n}^{1}&=&\frac{-\psi_{n-3}^{1}-\left[\epsilon+C(n-2)\right]\psi_{n}^{1}}{1-\left[\epsilon+C(n-2)\right]\left[\epsilon+A(n-1)\right]}+\frac{-\psi_{n+3}^{1}-\left[\epsilon+A(n+2)\right]\psi_{n}^{1}}{1-\left[\epsilon+A(n+2)\right]\left[\epsilon+C(n+1)\right]}\\
\left[\epsilon+C(n)\right]\psi_{n}^{2}&=&\frac{-\psi_{n-3}^{2}-\left[\epsilon+A(n-2)\right]\psi_{n}^{2}}{1-\left[\epsilon+A(n-2)\right]\left[\epsilon+B(n-1)\right]}+\frac{-\psi_{n+3}^{2}-\left[\epsilon+B(n+2)\right]\psi_{n}^{2}}{1-\left[\epsilon+B(n+2)\right]\left[\epsilon+A(n+1)\right]}.\\
\end{eqnarray*}
\end{widetext}
As before, we must interchange $B\leftrightarrow C$ if we wish to consider the high energy states of $\phi=1/3-M/N$. Expansions near to intermediate band extrema may be obtained by choosing the correct set of equations from above and selecting the desired intermediate solution for $\epsilon$ at zeroth order.
\section{Perturbative Wavefunctions and Energies for some $P/Q$ \label{example_wavefunctions}}
\subsection{$\phi=M/N$\label{example_wavefunctions_1N}}
To second order the energy levels of the lowest subbands are (setting $\delta=M/N$)
\be
\epsilon_l&=&-4+4\left(\pi\delta\right)\left(l+\frac{1}{2}\right)-\frac{1}{2}\left(\pi\delta\right)^2\left(2l^2+2l+1\right)
\ee
and the unitary operator that generates the second-order wavefunctions from the unperturbed states is given by
\be
U^\dagger&=&\exp\left[\left(\frac{1}{96}\left(\pi\delta\right)+\frac{1}{128}\left(\pi\delta\right)^2\right)\left(\ad{4}-\an{4}\right)\right.\\
&&\left.+\frac{1}{320}\left(\pi\delta\right)^2\left(\ad{5}a-a^\dagger\an{5}\right)\right]
\ee
such that
\be
\ket{\tilde{l}}&=&U^\dagger\ket{l}.
\ee
The highest energy wavefunctions are identical, and the corresponding energies are given by setting $\epsilon_l\to-\epsilon_l$.
\subsection{$\phi=1/2\pm M/N$\label{example_wavefunctions_12}}
The lowest energy levels and wavefunctions are symmetric about $\phi=1/2$, with
\be
\epsilon_l&=&-2\sqrt{2}+2\sqrt{2}\left(\pi\delta\right)\left(l+\frac{1}{2}\right)-\sqrt{2}\left(\pi\delta\right)^2\left(l^2+l+\frac{3}{4}\right)
\ee
(and where we have again set $\delta=M/N$). The unitary operators are cumbersome to write out, so here we just give the perturbations to the LLL ($\ket{\tilde{l};j(\lambda,s)}$).
\be
\ket{\tilde{0};0}&=&\ket{0}+\left(\pi\delta\right)\left[\left(\frac{\sqrt{2}+2}{4}\right)\ket{2}+\frac{\sqrt{24}}{24}\ket{4}\right]\\
&&+\left(\pi\delta\right)^2\left[\frac{3\sqrt{2}+6}{8}\ket{2}+\frac{3\sqrt{6}}{16}\ket{4}\right.\\
&&\left.+\frac{\sqrt{5}+\sqrt{10}}{8}\ket{6}+\frac{\sqrt{70}}{48}\ket{8}-\frac{5+3\sqrt{2}}{24}\ket{0}\right]
\ee
\be
\ket{\tilde{0};1}&=&\ket{0}+\left(\pi\delta\right)\left[-\left(\frac{2-\sqrt{2}}{4}\right)\ket{2}+\frac{\sqrt{24}}{24}\ket{4}\right]\\
&&+\left(\pi\delta\right)^2\left[-\left(\frac{6-3\sqrt{2}}{8}\right)\ket{2}+\frac{3\sqrt{6}}{16}\ket{4}\right.\\
&&\left.-\left(\frac{\sqrt{10}-\sqrt{5}}{8}\right)\ket{6}+\frac{\sqrt{70}}{48}\ket{8}\right.\\
&&\left.-\left(\frac{5-3\sqrt{2}}{24}\right)\ket{0}\right]
\ee
In the Landau gauge, a ground state wavefunction is labelled by $s$ and involves contributions from all sublattices. We can write this in ket notation as
\be
\ket{\tilde{0};s}&=&\sum_{\lambda\in\{0,1\}}A_{j(\lambda,s)}\delta_{x,\lambda}^{(2)}\ket{\tilde{0};j(\lambda,s)}
\ee
with $j(\lambda,s)=(\lambda + s)\mod{2}$ and where the normalised amplitudes are found to be
\be
A_0&=&\sqrt{\frac{2-\sqrt{2}}{2}}\\
A_1&=&(1+\sqrt{2})A_0.
\ee
The real space wavefunction is given by
\be
\tilde{\psi}^{s}_{0,k_y}(x')&=&\big\langle x'\big|\tilde{0};s\big\rangle
\ee
with $x'$ an offset coordinate defined in the main text. We can form Bloch and Wannier states from these $x$-wavefunctions as outlined in Section~\ref{genfluxperturb}.

The highest energy states ($+$) are related to these low energy states ($-$) by the transformations
\be
\e_l^+&=&-\e_l^-\\
\ket{\tilde{0};j}^+&=&\ket{\tilde{0};j}^-\\
A_0^+&=&- A_1^-\\
A_1^+&=& A_0^-.
\ee
\subsection{$\phi=1/3\pm M/N$}
The energy levels and wavefunctions are asymmetric about one third and now have additional wavefunction corrections at order $\sqrt{M/N}$. 

The lowest energy states at $\phi=1/3+M/N$ have energy corrections
\be
\epsilon_l&=&-\left(1+\sqrt{3}\right)+\left(\pi\delta\right)\left[3\left(\sqrt{3}-1\right)l+\frac{(1+\sqrt{3})}{2}\right]\\
&&-\left(\pi\delta\right)^{2}\left[\frac{(1+5\sqrt{3})}{8}+\frac{3(\sqrt{3}-1)l}{4}\right.\\
&&\left.+\frac{9(-11+7\sqrt{3})l^2}{4}\right]
\ee
and corrections to the ground-state wavefunction:
\be
\ket{\tilde{0};0}&=&\ket{0}+\left(\pi\delta\right)\left[-\frac{(\sqrt{6}-\sqrt{2})}{4}\ket{2}+\frac{3\sqrt{6}}{16}\ket{4}\right]\\
\ket{\tilde{0};1}&=&\ket{0}+\left(\pi\delta\right)\left[\frac{(\sqrt{6}+\sqrt{2})}{4}\ket{2}+\frac{3\sqrt{6}}{16}\ket{4}\right]\\
\ket{\tilde{0};2}&=&\ket{0}+\left(\pi\delta\right)\left[\frac{(\sqrt{6}+\sqrt{2})}{4}\ket{2}+\frac{3\sqrt{6}}{16}\ket{4}\right]\\
\ee
(the half-order corrections to state $\ket{l}$ are proportional to $\ket{l-1}$ and so do not affect the ground state).

The lowest energy states at $\phi=1/3-M/N$ have energy corrections
\be
\epsilon_l&=&-\left(1+\sqrt{3}\right)+\left(\pi\delta\right)\left[3\left(\sqrt{3}-1\right)l+\frac{(-7+5\sqrt{3})}{2}\right]\\
&&-\left(\pi\delta\right)^{2}\left[\frac{(-191+125\sqrt{3})}{8}+\frac{3(41\sqrt{3}-65)l}{4}\right.\\
&&\left.+\frac{9(-11+7\sqrt{3})l^2}{4}\right]
\ee
and corrections to the ground state:
\be
\ket{\tilde{0};0}&=&\ket{0}+\left(\pi\delta\right)\left[\frac{(15\sqrt{2}-11\sqrt{6})}{12}\ket{2}+\frac{3\sqrt{6}}{16}\ket{4}\right]\\
\ket{\tilde{0};1}&=&\ket{0}+\left(\pi\delta\right)^{\frac{1}{2}}(\sqrt{3}-1)\ket{1}+\left(\pi\delta\right)\left[\frac{3\sqrt{6}}{16}\ket{4}\right.\\
&&\left.+\frac{(7\sqrt{6}+3\sqrt{2})}{12}\ket{2}+(-2+\sqrt{3})\ket{0}\right]\\
\ket{\tilde{0};2}&=&\ket{0}-\left(\pi\delta\right)^{\frac{1}{2}}(\sqrt{3}-1)\ket{1}+\left(\pi\delta\right)\left[\frac{3\sqrt{6}}{16}\ket{4}\right.\\
&&\left.+\frac{(7\sqrt{6}+3\sqrt{2})}{12}\ket{2}+(-2+\sqrt{3})\ket{0}\right].
\ee
These states \emph{do} have corrections at half order in $\delta$. In the Landau gauge, assuming the Harmonic oscillator states are normalised, the full wavefunction then takes the form
\be
\ket{\tilde{0};s}&=&\sum_{\lambda\in\{0,1,2\}}A_{j(\lambda,s)}\delta_{x,\lambda}^{(3)}\ket{\tilde{0};j(\lambda,s)}
\ee
with
\be
A_0&=&\sqrt{\frac{3+\sqrt{3}}{2}}\\
A_1=A_2&=&\frac{\sqrt{3}-1}{2}A_0.
\ee
The real space wavefunction is given by
\be
\tilde{\psi}^{s}_{0,k_y}(x')&=&\big\langle x'\big|\tilde{0};s\big\rangle.
\ee

The highest energy states may be found by expanding about a different point in the Brillouin zone, as described in Appendix~\ref{otherkpoints}. The high energy states ($+$) are related to the low energy states given above ($-$) through the substitutions
\be
\e_l^-&=&-\e_l^+\\
\ket{\tilde{0};j}^+&=&\ket{\tilde{0};j+1}^{-}\\
A_1^-&=&\sqrt{\frac{3+\sqrt{3}}{2}}\\
A_0^-&=&-e^{i\pi/3}\frac{\sqrt{3}-1}{2}A_1^-\\
A_2^-&=&-e^{-i\pi/3}\frac{\sqrt{3}-1}{2}A_1^-.
\ee
\section{Chern Number of the Mini-bands and Subbands\label{perturbative_chern}}
The Chern number of a mini-band can be extracted from the TKNN diophantine equation (\ref{tknn}) as described previously. If we increase the denominator of the magnetic field fraction by increasing $N$, then the Chern numbers of the mini-bands generically also increase. However, the total Chern number of a subband stays fixed and equal to $\pm Q$---this is found to be the case numerically and is justified below.

In our perturbative approximation, we ignore the tunnelling that endows the mini-bands with such large Chern numbers, and instead average over the Berry curvatures of the constituent mini-bands. We therefore expect the Chern numbers of our perturbative wavefunctions to be different from the true values, although the total Chern number of a subband should be the same.

Following the approach of Thouless et al in Ref.~\onlinecite{Thouless:1982wi}, we can understand the Chern number of a mini-band intuitively by imagining pumping $k_y$ (e.g., through an adiabatically applied $\mathbf{E}$-field) and observing how the weight of the wavefunction moves in the $x$-direction. Recalling the simple Landau gauge picture of a Schr\"{o}dinger equation with a cosine potential of period $q$, we see that increasing $k_y$ moves the potential across the unit cell, and changes the potential energy on each lattice site. We expect most of the weight of the $r$th band to be centred on the lattice site with the $r$th lowest energy, and so the dominant weight of each band moves with the cosine potential: whenever two sites reach the same potential energy, the weight of the band hops from one site to the other. 

\begin{figure}[t!]
\begin{overpic}[scale=0.4]{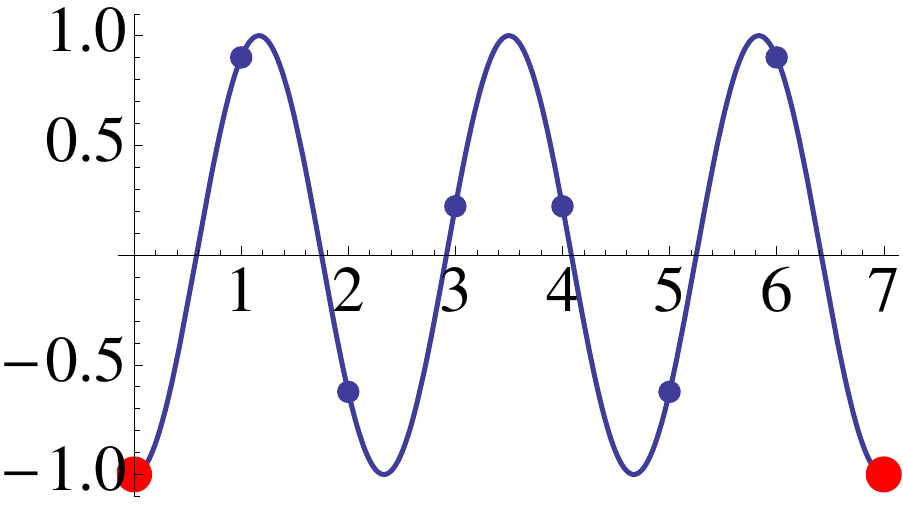}
\put(6,18){\color{black} (a)}
\end{overpic}
\begin{overpic}[scale=0.4]{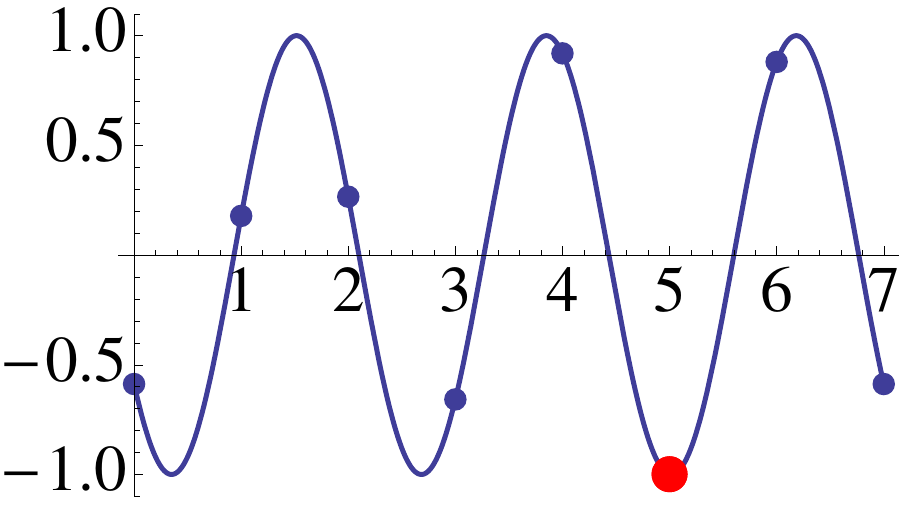}
\put(6,18){\color{black} (b)}
\end{overpic}
\begin{overpic}[scale=0.4]{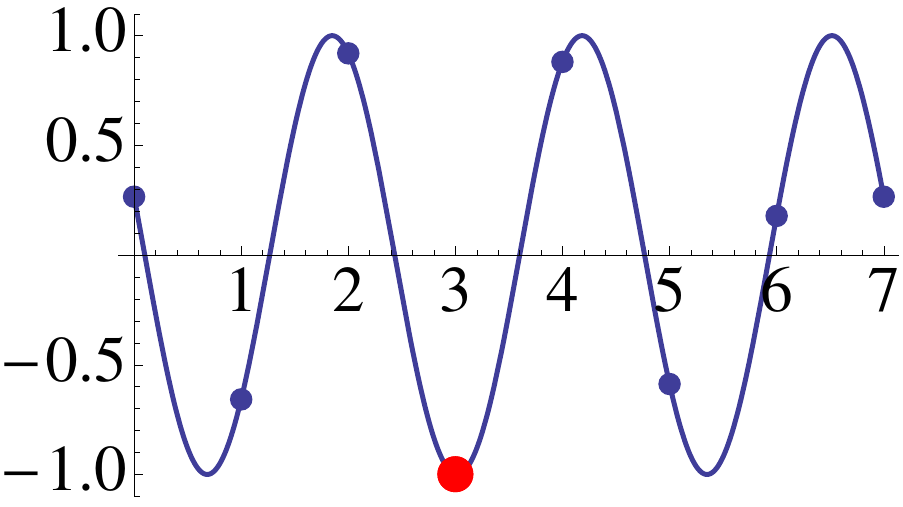}
\put(6,18){\color{black} (c)}
\end{overpic}
\begin{overpic}[scale=0.4]{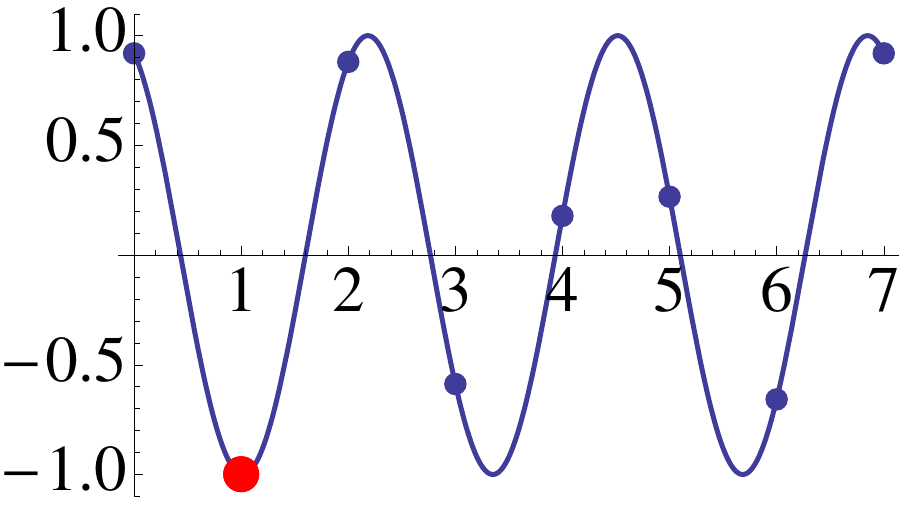}
\put(6,18){\color{black} (d)}
\end{overpic}
\caption{Cosine potential $-\cos(2\pi\phi n -k_y)$ for $\phi=3/7$ and $k_y = \{0, 2\pi/7, 4\pi/7, 6\pi/7\}$ for (a), (b), (c), and (d), respectively (we have actually chosen $k_y$ to be slightly greater than these values so that the site energies are not degenerate). The solid line shows the background cosine potential, whilst the points points give the potential energy on each lattice site. The weight of the lowest band is centred on the lattice point with the lowest potential energy, and is indicated by the large red points. This hops to the left by two sites whenever $k_y$ is changed by $2\pi/7$. When $k_y$ has changed by $2\pi$, the weight of the lowest band will have traversed 2 complete unit cells to left, corresponding to a Chern number of $-2$.}\label{tknnchern}
\end{figure}

Owing to the shape of the cosine potential and the fact that this is `sampled' by the discrete lattice sites, the weight may jump across several lattice sites in the unit cell or even into a neighbouring unit cell, as shown in Figure \ref{tknnchern}. Once $k_y$ has been increased by $2\pi$, however, the weight of the band must be back on the lattice site it started on. The Chern number is given by the number of complete unit cells traversed in the $x$-direction by the dominant weight of the $r$th band during this change in $k_y$.

\begin{figure}[t]
\includegraphics[scale=0.42]{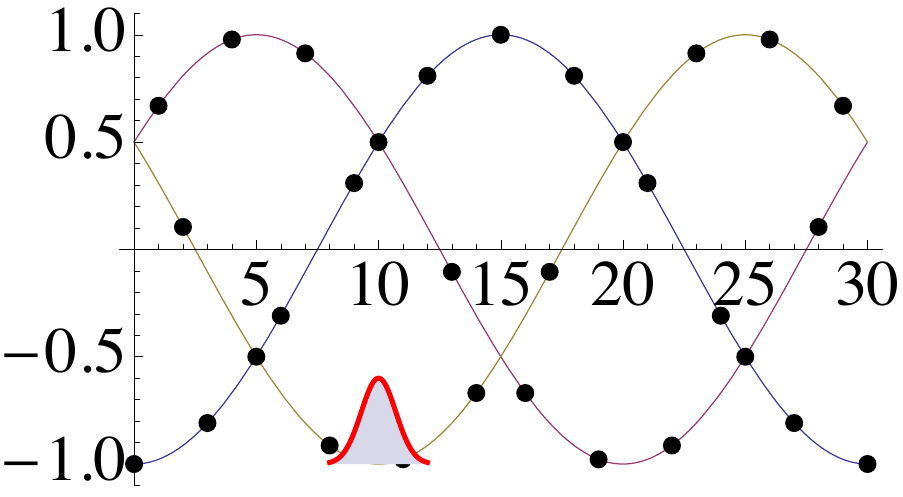}
\includegraphics[scale=0.42]{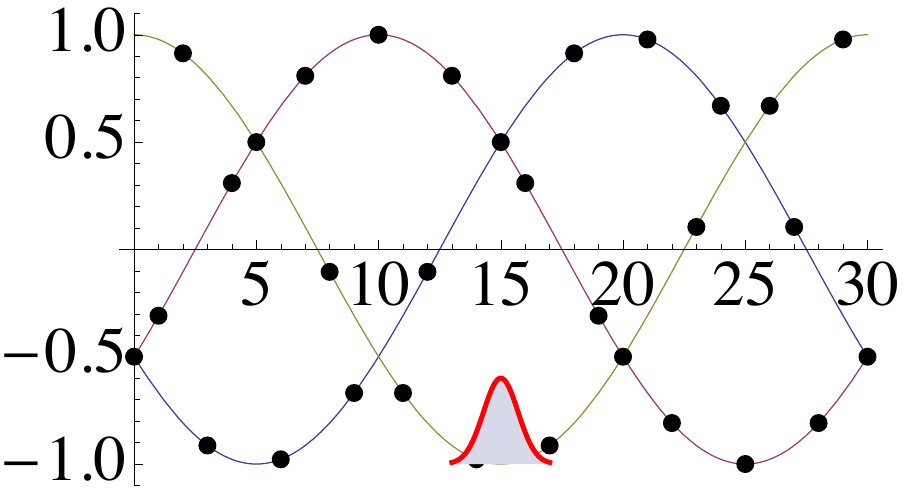}
\caption{Effective potentials and one perturbative wavefunction for $\phi=1/3+1/30$ and (left) $k=0$, (right) $k=\pi/3$. As we pump $k_y$, the wavefunction weight moves monotonically to the right. When $k_y=2\pi$ the wavefunction weight will have traversed one complete unit cell, corresponding to a Chern number of 1.}\label{tknnchern2}
\end{figure}

In Section~\ref{genflux} we assumed the oscillator-like states were isolated and saw that a change in $k_y$ acted as a pure translation in the $x$-direction. If we focus on the wavefunction near to $n=0$ and follow it as we increase $k_y$ from $0$ to $2\pi$, it moves uniformly in one direction (even though it has different amplitudes on different sublattices). This motion is to the right (left) if the plus (minus) sign is chosen in the fraction $\phi=P/Q\pm M/N$ (see Figure~\ref{tknnchern2}). When $k_y$ reaches $2\pi$, the weight of the wavefunction will have moved a total of $N/M$ lattice sites.

The Chern number this corresponds to depends on the magnetic unit cell size, which in turn depends on the cancellation in the fraction $\phi$. For the three cancellation possibilities considered in Eq.~\eqref{cancellation} with $Q$ prime and $M=1$, we find that in case (a) there is no cancellation and the unit cell is $QN$ lattice sites long. A translation of $N$ units then corresponds to a Chern number of $\pm 1/Q$. There are $Q^2$ mini-bands (spatial centres) in a subband, and so the total Chern number of the subband is $\pm Q$. In case (b) the unit cell size is $N$ and we find $Q$ mini-bands each with Chern number $\pm1$. Finally, in the maximally cancelling case (c), the unit cell size is $N/Q$ and we have a single mini-band with Chern number $\pm Q$. In all three cases, the total Chern number of the subband ($\pm Q$) is shared equally between the component mini-bands---this is also true for general $\phi=P/Q\pm M/N$ when $Q$ is not necessarily prime and $M\neq 1$. Higher order perturbative corrections do not change this picture, since the centre of mass of the wavefunction (and its position as a function of $k_y$) is unaffected by the addition of higher Landau level contributions.

Of course, the fractional Chern numbers of the mini-bands given above are unphysical and are just an artefact of our approximation. We emphasise that the set of mini-bands within a subband should be considered together and that it is the total Chern number ($\pm Q$) that is of significance: we would not be able to resolve the Chern number of the individual mini-bands as they are exponentially close in energy.

We can recover the true (but unresolvable) Chern number of each mini-band by allowing the band weight to hop between troughs in the unit cell in the same way that the band weight hops between lattice sites in Figure~\ref{tknnchern}. We leave a discussion of this to elsewhere.

Finally, we mention that a system with Chern number $C\geq2$ can in general have wavefunctions which are subtly different in character: when we go around the Brillouin zone by changing $k_y$, each species can stay the same, or the species can be interchanged. This is studied in Ref.~\onlinecite{Wu:2013ii}, where it is shown that this effect depends sensitively on the system size. In our case, in the large $N$ limit, the unit cell size is much larger than both the magnetic length and the interaction distances, so these small effects should not be important.

\section{Symmetric Gauge Two-particle Wavefunctions\label{twoparticles}}
In the symmetric gauge we define the centre of mass and relative coordinates
\be
z_C&=&\frac{z_1+z_2}{2}\\
z_R&=&z_1-z_2\\
\partial_{z_C}&=&\partial_{z_1}+\partial_{z_2}\\
\partial_{z_R}&=&\frac{\partial_{z_1}-\partial_{z_2}}{2}\\
\ee
and magnetic lengths
\be
l_{R}&=&\sqrt{2}l_B\\
l_{C}&=&\frac{l_B}{\sqrt{2}}.
\ee
The conjugate variables are defined similarly, but we note that $\left(\partial_z\right)^\dagger=-\partial_{\bar{z}}$. In terms of these new coordinates, we define Landau level raising and lowering operators,
\be
\hat{Z}^\dagger&=&\sqrt{2}\left(\frac{1}{4l_C}\bar{z}_C-l_C\partial_{z_C}\right)\\
\hat{Y}^\dagger&=&\sqrt{2}\left(\frac{1}{4l_R}\bar{z}_R-l_R\partial_{z_R}\right)
\ee
etc. These are analogous to the standard Landau level raising and lowering operators, and are related to the single particle operators through
\be
a_1^\dagger&=&\frac{1}{\sqrt{2}}\left(\hat{Z}^\dagger+\hat{Y}^\dagger\right)\\
a_2^\dagger&=&\frac{1}{\sqrt{2}}\left(\hat{Z}^\dagger-\hat{Y}^\dagger\right).\\
\ee
We also define the centre of mass and relative momentum raising and lowering operators
\be
\hat{M}^\dagger&=&\sqrt{2}\left(\frac{1}{4l_C}z_C-l_C\partial_{\bar{z}_C}\right)\\
\hat{L}^\dagger&=&\sqrt{2}\left(\frac{1}{4l_R}z_R-l_R\partial_{\bar{z}_R}\right).\\
\ee
In terms of these, a general two-particle state then takes the form
\begin{widetext}
\be
\ket{N_L,N_M;N_Z,N_Y}=\frac{\left(\hat{Z}^\dagger\right)^{N_Z}}{\sqrt{N_Z!}}\frac{\left(\hat{Y}^\dagger\right)^{N_Y}}{\sqrt{N_Y!}}\frac{\left(\hat{L}^\dagger\right)^{N_L}}{\sqrt{N_L!}}\frac{\left(\hat{M}^\dagger\right)^{N_M}}{\sqrt{N_M!}}\frac{1}{2\pi l_Rl_C}\exp\left[-\frac{1}{4l_C^2}|z_C|^2-\frac{1}{4l_R^2}|z_R|^2\right],
\ee
which can be written in terms of Laguerre polynomials through
\be
\left\langle z_R,\bar{z}_R;z_C,\bar{z}_C\right.\ket{N_L,N_M;N_Z,N_Y}&=&(-1)^{N_Y}\sqrt{\frac{N_Y!}{2\pi l_R^2 N_L!}}\left(\frac{z_R}{\sqrt{2}l_R}\right)^{N_L-N_Y}L_{N_Y}^{N_L-N_Y}\left[\frac{\bar{z}_Rz_R}{2l_R^2}\right] e^{-\frac{\left|z_R\right|^2}{4l_R^2}}\times\\
&&(-1)^{N_Z}\sqrt{\frac{N_Z!}{2\pi l_C^2 N_M!}}\left(\frac{z_C}{\sqrt{2}l_C}\right)^{N_M-N_Z}L_{N_Z}^{N_M-N_Z}\left[\frac{\bar{z}_Cz_C}{2l_C^2}\right]e^{-\frac{\left|z_C\right|^2}{4l_C^2}}.
\ee
In the text we are interested in the overlap integral $\bra{N_L',N_M';N_Y',N_Z'}\delta(z_R)\ket{N_L,N_M;N_Y,N_Z}$, where the $\delta$ function only affects the $z_R$-dependent terms. Integration over $z_C,\bar{z}_C$ just leads to the usual orthonormality condition,
\be
\bra{N_L',N_M';N_Y',N_Z'}\delta(z_R)\ket{N_L,N_M;N_Y,N_Z}&=&\delta_{N_Z,N_Z'}\delta_{N_M,N_M'}\int\mathrm{d}z_R\mathrm{d}\bar{z}_R\,(-1)^{N_Y}\sqrt{\frac{N_Y!}{2\pi l_R^2 N_L!}}\left(\frac{z_R}{\sqrt{2}l_R}\right)^{N_L-N_Y}\times\\
&&L_{N_Y}^{N_L-N_Y}\left[\frac{\bar{z}_Rz_R}{2l_R^2}\right] e^{-\frac{\bar{z}_Rz_R}{4l_R^2}}(-1)^{N_Y'}\sqrt{\frac{N_Y'!}{2\pi l_R^2 N_L'!}}\left(\frac{\bar{z}_R}{\sqrt{2}l_R}\right)^{N_L'-N_Y'}\times\\
&&L_{N_Y'}^{N_L'-N_Y'}\left[\frac{\bar{z}_Rz_R}{2l_R^2}\right] e^{-\frac{\bar{z}_Rz_R}{4l_R^2}}\delta(x_R)\delta(y_R).
\ee
For the remaining terms to be non-zero we must have $N_L=N_Y$ and $N_L'=N_Y'$---otherwise the factors in round brackets will vanish when the delta function is enacted. Then,
\be
\bra{N_L',N_M';N_Y',N_Z'}\delta(z_R)\ket{N_L,N_M;N_Y,N_Z}&=&\delta_{N_Z,N_Z'}\delta_{N_M,N_M'}\delta_{N_Y,N_L}\delta_{N_Y',N_L'}\times\\
&&(-1)^{N_Y}\sqrt{\frac{1}{2\pi l_R^2}}L_{N_Y}^{0}(0)(-1)^{N_Y'}\sqrt{\frac{1}{2\pi l_R^2}}L_{N_Y'}^{0}(0)\\
&=&\delta_{N_Z,N_Z'}\delta_{N_M,N_M'}\delta_{N_Y,N_L}\delta_{N_Y',N_L'}\frac{(-1)^{N_Y+N_Y'}}{4\pi l_B^2},
\ee
\end{widetext}
where in the last line we have used the property of Laguerre polynomials that
\be
L_n^\alpha(0)&=&\left(\begin{array}{c}
n+\alpha\\
n
\end{array}\right)\\
&=&\frac{(n+\alpha)!}{n!\alpha!}.
\ee
Finally, we remark that in general we calculate overlap integrals between particular components of the single particle wavefunctions. If a single particle wavefunction is split into $Q$ components, integration over one component only samples $1/Q$ of space and (with lattice spacing fixed at one) the integration should only return $1/Q$ of its usual value.
\section{Further Details on the Symmetric Gauge Pseudopotentials\label{symmpseudo}}
We showed in Section~\ref{symmgauge} that we could transform between Landau gauge and symmetric gauge states using the functions $B_m(k)$. We mention briefly that we can equivalently transform the interaction matrix elements themselves through
\begin{widetext}
\be
V_{s_1s_2s_3s_4}^{m_1m_2m_3m_4}&=&\int\mathrm{d}k_{y1}'\mathrm{d}k_{y2}'\mathrm{d}k_{y3}'\mathrm{d}k_{y4}'B_{m_1}(k_{y1}')B_{m_2}(k_{y2}')B_{m_3}(k_{y3}')B_{m_4}(k_{y4}')V_{s_1s_2s_3s_4}^{\mathbf{k}_1\mathbf{k}_2\mathbf{k}_3\mathbf{k}_4}.
\ee
To find the pseudopotentials $V^{LL'}$ from these, we can use the appropriate Clebsch-Gordan coefficients,
\be
D_{m_1m_2}^{LM}&=&\bra{L,M}\left.\!m_1,m_2\right\rangle=\delta_{m_1+m_2,L+M}\sqrt{\frac{m_1!m_2!}{2^{L+M}L!M!}}\sum^L_{\alpha=0}(-1)^\alpha\left(\begin{array}{c}
L\\
\alpha
\end{array}\right)\left(\begin{array}{c}
M\\
m_1-\alpha
\end{array}\right).
\ee
\end{widetext}
In practice it is easiest to interpret the wavefunction perturbation series in the symmetric gauge and compute the overlap integral in Eq.~\eqref{pseudosym}, making sure that $\Sigma s$ is conserved modulo $Q$.

As an example, we calculate the zeroth and first order pseudopotential matrix elements for $\phi=1/2+1/N$ with $s_1=s_2=1$ and $s_3=s_4=0$, i.e. $V^{LL'}_{1100}$. We see from Appendix \ref{example_wavefunctions} that the two possible perturbation series are given by the unitary operators
\be
U^\dagger_0&=&1+\pionn\left(\left(\frac{1+\sqrt{2}}{4}\right)\ad{2}+\frac{1}{24}\ad{4}\right)\\
U^\dagger_1&=&1+\pionn\left(\left(\frac{1-\sqrt{2}}{4}\right)\ad{2}+\frac{1}{24}\ad{4}\right).
\ee
The two possible LLL one-particle wavefunctions are then
\be
\ket{\tilde{0};0}&=&\sqrt{\frac{2-\sqrt{2}}{2}}\left[\delta_{x,0}^{(2)}U^\dagger_0+(1+\sqrt{2})\delta_{x,1}^{(2)}U^\dagger_1\right]\ket{0}\\
\ket{\tilde{0};1}&=&\sqrt{\frac{2-\sqrt{2}}{2}}\left[(1+\sqrt{2})\delta_{x,0}^{(2)}U^\dagger_1+\delta_{x,1}^{(2)}U^\dagger_0\right]\ket{0},
\ee
where the kets on the left have perturbed Landau level index $\tilde{l}$ and species index $s$ labelled through $\ket{\tilde{l};s}$, and the kets on the right are labelled only by unperturbed Landau level index $l$ (we have suppressed the angular momentum quantum numbers).

For $V^{LL'}_{1100}$, the initial two particle state is formed from two single-particle states with $s_1=s_2=0$,
\be
&&\ket{\tilde{0};0}\otimes\ket{\tilde{0};0}\\
&&=\left(\frac{2-\sqrt{2}}{2}\right)\bigg\{\left[\delta_{x_1,0}^{(2)}U^\dagger_{0}+(1+\sqrt{2})\delta_{x_1,1}^{(2)}U^\dagger_{1}\right]\otimes\\
&&\left[\delta_{x_2,0}^{(2)}U^\dagger_{0}+(1+\sqrt{2})\delta_{x_2,1}^{(2)}U^\dagger_{1}\right]\bigg\}\ket{0}\otimes\ket{0}.
\ee
Here the left factors of the tensor products refer to particle 1 and the right factors to particle 2. As we will eventually be enacting a delta function which sets $x_1=x_2$, the cross terms in the above expression will vanish. Therefore, keeping only diagonal terms to first order, we have
\begin{widetext}
\be
\ket{\tilde{0},0}\otimes\ket{\tilde{0};0}&=&\left(\frac{2-\sqrt{2}}{2}\right)\left[\left\{1+\pionn\left(\frac{1+\sqrt{2}}{4}\right)\left(\adon{2}+\adtw{2}\right)+\frac{1}{24}\left(\adon{4}+\adtw{4}\right)\right\}\delta_{x_1,0}^{(2)}\delta_{x_2,0}^{(2)}\right.\\
&&\left.+(3+2\sqrt{2})\left\{1+\pionn\left(\frac{1-\sqrt{2}}{4}\right)\left(\adon{2}+\adtw{2}\right)+\frac{1}{24}\left(\adon{4}+\adtw{4}\right)\right\}\delta_{x_1,1}^{(2)}\delta_{x_2,1}^{(2)}\right]\\
&&\ket{0}\otimes\ket{0}.
\ee
The operator subscripts indicate which particle (1 or 2) is being raised in Landau level. Next we convert to the relative and centre of mass Landau level raising operators $\{a^\dagger_1,a^\dagger_2\}\to\{Y^\dagger,Z^\dagger\}$, and write the states in this basis through $\ket{\tilde{l}_1;s_1}\otimes\ket{\tilde{l}_2;s_2}\to\ket{N_L,N_M;\tilde{N}_Y,\tilde{N}_Z;s_1;s_2}$. We have also now included relative and centre of mass angular momentum labels, $N_L$ and $N_M$. The perturbation series is valid for any angular momentum state, so we leave $N_L$ and $N_M$ unspecified.
\be
\ket{N_L,N_M;\tilde{0},\tilde{0};0,0}&=&\left(\frac{2-\sqrt{2}}{2}\right)\left[\left\{1+\pionn\left(\frac{1+\sqrt{2}}{4}\right)\left(\zup{2}+\yup{2}\right)\right.\right.\\
&&\left.+\frac{1}{48}\left(\zup{4}+6\zup{2}\yup{2}+\yup{4}\right)\right\}\delta_{x_1,0}^{(2)}\delta_{x_2,0}^{(2)}\\
&&+(3+2\sqrt{2})\left\{1+\pionn\left(\frac{1-\sqrt{2}}{4}\right)\left(\zup{2}+\yup{2}\right)\right.\\
&&\left.\left.+\frac{1}{48}\left(\zup{4}+6\zup{2}\yup{2}+\yup{4}\right)\right\}\delta_{x_1,1}^{(2)}\delta_{x_2,1}^{(2)}\right]\ket{N_L,N_M;0,0}.
\ee
Enacting the operators leaves us with
\be
\ket{N_L,N_M;\tilde{0},\tilde{0};0,0}&=&\left(\frac{2-\sqrt{2}}{2}\right)\left[\left\{\ket{N_L,N_M;0,0}+\pionn\bigg(\frac{\sqrt{2}+2}{4}\right)\left(\ket{N_L,N_M;0,2}+\ket{N_L,N_M;2,0}\bigg)\right.\right.\\
&&\left.+\frac{1}{48}\bigg(\sqrt{24}\ket{N_L,N_M;0,4}+12\ket{N_L,N_M;2,2}+\sqrt{24}\ket{N_L,N_M;4,0}\bigg)\right\}\delta_{x_1,0}^{(2)}\delta_{x_2,0}^{(2)}\\
&&+(3+2\sqrt{2})\left\{\ket{N_L,N_M;0,0}+\pionn\bigg(\frac{\sqrt{2}-2}{4}\right)\left(\ket{N_L,N_M;0,2}+\ket{N_L,N_M;2,0}\bigg)\right.\\
&&\left.\left.+\frac{1}{48}\bigg(\sqrt{24}\ket{N_L,N_M;0,4}+12\ket{N_L,N_M;2,2}+\sqrt{24}\ket{N_L,N_M;4,0}\bigg)\right\}\delta_{x_1,1}^{(2)}\delta_{x_2,1}^{(2)}\right].
\ee
Similarly, the final two-particle state is given by
\be
\bra{N_L,N_M;\tilde{0},\tilde{0};1,1}&=&\left(\frac{2-\sqrt{2}}{2}\right)\left[(3+2\sqrt{2})\left\{\bra{N_L,N_M;0,0}\right.\right.\\
&&+\pionn\bigg(\frac{\sqrt{2}-2}{4}\bigg)\bigg(\bra{N_L,N_M;0,2}+\bra{N_L,N_M;2,0}\bigg)\\
&&\left.+\frac{1}{48}\bigg(\sqrt{24}\bra{N_L,N_M;0,4}+12\bra{N_L,N_M;2,2}+\sqrt{24}\bra{N_L,N_M;4,0}\bigg)\right\}\delta_{x_1,0}^{(2)}\delta_{x_2,0}^{(2)}\\
&&+\left\{\bra{N_L,N_M;0,0}+\pionn\bigg(\frac{\sqrt{2}+2}{4}\bigg)\bigg(\bra{N_L,N_M;0,2}+\bra{N_L,N_M;2,0}\bigg)\right.\\
&&\left.\left.+\frac{1}{48}\bigg(\sqrt{24}\bra{N_L,N_M;0,4}+12\bra{N_L,N_M;2,2}+\sqrt{24}\bra{N_L,N_M;4,0}\bigg)\right\}\delta_{x_1,1}^{(2)}\delta_{x_2,1}^{(2)}\right].
\ee
Finally, we calculate the delta function overlap integral between these two states using Equation \ref{overlap_integral}, (including the factor of $1/2$ since each integral is over only half of space). We now also enact the modulo-2 Kronecker deltas which would remove the cross-terms we have chosen to ignore. We set $N_M=N_M'=0$ without loss of generality and keep terms only to first order:
\be
\bra{N_L,0;\tilde{0},\tilde{0};1,1}\delta\left(z_R\right)\ket{N_L',0;\tilde{0},\tilde{0};0,0}&=&\frac{1}{8\pi l_B^2}\left[\delta_{N_L0}\delta_{N_L'0}+\pionn\left(\frac{\sqrt{2}}{4}\delta_{N_L'0}\delta_{N_L2}+\frac{\sqrt{2}}{4}\delta_{N_L'2}\delta_{N_L0}\right.\right.\\
&&\left.\left.+\frac{\sqrt{6}}{24}\delta_{N_L'0}\delta_{N_L4}+\frac{\sqrt{6}}{24}\delta_{N_L'4}\delta_{N_L0}\right)\right]+O\pion{2}.
\ee
\end{widetext}
This pseudopotential matrix then looks at this order like
\be
V_{1100}^{LL'}&=&\frac{1}{8\pi l_B^2}\left(
\renewcommand\arraystretch{1.5}\begin{array}{ccc}
1 & \frac{\sqrt{2}}{4}\left(\pi\delta\right) &  \frac{\sqrt{6}}{24}\left(\pi\delta\right) \\
\frac{\sqrt{2}}{4}\left(\pi\delta\right) & 0 & 0 \\
\frac{\sqrt{6}}{24}\left(\pi\delta\right) & 0 & 0
\end{array}\right)
\ee
where we have only included even rows and columns and where $\delta=1/N$.

\section{Summary of energy and wavefunction corrections for other lattice types\label{otherlattice}}
\subsection{Anisotropic square lattice}
\begin{figure}[h]

\includegraphics[scale=0.5]{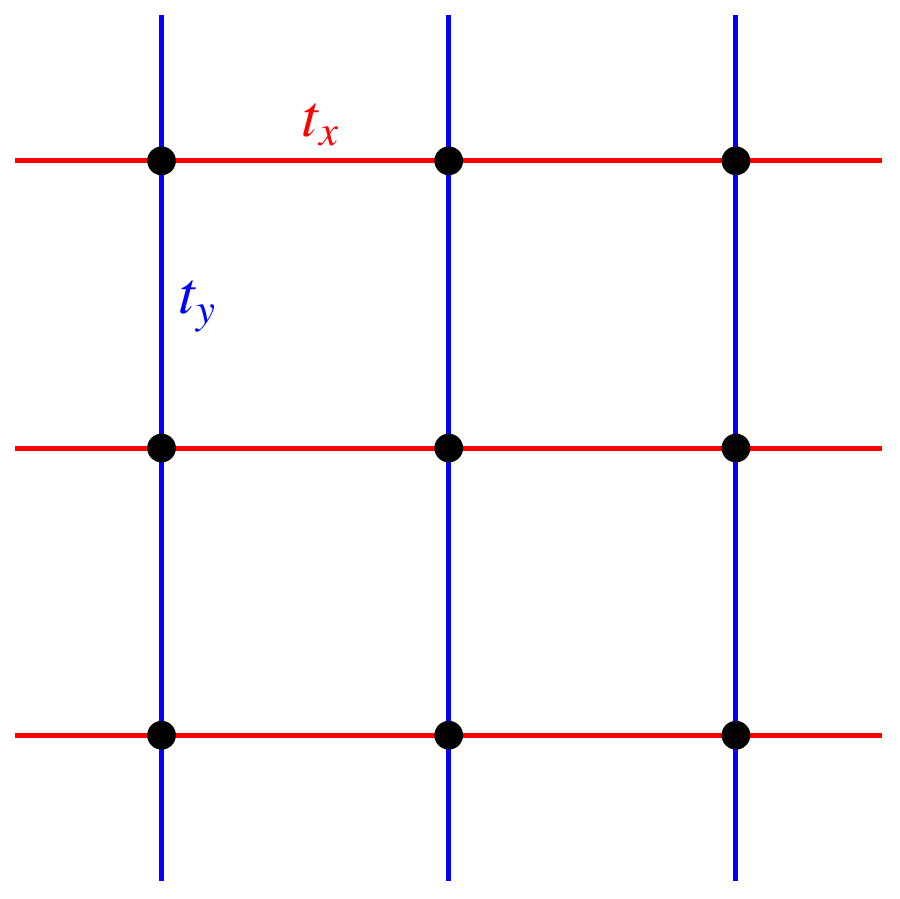}

\caption{Anisotropic square lattice.}\label{sqanis}
\end{figure}
On the anisotropic square lattice with hopping parameters $t_x$ and $t_y$ (see Fig.\ref{sqanis}), the discrete Schr\"{o}dinger equation for the symmetric gauge may be written as
\be
\epsilon \psi(m,n)&=&-t_xe^{i\pi\phi n}\psi(m+1,n)-t_xe^{-i\pi\phi n }\psi(m-1,n)\\
&&-t_ye^{-i\pi\phi m}\psi(m,n+1)-t_ye^{i\pi\phi m}\psi(m,n-1).
\ee
If we expand this in the $\phi=1/N$ limit, we may write the ladder operators
\be
a^\dagger&=&\sqrt{\frac{\omega}{2}}z^*-\frac{\partial_{z}}{\sqrt{2\omega}}
\ee
with $\omega=\sqrt{2\pi/N}$ as before but now
\be
z&=&\frac{1}{\sqrt{2}l_B}\sqrt[4]{\frac{t_y}{t_x}}m+i\sqrt[4]{\frac{t_x}{t_y}}n.
\ee
This leads to the energy corrections
\be
\epsilon_l&=&-2(t_x+t_y)+\frac{4\pi\sqrt{t_xt_y}}{N}\left(l+\frac{1}{2}\right)\\
&&-\frac{t_x+t_y}{4}\pion{2}\left(1+2l+2l^2\right)
\ee
and the wavefunction corrections
\be
&&\ket{\tilde{l}}=\ket{l}\\
&&-\frac{1}{192}\pionn\frac{t_x+t_y}{\sqrt{t_xt_y}}\left[\sqrt{l(l-1)(l-2)(l-3)}\ket{l-4}\right.\\
&&\left.-\sqrt{(l+1)(l+2)(l+3)(l+4)}\ket{l+4}\right]\\
&&-\frac{1}{16}\pionn\frac{t_x-t_y}{\sqrt{t_xt_y}}\left[\sqrt{l(l-1)}\ket{l-2}\right.\\
&&\left.-\sqrt{(l+1)(l+2)}\ket{l+2}\right]\\
&&-\frac{1}{24}\pionn\frac{t_x-t_y}{\sqrt{t_xt_y}}\left[(l-2)\sqrt{l(l-1)}\ket{l-2}\right.\\
&&\left.-l\sqrt{(l+1)(l+2)}\ket{l+2}\right].
\ee
With anisotropic hopping, the wavefunction takes on the new symmetry of the system, which reduces from $C_4$ to $C_2$. The isotropic case can be recovered if we set $t_x=t_y$. We note that if we take the hopping amplitude in one direction to be much larger than the amplitude in the other direction (by a factor $O(N)$), then our perturbation theory becomes invalid. 
\subsection{Square lattice with next-nearest-neighbour hopping}
\begin{figure}[h]

\includegraphics[scale=0.45]{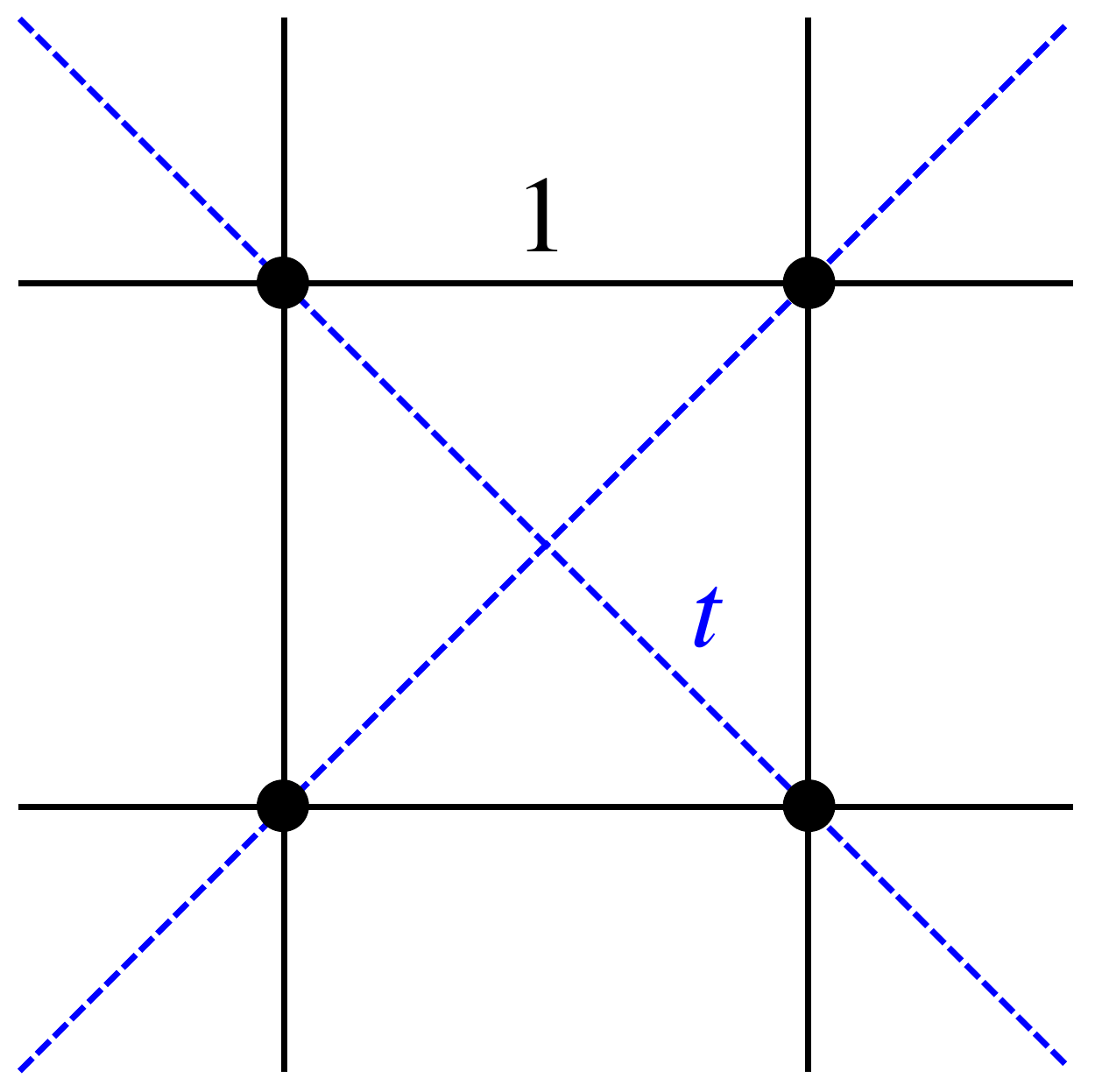}

\caption{Section of square lattice showing both types of hopping.}\label{squarennn}
\end{figure}

We also consider next-nearest-neighbour hopping on the square lattice, defining the horizontal and vertical hopping amplitude to be one and the diagonal hopping amplitude to be $t$ (as in Fig.~\ref{squarennn}).
The discrete difference equation in the Landau gauge (with $\mathbf{k}=0$) is given by 
\be
\epsilon \psi(n)&=&-\psi(n-1)-\psi(n+1)-2\cos\left(2\pi m\phi\right)\psi(n)\\
&&-2t\cos\left(2\pi n\phi +\pi\phi\right)\psi(n+1)\\
&&-2t\cos\left(2\pi n\phi-\pi\phi\right)\psi(n-1).
\ee
Expanding as usual, we find that the energy bands are
\be
\epsilon_l&=&-4(1+t)+4(1+2t)\pionn\left(l+\frac{1}{2}\right)\\
&&-\pion{2}\left(1+4t\right)\left(\frac{1}{2}+l+l^2\right)
\ee
and the wavefunction corrections are given by
\be
&&\ket{\tilde{l}}=\ket{l}-\frac{1}{96}\pionn\frac{1-4t}{1+2t}\left[\sqrt{l(l-1)(l-2)(l-3)}\ket{l-4}\right.\\
&&\left.+\sqrt{(l+1)(l+2)(l+3)(l+4)}\ket{l+4}\right].
\ee
We note that increasing $t$ from zero only decreases the relative wavefunction correction, so we expect the pseudopotential corrections to be even smaller than they are for the ordinary square lattice---the results may be interesting when $t$ is negative, however. We recover the original square lattice case if we set $t=0$.
\subsection{Triangular lattice}
\begin{figure}[h]
\includegraphics[scale=0.6]{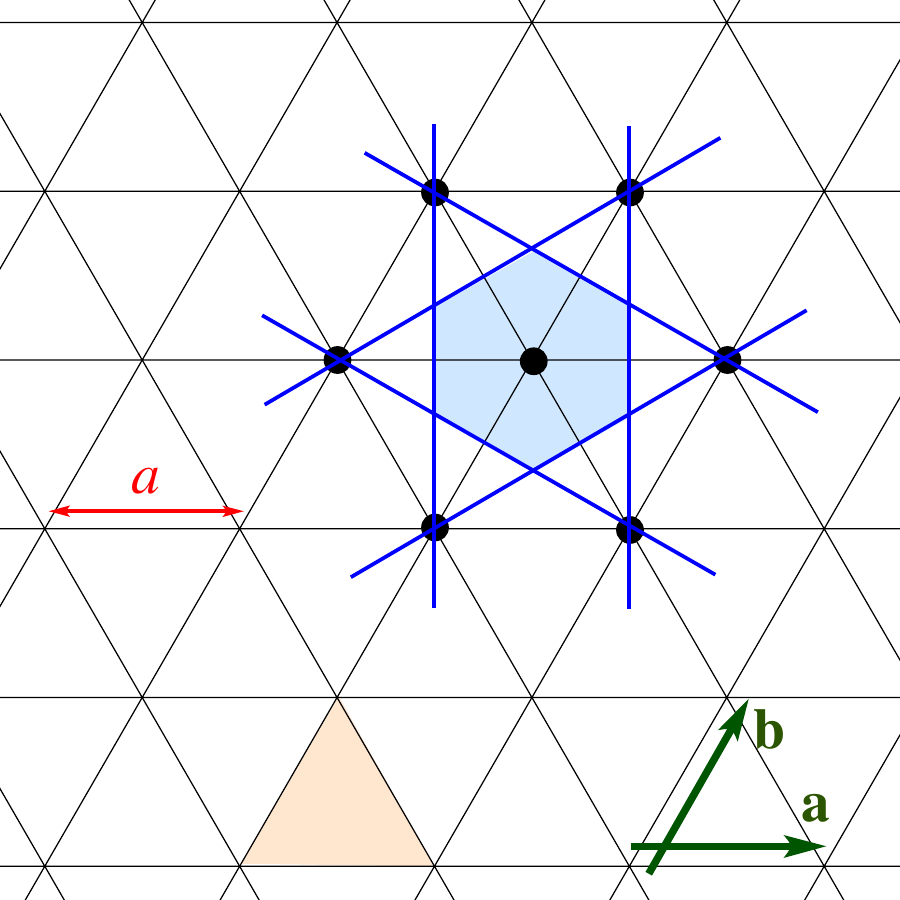}
\caption{Triangular lattice with Wigner-Seitz Cell shaded in blue and a single plaquette shaded in red. We set the lattice spacing $a=1$.}\label{triangle}
\end{figure}

For the triangular lattice we define $\phi=1/N$ to be the magnetic flux per unit cell rather than per plaquette, as shown in Figure \ref{triangle}.

In the Landau gauge, the discrete difference equation is
\be
\epsilon \psi(n)&=&-\psi(n-1)-\psi(n+1)\\
&&-2\cos\left(\pi\phi\left(n+\frac{1}{2}\right)+\frac{\sqrt{3}k_y}{2}\right)\psi(n+1/2)\\
&&-2\cos\left(\pi\phi\left(n-\frac{1}{2}\right)+\frac{\sqrt{3}k_y}{2}\right)\psi(n-1/2).
\ee
If we substitute $\phi=p/q$, we notice that this is cyclic under the substitution $n\rightarrow n+q$ if $p$ is even, and it is cyclic under the substitution $n\rightarrow n+2q$ if $p$ is odd. Bloch periodicity therefore requires that
\be
\left.\begin{array}{c}
\ket{n+q}\\
\ket{n+2q}
\end{array}\right\}=\left\{\begin{array}{c}
e^{ik_xq/2}\ket{n}\\
e^{ik_xq}\ket{n}
\end{array}\right.\,\,\,\,\,\,\,\,\,\begin{array}{c}
p\mbox{ even}\\
p\mbox{ odd}
\end{array}
\ee
Setting $\phi=1/N$ and expanding as usual, we find that the characteristic frequency is now $\omega=2\pi/(\sqrt{3}N)$ and the energy bands are given by
\be
\epsilon_l&=&-6+4\sqrt{3}\left(\frac{\pi}{N}\right)\left(l+\frac{1}{2}\right)-\pion{2}\left(2l^2+2l+1\right).
\ee
There is no correction to the wavefunction at first order, but at second order we find
\be
\ket{\tilde{l}}&=&\ket{l}-\frac{1}{3240}\pion{2}\left[\ad{6}-\an{6}\right]\ket{l}
\ee
The addition of $\ket{n\pm6}$ allows the wavefunction to adopt the six-fold symmetry of the lattice---but we notice that these corrections come with a factor of $1/N^2$ and so will be highly suppressed. In this sense, the triangular lattice is a better approximation to the continuum than the square lattice.

We find that the hexagonal lattice gives the same results as the triangular lattice but with $\omega=4\pi/(3\sqrt{3}N)$ and with two copies of each wavefunction per unit cell. 
\begin{widetext}
\section{Derivation of Formula for Site-Site Interactions\label{genint}}
We begin with the expression for a pseudopotential element in terms of Landau gauge wavefunctions
\be
V^{LL',u+iv}_{s_1s_2s_3s_4}&=&\sum_{m_1m_2m_3m_4}D^{L*}_{m_1m_2}D^{L'}_{m_3m_4}\int\mathrm{d}x_1\mathrm{d}x_2\mathrm{d}y_1\mathrm{d}y_2\,\delta\left(x_1-x_2-u\right)\delta\left(y_1-y_2-v\right)\times\\
&&e^{-2\pi i{s_1}y_1/Q}e^{-2\pi i{s_2}y_2/Q}e^{2\pi i{s_3}y_1/Q}e^{2\pi i{s_4}y_2/Q}\int\{\mathrm{d}k\}B_{m_1}(k_1')B_{m_2}(k_2')B_{m_3}(k_3')B_{m_4}(k_4')\times\\
&&e^{ik_{1}'y_1}e^{ik_{2}'y_2}e^{-ik_{3}'y_2}e^{-ik_{4}'y_1}\left[\tilde{\psi}_{0,k_{1}'}^{s_1}(x_1)\right]^*\left[\tilde{\psi}_{0,k_{2}'}^{s_2}(x_2)\right]^*\tilde{\psi}_{0,k_{3}'}^{s_3}(x_1)\tilde{\psi}_{0,k_{4}'}^{s_4}(x_2)\\
\ee
where the definitions follow from Section~\ref{symmgauge} and the initial sum is over the appropriate Clebsch-Gordan coefficients $D_{m_1m_2}^{L0}$ defined in Appendix~\ref{symmpseudo} . The final integral transforms to the symmetric gauge, giving
\be
V^{LL',u+iv}_{s_1s_2s_3s_4}&=&\sum_{m_1m_2m_3m_4}D^{L*}_{m_1m_2}D^{L'}_{m_3m_4}\int\mathrm{d}x_1\mathrm{d}x_2\mathrm{d}y_1\mathrm{d}y_2\delta\left(x_1-x_2-u\right)\delta\left(y_1-y_2-v\right)\times\\
&&e^{2\pi iy_1({s_3}-{s_1})/Q}e^{2\pi iy_2({s_4}-{s_2})/Q}\left[\psi_{0,m_1}^{s_1}(z_1)\right]^*\left[\psi_{0,m_2}^{s_2}(z_2)\right]^*\psi_{0,m_3}^{s_3}(z_1)\psi_{0,m_4}^{s_4}(z_2).
\ee
We convert to relative coordinates and fix $s_1+s_2=(s_3+s_4)\mod{Q}$ to give
\be
V^{LL',u+iv}_{s_1s_2s_3s_4}&=&\int\mathrm{d}x_R\mathrm{d}x_C\mathrm{d}y_R\mathrm{d}y_C\delta\left(x_R-u\right)\delta\left(y_R-v\right)e^{2\pi iy_C\left(s_3+s_4-s_1-s_2\right)/Q}e^{\pi iy_R\left(s_3-s_4-s_1+s_2\right)/Q}\times\\
&&\left[\psi_L^{s_1s_2}(z_R,z_C)\right]^*\delta_{x_1,x_2+u}\psi_{L'}^{s_3s_4}(z_R,z_C)\\
&=&e^{-\pi iv\left(s_3-s_4-s_1+s_2\right)/Q}\int\mathrm{d}x_C\mathrm{d}y_C\,\left[\psi_L^{s_1s_2}(u+iv,z_C)\right]^*\delta_{\lambda_1,\lambda_2+u}\psi_{L'}^{s_3s_4}(u+iv,z_C).
\ee
Finally, we introduce the translation operators to give the final expression
\be
V^{LL',u+iv}_{s_1s_2s_3s_4}&=&e^{\pi i v(s_3-s_4+s_2-s_1)}\left[\hat{T}_{u+iv}U^\dagger_{s_1s_2}\ket{L',0;0,0}\right]^\dagger\delta_{\lambda_1,\lambda_2+u}\delta(z_R)\delta(\bar{z}_R)\left[\hat{T}_{u+iv}U^\dagger_{s_3s_4}\ket{L,0;0,0}\right],
\ee
where the explicit form of the $\hat{T}$ operators is given in Eq.~\eqref{tuiv}. In this way, we transfer the spatial offset from the delta function to the wavefunctions themselves.

As an example we consider the nearest neighbour interaction,
\be
\hat{V}&=&V\left[\delta_{z_2,z_1+1}+\delta_{z_2,z_1-1}+\delta_{z_2,z_1+i}+\delta_{z_2,z_1-i}\right],
\ee
for small flux $\phi=M/N\ll1$ on the square lattice. The pseudopotential matrix in this case is
\be
\frac{V}{4\pi l_B^2}\left(\renewcommand\arraystretch{1.5}\begin{array}{ccccccccc}
4-2\left(\pi\delta\right)+\frac{185}{384}\left(\pi\delta\right)^2 & 0 & 0 & 0 & \frac{\sqrt{6}}{24}\left(\pi\delta\right)+\frac{\sqrt{6}}{96}\left(\pi\delta\right)^2& 0 & 0 & 0  & \frac{\sqrt{70}}{768}\left(\pi\delta\right)^2\\
0 & 2\left(\pi\delta\right)-\left(\pi\delta\right)^2 & 0 & 0 & 0 & \frac{\sqrt{30}}{48}\left(\pi\delta\right)^2 & 0 & 0 & 0\\
0 & 0 & \frac{33}{64}\left(\pi\delta\right)^2 & 0 & 0 & 0 & 0 & 0 & 0\\
0 & 0 & 0 & 0 & 0 & 0 & 0 & 0 & 0\\
\frac{\sqrt{6}}{24}\left(\pi\delta\right)+\frac{\sqrt{6}}{96}\left(\pi\delta\right)^2 & 0 & 0 & 0  & \frac{1}{384}\left(\pi\delta\right)^2 & 0 & 0 & 0 & 0\\
0 & \frac{\sqrt{30}}{48}\left(\pi\delta\right)^2 & 0 & 0 & 0 & 0 & 0 & 0 & 0\\
0 & 0 & 0 & 0 & 0 & 0 & 0 & 0 & 0\\
0 & 0 & 0 & 0 & 0 & 0 & 0 & 0 & 0\\
\frac{\sqrt{70}}{768}\left(\pi\delta\right)^2 & 0 & 0 & 0 & 0 &0 & 0 & 0 & 0
\end{array}\right),
\ee
where $\delta=M/N$ and where all rows and columns for $L\in\{0,1,2,3,4,5,6,7,8\}$ are shown. Unlike the pure delta function interaction, the nearest neighbour interaction could be experienced by both fermions and bosons.
\end{widetext}

\clearpage
\bibliography{writeup16}
\end{document}